\documentclass[smallcondensed]{svjour3}

\smartqed

\usepackage{booktabs}   
\usepackage{subcaption,wrapfig} 
                   
\usepackage[usenames,dvipsnames]{xcolor}  
\usepackage{tikz} 
\usepackage{tikz-dimline}
\usetikzlibrary{arrows,automata,shapes} 
\tikzstyle{every picture}+=[remember picture]
              
\usepackage{stfloats}
\usepackage{fancyvrb}
\usepackage{etoc}
\usepackage{stmaryrd}
\usepackage{mathpartir}
\usepackage{amsmath}
\usepackage[noend]{algpseudocode}
\usepackage{algorithm}
\usepackage[procnames]{listings}
\usepackage{lstautogobble}

\algnewcommand{\LineComment}[1]{\Statex \(\triangleright\) #1}

\DeclareMathOperator{\dom}{dom}

\usepackage{pgfplots}
\pgfplotsset{width=7cm,compat=1.8}
\usepackage{pgfplotstable}

\DefineVerbatimEnvironment{scenario}{Verbatim}{numbers=left,xleftmargin=5mm,numbersep=2mm}

\graphicspath{{pics/}}

\usepackage{hyperref}

\newcommand{\scenic}{\textsc{Scenic}}
\newcommand{\verifai}{\textsc{VerifAI}}

\DeclareMathOperator{\erode}{erode}
\DeclareMathOperator{\dilate}{dilate}

\newcommand{\Is}{\ensuremath{:=}}
\newcommand{\Or}{\ensuremath{|} }
\newcommand{\nt}[1]{\textrm{\textit{#1}}}
\newcommand{\tm}[1]{\texttt{#1}}
\newcommand{\csl}[1]{#1,\ \textrm{\dots}}
\newcommand{\opt}[1]{\textrm{[}#1\textrm{]}}
\renewcommand{\star}[1]{\textrm{(}#1\textrm{)}$^*$}
\newcommand{\either}[2]{\textrm{(}#1 \Or #2\textrm{)}}
\newcommand{\noreqs}{\textrm{---}}

\newcommand{\sem}[1]{\llbracket#1\rrbracket}
\newcommand{\objects}{\mathcal{O}}

\newcommand{\pstate}[4]{\langle #1, #2, #3, #4 \rangle}
\newcommand{\pexpr}[1]{\langle #1, \sigma, \pi, \objects \rangle}

\newcommand{\pass}{\texttt{pass}}
\newcommand{\subst}[2]{[#1 / #2]}

\newcommand{\iou}[2]{IoU(#1,#2)}
\newcommand{\area}[1]{\nt{area}\,(#1)}

\newcommand{\vx}{\mathbf{x}}
\newcommand{\vy}{\mathbf{y}}

\newcounter{myctr}
\newenvironment{mylist}{\begin{list}{\arabic{myctr}.}
{\usecounter{myctr}
\setlength{\topsep}{1mm}\setlength{\itemsep}{0.25mm}
\setlength{\parsep}{0.1mm}
\setlength{\itemindent}{0mm}\setlength{\partopsep}{0mm}
\setlength{\labelwidth}{15mm}
\setlength{\leftmargin}{4mm}}}{\end{list}}

\newenvironment{myitemize}{\begin{list}{$\bullet$}
{\setlength{\topsep}{1mm}\setlength{\itemsep}{0.25mm}
\setlength{\parsep}{0.1mm}
\setlength{\itemindent}{0mm}\setlength{\partopsep}{0mm}
\setlength{\labelwidth}{15mm}
\setlength{\leftmargin}{4mm}}}{\end{list}}

\newcommand{\myparagraph}{\paragraph}

\journalname{Submitted}

\begin{document}

\title{\scenic{}: A Language for Scenario Specification \\ and Data Generation\thanks{Preliminary versions of this article appeared in the \emph{Proceedings of the 40th {ACM} {SIGPLAN} Conference on Programming Language Design and Implementation (PLDI 2019)}~\cite{scenic-pldi} and as an April 2018 UC Berkeley technical report~\cite{fremont-tr18}.}}

\author{Daniel J. Fremont \and Edward Kim \and Tommaso Dreossi \and Shromona Ghosh \and Xiangyu Yue \and Alberto L. Sangiovanni-Vincentelli \and Sanjit A. Seshia}

\institute{Daniel J. Fremont \at
              University of California, Santa Cruz \\
              \email{dfremont@ucsc.edu}
           \and
           Edward Kim \and Tommaso Dreossi \and Shromona Ghosh \and Xiangyu Yue \and Alberto L. Sangiovanni-Vincentelli \and Sanjit A. Seshia \at
              University of California, Berkeley
}

\date{Received: date / Accepted: date}

\maketitle

\begin{abstract}
We propose a new probabilistic programming language for the design and analysis of cyber-physical systems, especially those based on machine learning.
Specifically, we consider the problems of training a system to be robust to rare events, testing its performance under different conditions, and debugging failures.
We show how a probabilistic programming language can help address these problems by specifying distributions encoding interesting types of inputs, then sampling these to generate specialized training and test data.
More generally, such languages can be used to write environment models, an essential prerequisite to any formal analysis.
In this paper, we focus on systems like autonomous cars and robots, whose environment at any point in time is a \emph{scene}, a configuration of physical objects and agents.
We design a domain-specific language, \scenic, for describing \emph{scenarios} that are distributions over scenes and the behaviors of their agents over time.
As a probabilistic programming language, \scenic{} allows assigning distributions to features of the scene, as well as declaratively imposing hard and soft constraints over the scene.
We develop specialized techniques for sampling from the resulting distribution, taking advantage of the structure provided by \scenic{}'s domain-specific syntax.
Finally, we apply \scenic{} in a case study on a convolutional neural network designed to detect cars in road images, improving its performance beyond that achieved by state-of-the-art synthetic data generation methods.
\end{abstract}

\begin{figure*}
\centering
\includegraphics[width=0.32\textwidth]{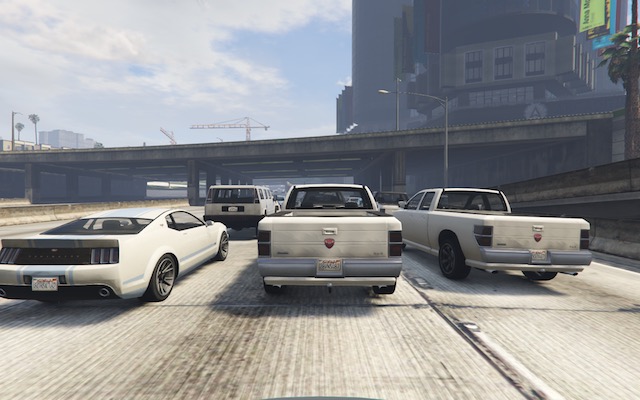}
\includegraphics[width=0.32\textwidth]{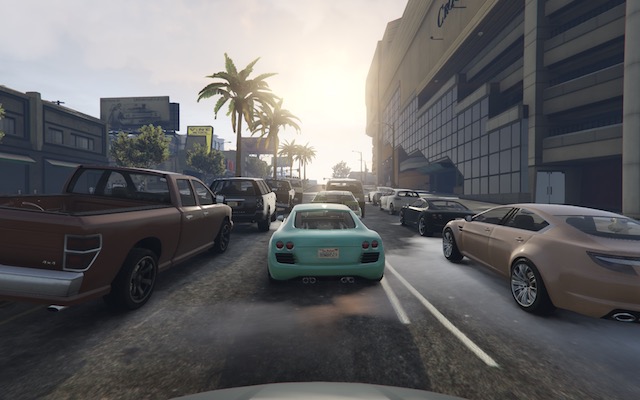}
\includegraphics[width=0.32\textwidth]{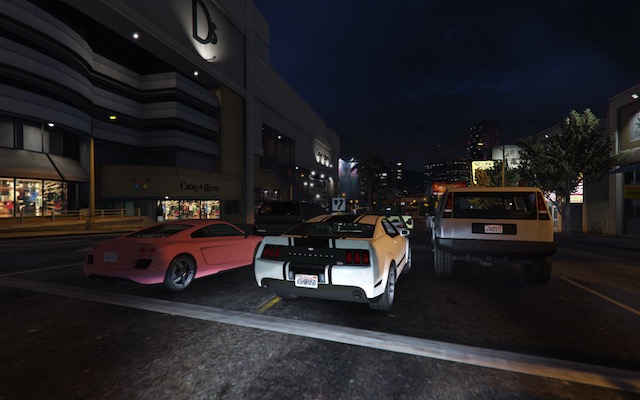}
\caption{Three scenes generated from a single $\sim$20-line \scenic{} program representing bumper-to-bumper traffic.}
\label{figure:bumper-to-bumper}
\end{figure*}

\section{Introduction}


Machine learning (ML) is increasingly used in safety-critical applications, thereby
creating an acute need for techniques to gain higher assurance in 
ML-based systems~\cite{russell2015letter,SeshiaS16,amodei2016concrete}.
ML has proved particularly effective at the difficult perceptual tasks (e.g., vision) arising in \emph{cyber-physical systems} like autonomous vehicles which operate in heterogeneous, complex physical environments.
Thus, there is a pressing need to tackle several important problems in the design of such ML-based cyber-physical systems, including:
\begin{myitemize}
\item \emph{training} the system to be robust, correctly responding to events that happen only rarely;
\item \emph{testing} the system under a variety of conditions, especially unusual ones, and
\item \emph{debugging} the system to understand the root cause of a failure and eliminate it.
\end{myitemize}
The traditional ML approach to these problems is to gather more data from the environment, retraining the system until its performance is adequate.
The major difficulty here is that collecting real-world data can be slow and expensive, since it must be preprocessed and correctly labeled before use.
Furthermore, it may be difficult or impossible to collect data for corner cases that are rare and even dangerous but nonetheless necessary to train and test against: for example, a car accident.
As a result, recent work has investigated training and testing systems with \emph{synthetically generated data}, which can be produced in bulk with correct labels and giving the designer full control over the distribution of the data~\cite{jaderberg2014synthetic,gupta2016synthetic,tobin2017domain,johnson2017driving}.

A challenge to the use of synthetic data is that it can be highly non-trivial to generate \emph{meaningful} data, since this usually requires modeling
complex environments~\cite{SeshiaS16}.
Suppose we wanted to train a neural network on images of cars on a road.
If we simply sampled uniformly at random from all possible configurations of, say, 12 cars, we would get data that was at best unrealistic, with cars facing sideways or backward, and at worst physically impossible, with cars intersecting each other.
Instead, we want scenes like those in Fig.~\ref{figure:bumper-to-bumper}, where the cars are laid out in a consistent and realistic way.
Furthermore, we may want scenes that are not only realistic but represent particular \emph{scenarios} of interest for training or testing, e.g., parked cars, cars passing across the field of view, or bumper-to-bumper traffic as in Fig.~\ref{figure:bumper-to-bumper}.
In general, we need a way to \emph{guide} data generation toward scenarios that make sense for our application.

We argue that probabilistic programming languages (PPLs)~\cite{prob-prog} provide a natural solution to this problem.
Using a PPL, the designer of a system can construct distributions representing different input regimes of interest, and sample from these distributions to obtain concrete inputs for training and testing.
More generally, the designer can model the system's environment, with the program becoming a specification of the distribution of environments under which the system is expected to operate correctly with high probability.
Such environment models are essential for any formal analysis: in particular, composing the system with the model, we obtain a closed program which we could potentially prove properties about to establish the correctness of the system.

In this paper, we focus on designing and analyzing ML-based cyber-physical systems. We refer to the environment of such a system at any point in time as a \emph{scene}, a configuration of objects in space (including dynamic agents, such as vehicles) along with their features.
We develop a domain-specific \emph{scenario description language}, \scenic, to specify such environments.
\scenic{} is a probabilistic programming language, and a \scenic{} scenario defines a distribution over both scenes and the behaviors of the dynamic agents in them over time.
As we will see, the syntax of the language is designed to simplify the task of writing complex scenarios, and to enable the use of specialized sampling techniques.
In particular, \scenic{} allows the user to both construct objects in a straightforward imperative style and impose hard and soft constraints declaratively.
It also provides readable, concise syntax for \emph{spatial} and \emph{temporal} relationships: constructs for common geometric relationships that would otherwise require complex non-linear expressions and constraints, as well as temporal constructs like interrupts for building complex dynamic behaviors in a modular way.
In addition, \scenic{} provides a notion of classes allowing properties of objects to be given default values depending on other properties: for example, we can define a \texttt{Car} so that by default it faces in the direction of the road at its position.
More broadly, \scenic{} uses a novel approach to object construction which factors the process into syntactically-independent \emph{specifiers} which can be combined in arbitrary ways, mirroring the flexibility of natural language.
Finally, \scenic{} provides constructs to \emph{generalize} simple scenarios by adding noise or by \emph{composing} multiple scenarios together.

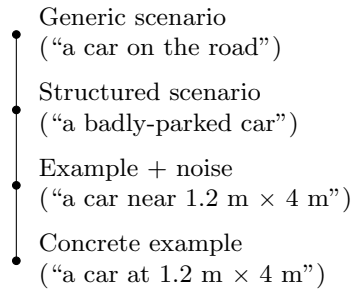
\begin{wrapfigure}[15]{r}{0.4\textwidth}
	\centering
	\vspace{-1em}
	\begin{tikzpicture}[scale=1]
		\draw[-] (0,0) -- (0,3); 
		\draw[fill=black] (0,0) circle (0.05) node[right=5pt,align=left] {Concrete example\\(``a car at 1.2 m $\times$ 4 m'')};
		\draw[fill=black] (0,1) circle (0.05) node[right=5pt,align=left] {Example + noise\\(``a car near 1.2 m $\times$ 4 m'')};
		\draw[fill=black] (0,2) circle (0.05) node[right=5pt,align=left] {Structured scenario\\(``a badly-parked car'')};
		\draw[fill=black] (0,3) circle (0.05) node[right=5pt,align=left] {Generic scenario\\(``a car on the road'')};
	\end{tikzpicture}
	\caption{Spectrum of scenarios, from general to specific.\label{fig:spectrum}}
\end{wrapfigure}

The variety of constructs in \scenic{} makes it possible to model scenarios anywhere on a spectrum from concrete scenes (i.e. individual test cases) to extremely
broad classes of abstract scenarios (see Fig.~\ref{fig:spectrum}). A scenario can be reached by moving along the spectrum from either end:
the top-down approach
is to progressively constrain a very general scenario, while the bottom-up approach is 
to generalize from a concrete example (such as a known failure case), for example by adding random noise.
Probably most usefully,
one can write a scenario in the middle which is far more general than simply adding noise to a single scene but has much more structure than a completely random scene: for example, the traffic scenario depicted in Fig.~\ref{figure:bumper-to-bumper}.
We will illustrate all three ways of developing a scenario, which as we will see are useful for different training, testing, and debugging tasks.

Generating scenarios from a \scenic{} program requires sampling from the probability distribution it implicitly defines.
This task is closely related to the inference problem for imperative PPLs with observations \cite{prob-prog}.
While \scenic{} could be implemented as a library on top of such a language, we found that clarity and concision could be significantly improved with new syntax (specifiers and interrupts in particular) difficult to implement as a library.
Furthermore, while \scenic{} could be translated into existing PPLs, using a new language allows us to impose restrictions enabling domain-specific sampling techniques not possible with general-purpose PPLs.
In particular, we develop algorithms which take advantage of the particular structure of distributions arising from \scenic{} programs to dramatically prune the sample space.

We also integrate \scenic{} as the environment modeling language for \verifai{}, a tool for the formal design and analysis of AI-based systems~\cite{verifai}.
\verifai{} allows writing system-level specifications in Metric Temporal Logic~\cite{mtl} and performing falsification, running simulations and monitoring for violations of the specifications.
\verifai{} provides several search techniques, including active samplers that use feedback from earlier simulations to try to drive the system towards violations.
We make these techniques available from \scenic{} using syntax to define \emph{external parameters} which are sampled by \verifai{} or another external tool.
Such parameters need not have a fixed distribution of values: in particular, we can define a \emph{prior} distribution, but then use cross-entropy optimization~\cite{cross-entropy} to drive the distribution towards one that is concentrated on values that tend to lead to system failures~\cite{fremont-cav20}.

We demonstrate the utility of \scenic{} in training, testing, and debugging ML-based cyber-physical systems. Our first case study is on SqueezeDet~\cite{squeezedet}, a convolutional neural network for object detection in autonomous cars.
For this task, it has been shown~\cite{johnson2017driving} that good performance on real images can be achieved with networks trained purely on synthetic images from the video game Grand Theft Auto V (GTAV \cite{gtav}).
We implemented a sampler for \scenic{} scenarios, using it to generate scenes which were rendered into images by GTAV.
Our experiments demonstrate using \scenic{} to:
\begin{myitemize}
\item evaluate the accuracy of the ML model under particular conditions, e.g.~in good or bad weather,\smallskip
\item improve performance in corner cases by emphasizing them during training: we use \scenic{} to both identify a deficiency in a state-of-the-art car detection data set~\cite{johnson2017driving} and generate a new training set of equal size but yielding significantly better performance, and \smallskip
\item debug a known failure case by generalizing it in many directions, exploring sensitivity to different features and developing a more general scenario for retraining: we use \scenic{} to find an image the network misclassifies, discover the root cause, and fix the bug, in the process improving the network's performance on its original test set (again, without increasing training set size).
\end{myitemize}
These experiments show that \scenic{} can be a very useful tool for understanding and improving perception systems.

While this case study is performed in the domain of visual perception for autonomous driving, and uses one particular simulator (GTAV), we stress that \scenic{} is not specific to either.
In Sec.~\ref{sec:motivating_examples} we give an example of a different domain, namely robotic motion planning (using the Webots simulator~\cite{webots}), and in Sec.~\ref{sec:carla-falsification} we use \scenic{} and \verifai{} to falsify an autonomous agent in the CARLA driving simulator~\cite{Dosovitskiy17}.
The latter experiment demonstrates \scenic{}'s usefulness applied not only to perception components in isolation but to entire closed-loop cyber-physical systems.
In fact, since the conference version of this paper we have successfully applied \scenic{} in two industrial case studies on large ML-based systems~\cite{fremont-cav20,fremont-itsc20}: an aircraft navigation system from Boeing (tested in the X-Plane flight simulator~\cite{xplane}) and the Apollo autonomous driving platform~\cite{apollo} (tested in the LGSVL driving simulator~\cite{lgsvl} and on an actual test track).
Generally, \scenic{} {\em can produce data of any desired type} (e.g. RGB images, LIDAR point clouds, or trajectories from dynamical simulations) by interfacing it to an appropriate simulator.
This requires only two steps: (1) writing a small \scenic{} library defining the types of objects supported by the simulator, as well as the geometry of the workspace; (2) writing an interface layer converting the configurations output by \scenic{} into the simulator's input format (and, for dynamic scenarios, transferring simulator state back into \scenic{}).
While the current version of \scenic{} is primarily concerned with geometry, leaving the details of rendering up to the simulator, the language allows putting distributions on any parameters the simulator exposes: for example, in GTAV the meshes of the various car models are fixed but we can control their overall color. We have also used \scenic{} to specify distributions over parameters on system dynamics, such as mass.

In summary, the main contributions of this work are:
\begin{itemize}
	\item \scenic, a domain-specific probabilistic programming language for describing \emph{scenarios}: distributions over spatio-temporal configurations of physical objects and agents; \smallskip
	\item a methodology for using PPLs to design and analyze cyber-physical systems, especially those based on ML; \smallskip
	\item domain-specific algorithms for sampling from the distribution defined by a \scenic{} program; \smallskip
	\item a case study using \scenic{} to analyze and improve the accuracy of a practical deep neural network used for perception in an autonomous driving context beyond what is achieved by state-of-the-art synthetic data generation methods.
\end{itemize}

The paper is structured as follows: we begin with an overview of our approach in Sec.~\ref{sec:overview}.
Section~\ref{sec:motivating_examples} gives examples highlighting the major features of \scenic{} and motivating various choices in its design.
In Sec.~\ref{sec:scenario_def_language} we describe the \scenic{} language in detail, and in Sec.~\ref{sec:improvisation} we discuss 
its formal semantics and our sampling algorithms.
Section~\ref{sec:experiments} describes the setup and results of our car detection case study and other experiments.
Finally, we discuss related work in Sec.~\ref{sec:related-work} and conclude in Sec.~\ref{sec:conclusion} with a summary and directions for future work.

An early version of this paper appeared as~\cite{fremont-tr18}, extended and published as~\cite{scenic-pldi}.
This paper further extends~\cite{scenic-pldi} by generalizing \scenic{} to dynamic scenarios (including new spatiotemporal pruning techniques), adding constructs for composing scenarios, and integrating \scenic{} within the broader \verifai{} toolkit.

\section{Using PPLs to Design and Analyze ML-Based Cyber-Physical Systems} \label{sec:overview}

\begin{figure}[tb]
\centering
\includegraphics[width=\columnwidth]{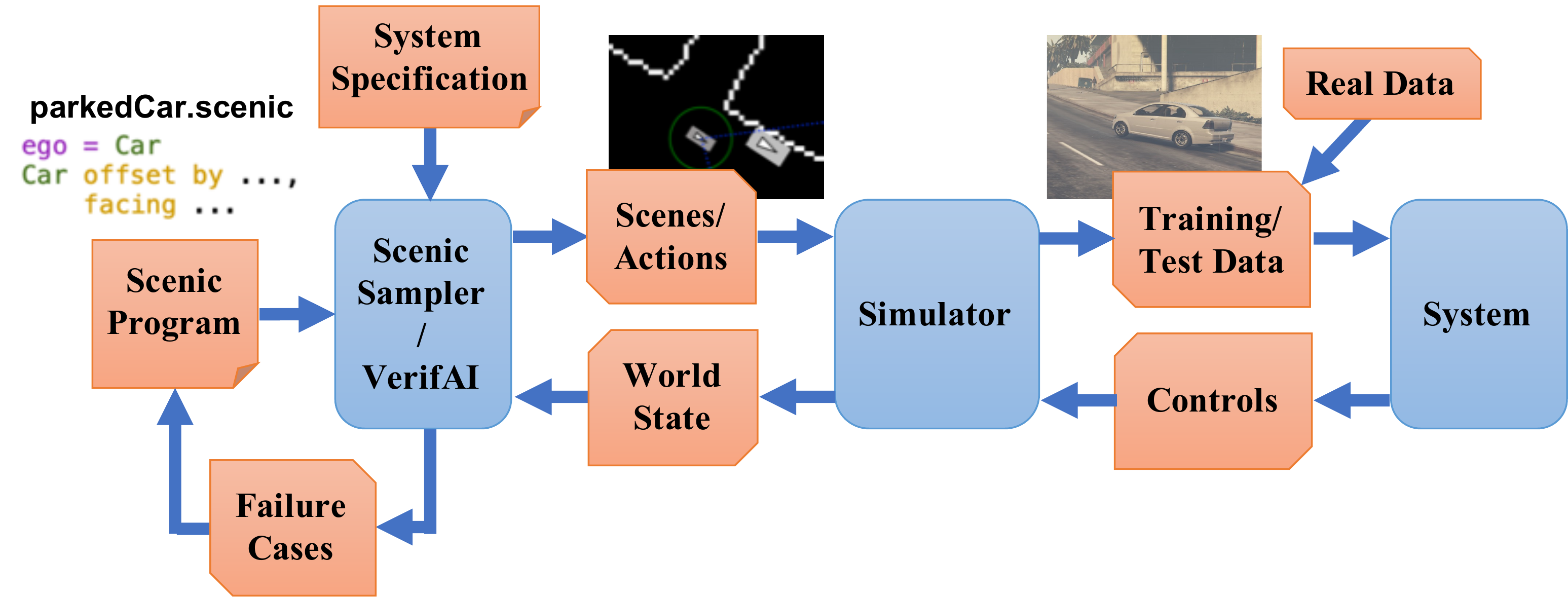}
\caption{Tool flow using \scenic{} to train, test, and debug a cyber-physical system.}
\label{fig:toolflow}
\end{figure}

We propose a methodology for training, testing, and debugging ML-based cyber-physical systems using probabilistic programming languages.
The core idea is to use PPLs to formalize general operation scenarios, then sample from these distributions to generate concrete environment configurations.
Putting these configurations into a simulator, we obtain images or other sensor data which can be used to test and train the system.
The general procedure is outlined in Fig.~\ref{fig:toolflow}.
For a demonstration of this paradigm on an industrial system, proceeding from falsification through failure analysis, retraining, and validation, see~\cite{fremont-cav20}.
Note that the training/testing datasets need not be purely synthetic: we can generate data to supplement existing real-world data (possibly mitigating a deficiency in the latter, while avoiding overfitting).
Furthermore, even for models trained purely on real data, synthetic data can still be useful for testing and debugging, as we will see below.
Now we discuss the three design problems from the Introduction in more detail.

\myparagraph{Testing under Different Conditions.}
The most straightforward problem is that of assessing system performance under different conditions.
We can simply write scenarios capturing each condition, generate a test set from each one, and evaluate the performance of the system on these.
Note that conditions which occur rarely in the real world present no additional problems: as long as the PPL we use can encode the condition, we can generate as many instances as desired.
If we do not have particular conditions in mind, we can write a very general scenario describing the expected operation regime of the system (e.g., the ``Operational Design Domain'' (ODD) of an autonomous vehicle~\cite{nhtsa-testing}) and perform falsification, looking for violations of the system's specification.

\myparagraph{Training on Rare Events.}
Extending the previous application, we can use this procedure to help ensure the system performs adequately even in unusual circumstances or particularly difficult cases.
Writing a scenario capturing these rare events, we can generate instances of them to augment or replace part of the original training set.
Emphasizing these instances in the training set can improve the system's performance in the hard case without impacting performance in the typical case.
In Sec.~\ref{sec:experiment-two} we will demonstrate this for car detection, where a hard case is when one car partially overlaps another in the image.
We wrote a \scenic{} program to generate a set of these overlapping images.
Training the car-detection network on a state-of-the-art synthetic dataset obtained by randomly driving around inside the simulated world of GTAV and capturing images periodically~\cite{johnson2017driving}, we find its performance is significantly worse on the overlapping images.
However, if we keep the training set size fixed but increase the proportion of overlapping images, performance on such images dramatically improves \emph{without harming performance on the original generic dataset}.

\myparagraph{Debugging Failures.}
Finally, we can use the same procedure to help understand and fix bugs in the system.
If we find an environment configuration where the system fails, we can write a scenario reproducing that particular configuration.
Having the configuration encoded as a program then makes it possible to explore the neighborhood around it in a variety of different directions, leaving some aspects of the scene fixed while varying others.
This can give insight into which features of the scene are relevant to the failure, and eventually identify the root cause.
The root cause can then itself be encoded into a scenario which generalizes the original failure, allowing retraining without overfitting to the particular counterexample.
We will demonstrate this approach in Sec.~\ref{sec:generalizing-failure}, starting from a single misclassification, identifying a general deficiency in the training set, replacing part of the training data to fix the gap, and ultimately achieving higher performance on the original test set.

\myparagraph{}
For all of these applications we need a PPL which can encode a wide range of general and specific environment scenarios.
In the next section, we describe the design of a language suited to this purpose.

\section{The \scenic{} Language\label{sec:motivating_examples}}


We use \scenic\ scenarios from our autonomous car case study to motivate and illustrate the main features of the language,
focusing on features that make \scenic\ particularly well-suited for the domain of specifying scenarios for cyber-physical systems.
We begin by describing how \scenic{} can define \emph{spatial} relationships between objects to model scenarios like ``a badly-parked car'', moving on to \emph{temporal} relationships for dynamic scenarios like ``a badly-parked car, which pulls into the road as you approach''.
Finally, we outline \scenic{}'s support for \emph{composing} multiple scenarios together to produce more complex ones.

\subsection{Basic Scenarios}

\begin{paragraph}{Classes, Objects, Geometry, and Distributions.}
To start, suppose we want scenes of one car viewed from another on the road.
We can simply write:
\begin{lstlisting}
from ~scenic.simulators.gta.model~ import *
ego = Car
Car
\end{lstlisting}
First, we import \scenic{}'s \emph{world model} for the GTAV simulator: a \scenic{} library containing everything specific to our case study, including the class \lstinline{Car} and information about the locations of roads (from now on we suppress this line).
Only general geometric concepts are built into \scenic{}.

The second line creates a \lstinline{Car} and assigns it to the special variable \lstinline{ego} specifying the \emph{ego object} which is the reference point for the scenario.
In particular, rendered images from the scenario are from the perspective of the ego object (it is a syntax error to leave \lstinline{ego} undefined).
Finally, the third line creates an additional \lstinline{Car}.
Note that we have not specified the position or any other properties of the two cars: this means they are inherited from the \emph{default values} defined in the \emph{class} \lstinline{Car}.
Object-orientation is valuable in \scenic{} since it provides a natural organizational principle for scenarios involving different types of physical objects.
It also improves compositionality, since we can define a generic \lstinline{Car} model in a library like the GTAV world model and use it in different scenarios.
Our definition of \lstinline{Car} begins as follows (slightly simplified):
\begin{lstlisting}
class Car:
    position: Point on road
    heading: roadDirection at self.position
\end{lstlisting}
Here \lstinline{road} is a \emph{region} (one of \scenic's primitive types) defined in the GTAV world model to specify which points in the workspace are on a road.
Similarly, \lstinline{roadDirection} is a \emph{vector field} specifying the prevailing traffic direction at such points.
The operator $F \texttt{ at } X$ simply gets the direction of the field $F$ at point $X$, so the default value for a car's \lstinline{heading} is the road direction at its \lstinline{position}.
The default \lstinline{position}, in turn, is a \lstinline{Point on road} (we will explain this syntax shortly), which means a \emph{uniformly random} point on the road.

The ability to make random choices like this is a key aspect of \scenic{}.
\scenic{}'s probabilistic nature allows it to model real-world stochasticity, for example encoding a distribution for the distance between two cars learned from data.
This in turn is essential for our application of PPLs to training perception systems: using randomness, a PPL can generate training data matching the distribution the system will be used under.
\scenic{} provides several basic distributions (and allows more to be defined).
For example, we can write
\begin{lstlisting}
Car offset by (Range(-10, 10), Range(20, 40))
\end{lstlisting}
to create a car that is 20--40 m ahead of the camera.
The notation \lstinline{Range($\nt{X}$, $\nt{Y}$)} creates a uniform distribution over the given continuous range, and \lstinline{($\nt{X}$, $\nt{Y}$)} creates a pair, interpreted here as a vector given by its $xy$ coordinates.
\end{paragraph}

\begin{paragraph}{Local Coordinate Systems.}
Using \lstinline{offset by} as above overrides the default position of the \lstinline{Car}, leaving the default orientation (along the road) unchanged.
Suppose for greater realism we don't want to require the car to be \emph{exactly} aligned with the road, but to be within say $5^\circ$.
We could try:
\begin{lstlisting}
Car offset by (Range(-10, 10), Range(20, 40)),
    facing Range(-5, 5) deg
\end{lstlisting}
but this is not quite what we want, since this sets the orientation of the \lstinline{Car} in \emph{global} coordinates (i.e. within $5^\circ$ of North).
Instead we can use \scenic's general operator \lstinline{$\nt{X}$ relative to $\nt{Y}$}, which can interpret vectors and headings as being in a variety of local coordinate systems:
\begin{lstlisting}
Car offset by (Range(-10, 10), Range(20, 40)),
  facing Range(-5, 5) deg relative to roadDirection
\end{lstlisting}
If we want the heading to be relative to the ego car's orientation, we simply write \lstinline{Range(-5, 5) deg relative to ego}.

Notice that since \lstinline{roadDirection} is a vector field, it defines a coordinate system at each point, and an expression like \lstinline{15 deg relative to field} does not define a unique heading.
The example above works because \scenic\ knows that \lstinline{Range(-5, 5) deg relative to roadDirection} depends on a reference position, and automatically uses the \lstinline{position} of the \lstinline{Car} being defined.
This is a feature of \scenic{}'s system of \emph{specifiers}, which we explain next.
\end{paragraph}

\begin{paragraph}{Readable, Flexible Specifiers.} 
The syntax \lstinline{offset by $\nt{X}$} and \lstinline{facing $\nt{Y}$} for specifying positions and orientations may seem unusual compared to typical constructors in object-oriented languages.
There are two reasons why \scenic\ uses this kind of syntax: first, readability.
The second is more subtle and based on the fact that in natural language there are many ways to specify positions and other properties, some of which interact with each other.
Consider the following ways one might describe the location of an object:
\begin{mylist}
\item ``is at position \emph{X}'' (absolute position);
\item ``is just left of position \emph{X}'' (position based on orientation);
\item ``is 3 m west of the taxi'' (relative position);
\item ``is 3 m left of the taxi'' (a local coordinate system);\label{position:left-taxi}
\item ``is one lane left of the taxi'' (another local coordinate system);\label{position:left-lane}
\item ``appears to be 10 m behind the taxi'' (relative to the line of sight);
\item ``is 10 m along the road from the taxi'' (following a vector field; consider a curving road).
\end{mylist}
These are all fundamentally different from each other: e.g., (\ref{position:left-taxi}) and (\ref{position:left-lane}) differ if the taxi is not parallel to the lane.

Furthermore, these specifications combine other properties of the object in different ways: to place the object ``just left of'' a position, we must first know the object's \lstinline{heading}; whereas if we wanted to face the object ``towards'' a location, we must instead know its \lstinline{position}.
There can be chains of such \emph{dependencies}: ``the car is 0.5 m left of the curb'' means that the \emph{right edge} of the car is 0.5 m away from the curb, not the car's \lstinline{position}, which is its center.
So the car's \lstinline{position} depends on its \lstinline{width}, which in turn depends on its \lstinline{~model~}.
In a typical object-oriented language, this might be handled by computing values for \lstinline{position} and other properties and passing them to a constructor.
For ``a car is 0.5 m left of the curb'' we might write:
\begin{lstlisting}
# hypothetical Python-like language
m = Car.defaultModelDistribution.sample()
pos = curb.offsetLeft(0.5 + m.width / 2)
car = Car(pos, ~model~=m)
\end{lstlisting}
Notice how \lstinline{m} must be used twice, because \lstinline{m} determines both the model of the car and (indirectly) its position.
This is inelegant and breaks encapsulation because the default model distribution is used outside of the \lstinline{Car} constructor.
The latter problem could be fixed by having a specialized constructor or factory function,
\begin{lstlisting}
car = CarLeftOfBy(curb, 0.5)
\end{lstlisting}
but these would proliferate since we would need to handle all possible combinations of ways to specify different properties (e.g. do we want to require a specific model? Are we overriding the width provided by the model for this specific car?).
Instead of having a multitude of such monolithic constructors, \scenic\ factors the definition of objects into potentially-interacting but syntactically-independent parts:
\begin{lstlisting}
Car left of spot by 0.5, &with model& BUS
\end{lstlisting}
Here \lstinline{left of $\nt{X}$ by $\nt{D}$} and \lstinline{&with model& $\nt{M}$} are \emph{specifiers} which do not have an order, but which \emph{together} specify the properties of the car.
\scenic\ works out the dependencies between properties (here, \lstinline{position} is provided by \lstinline{left of}, which depends on \lstinline{width}, whose default value depends on \lstinline{~model~}) and evaluates them in the correct order.
To use the default model distribution we would simply leave off \lstinline{&with model& BUS}; keeping it affects the \lstinline{position} appropriately without having to specify \lstinline{BUS} more than once.
\end{paragraph}

\begin{paragraph}{Specifying Multiple Properties Together.}
Recall that we defined the default \lstinline{position} for a \lstinline{Car} to be a \lstinline{Point on road}: this is an example of another specifier, \lstinline{on $\nt{region}$}, which specifies \lstinline{position} to be a uniformly random point in the given region.
This specifier illustrates another feature of \scenic, namely that specifiers can specify multiple properties simultaneously.
Consider the following scenario, which creates a parked car given a region \lstinline{curb} defined in the GTAV world model:
\begin{lstlisting}
spot = OrientedPoint on `visible` curb
Car left of spot by 0.25
\end{lstlisting}
The function \lstinline{`visible` $\nt{region}$} returns the part of the region that is visible from the ego object.
The specifier \lstinline{on `visible` curb} will then set \lstinline{position} to be a uniformly random visible point on the curb.
We create \lstinline{spot} as an \lstinline{OrientedPoint}, which is a built-in class that defines a local coordinate system by having both a \lstinline{position} and a \lstinline{heading}.
The \lstinline{on $\nt{region}$} specifier can also specify \lstinline{heading} if the region has a preferred orientation (a vector field) associated with it: in our example, \lstinline{curb} is oriented by \lstinline{roadDirection}.
So \lstinline{spot} is, in fact, a uniformly random visible point on the curb, oriented along the road.
That orientation then causes the car to be placed 0.25 m left of \lstinline{spot} in \lstinline{spot}'s local coordinate system, i.e. away from the curb, as desired.

In fact, \scenic\ makes it easy to elaborate the scenario without needing to alter the code above.
Most simply, we could specify a particular model or non-default distribution over models by just adding \lstinline{&with model& $\nt{M}$} to the definition of the \lstinline{Car}.
More interestingly, we could produce a scenario for \emph{badly}-parked cars by adding two lines:
\begin{lstlisting}
spot = OrientedPoint on `visible` curb
badAngle = Uniform(1.0, -1.0) * Range(10, 20) deg
Car left of spot by 0.5,
    facing badAngle relative to roadDirection
\end{lstlisting}
This will yield cars parked 10-20$^\circ$ off from the direction of the curb, as seen in Fig.~\ref{fig:badly-parked}.
This illustrates how specifiers greatly enhance \scenic{}'s flexibility and modularity.

\begin{figure}[tb]
\centering
\includegraphics[width=0.7\columnwidth]{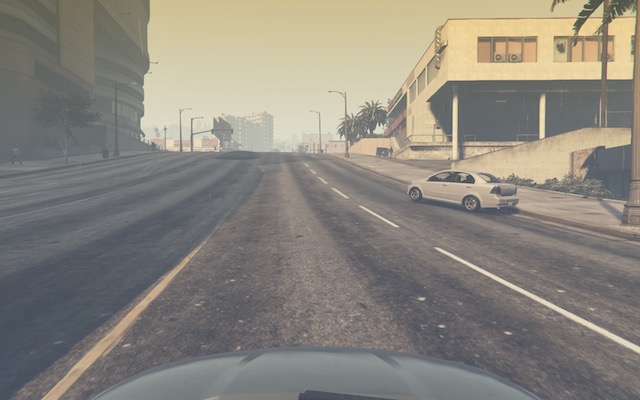}
\caption{A scene of a badly-parked car.}
\label{fig:badly-parked}
\end{figure}
\end{paragraph}

\begin{paragraph}{Declarative Specifications of Hard and Soft Constraints.}
Notice that in the scenarios above we never explicitly ensured that the two cars will not intersect each other.
Despite this, \scenic{} will never generate such scenes.
This is because \scenic{} enforces several \emph{default requirements}: all objects must be contained in the workspace, must not intersect each other, and must be visible from the ego object.\footnote{The last requirement ensures that the object will affect the rendered image. It can be disabled on a per-object basis, for example in dynamic scenarios where the object is initially out of sight but may interact with the ego object later on.}
\scenic\ also allows the user to define custom requirements checking arbitrary conditions built from various geometric predicates.
For example, the following scenario produces a car headed roughly towards us, while still facing the nominal road direction:
\begin{lstlisting}
carB = Car offset by (Range(-10, 10), Range(20, 40)),
           &with viewAngle& 30 deg
require carB can see ego
\end{lstlisting}
Here we have used the \lstinline{$\nt{X}$ can see $\nt{Y}$} predicate, which in this case is checking that the ego car is inside the $30^\circ$ view cone of the second car.
If we only need this constraint to hold part of the time, we can use a \emph{soft requirement} specifying the minimum probability with which it must hold:
\begin{lstlisting}
require[0.5] carB can see ego
\end{lstlisting}
Hard requirements, called ``observations'' in other PPLs (see, e.g., \cite{prob-prog}), are very convenient in our setting because they make it easy to restrict attention to particular cases of interest.
They also improve encapsulation, since we can restrict an existing scenario without altering it (we can simply import it in a new \scenic{} program that includes additional \lstinline{require} statements).
Finally, soft requirements are useful in ensuring adequate representation of a particular condition when generating a training set: for example, we could require that at least 90\% of the images have a car driving on the right side of the road.
\end{paragraph}

\begin{paragraph}{Mutations.} 
\scenic\ provides a simple \emph{mutation} system that improves compositionality by providing a mechanism to add variety to a scenario without changing its code.
This is useful, for example, if we have a scenario encoding a single concrete scene obtained from real-world data and want to quickly generate variations.
For instance:
\begin{lstlisting}
taxi = Car at (120, 300), facing 37 deg, ...
...
mutate taxi
\end{lstlisting}
This will add Gaussian noise to the \lstinline{position} and \lstinline{heading} of \lstinline{taxi}, while still enforcing all built-in and custom requirements.
The standard deviation of the noise can be scaled by writing, for example, \lstinline{mutate taxi by 2} (which adds twice as much noise), and we will see later that it can be controlled separately for \lstinline{position} and \lstinline{heading}.
\end{paragraph}

\begin{paragraph}{Multiple Domains and Simulators.}

We conclude this section by illustrating a second application domain, namely generating workspaces to test motion planning algorithms, and \scenic{}'s ability to work with different simulators.
A robot like a Mars rover able to climb over rocks can have very complex dynamics, with the feasibility of a motion plan depending on exact details of the robot's hardware and the geometry of the terrain.
We can use \scenic{} to write a scenario generating challenging cases for a planner to solve.
Figure~\ref{fig:mars} shows a scene, visualized using an interface we wrote between \scenic{} and the Webots robotics simulator~\cite{webots}, with a bottleneck between the robot and its goal that forces the planner to consider climbing over a rock.
The \scenic{} code for this scenario is given in Appendix~\ref{sec:gallery}.

\begin{figure}[tb]
\centering
\includegraphics[width=0.7\columnwidth]{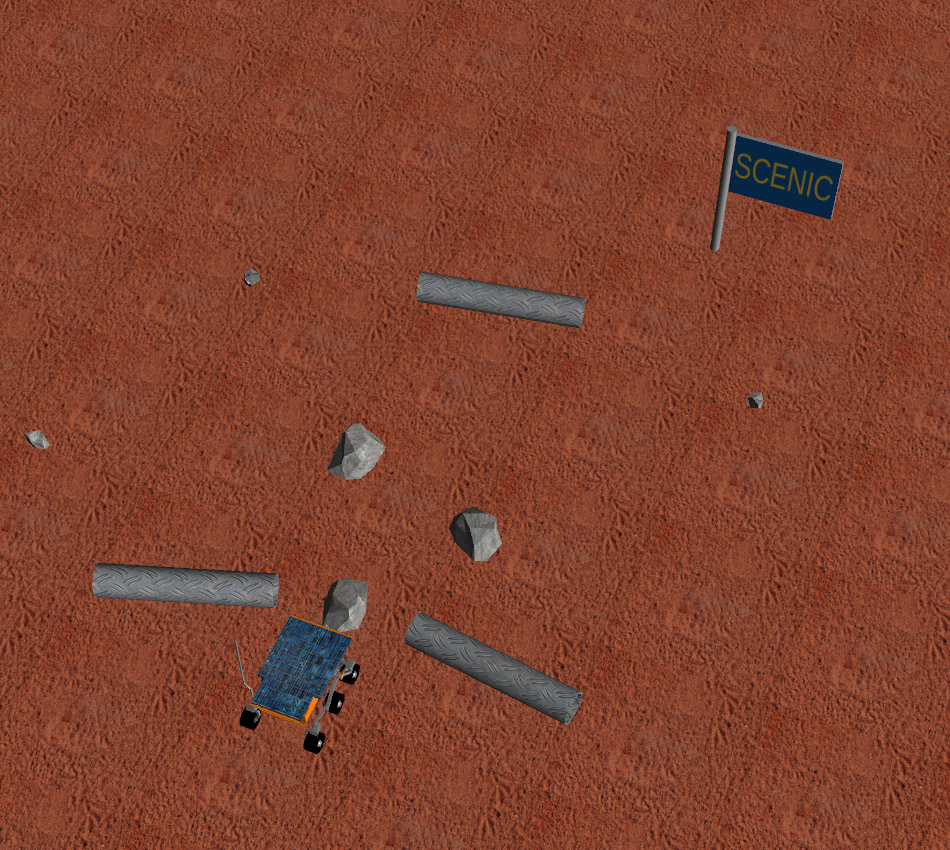}
\caption{Webots scene of a Mars rover in a debris field with a bottleneck.}
\label{fig:mars}
\end{figure}

Even within a single application domain, such as autonomous driving, \scenic{} enables writing \emph{cross-platform} scenarios that will work without change in multiple simulators.
This is made possible by what we call \emph{abstract application domains}: \scenic{} world models which define object classes and other world information like our GTAV world model, but which are abstract, simulator-agnostic protocols that can be implemented by models for particular simulators.
For example, \scenic{} includes an abstract domain for autonomous driving, \texttt{scenic.domains.driving}, which loads road networks from standard formats, providing a uniform API for referring to lanes, maneuvers, and other aspects of road geometry.
The driving domain also provides generic \lstinline{Car} and \lstinline{Pedestrian} classes, complete with implementations of common dynamic behaviors (covered in the next section) like lane following.
These make it straightforward to implement complex driving scenarios, which are then guaranteed to work in any simulator supporting the driving domain.
Figure~\ref{fig:oas_5} illustrates this, showing the exact same \scenic{} code being used to generate scenarios in both the CARLA~\cite{Dosovitskiy17} and LGSVL~\cite{lgsvl} simulators.

\begin{figure}[tb]
\centering
\includegraphics[width=0.49\columnwidth]{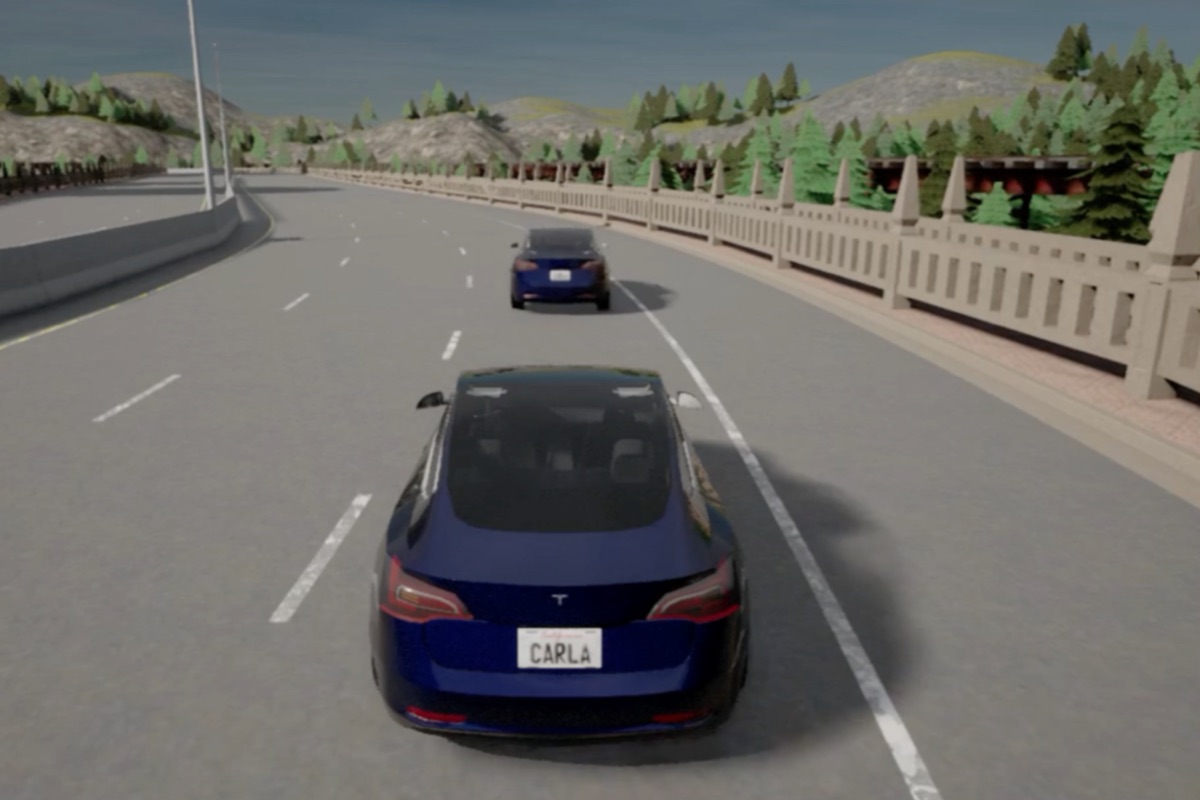}
\includegraphics[width=0.49\columnwidth]{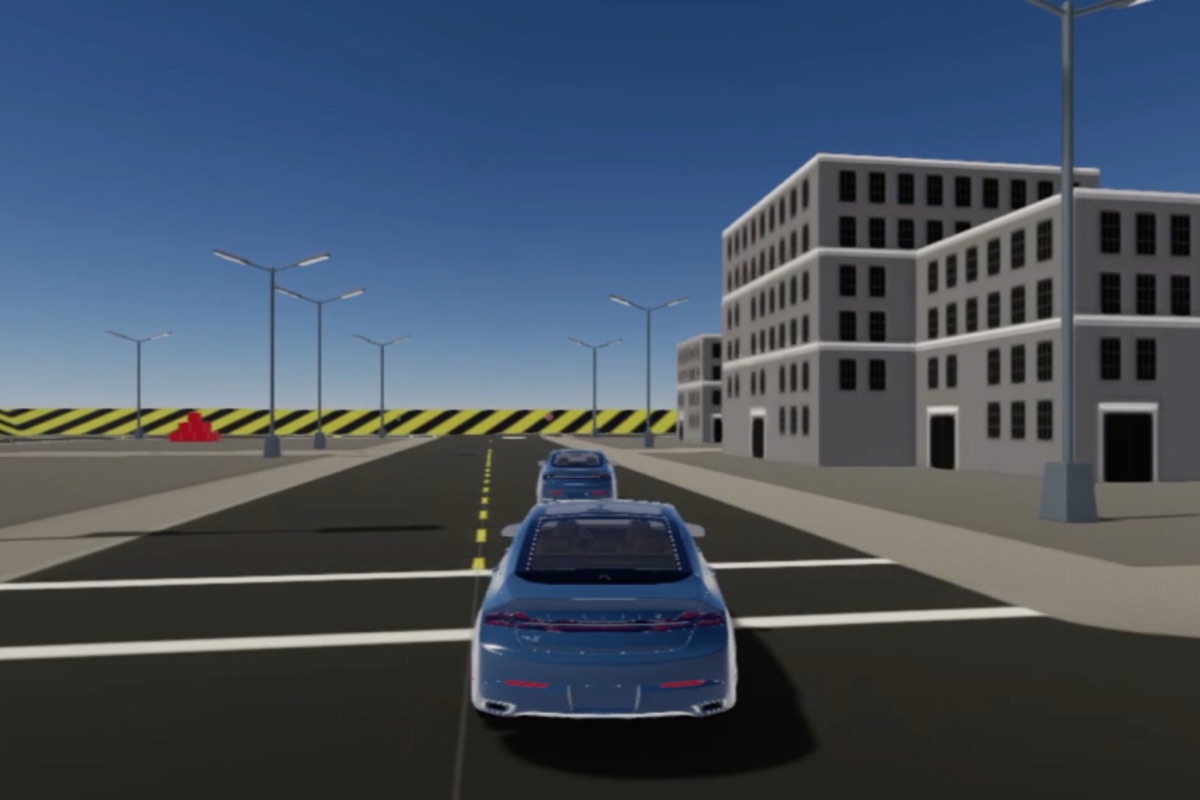}
\caption{Scenes sampled from the same \scenic{} program in CARLA and LGSVL.}
\label{fig:oas_5}
\end{figure}
\end{paragraph}

\subsection{Dynamic Scenarios}
\label{sec:dynamic-scenarios}

Having seen the basic constructs \scenic{} provides for defining objects and their spatial relationships, we now outline \scenic{}'s support for \emph{dynamic} scenarios which also define the \emph{temporal} properties of objects. 

\begin{paragraph}{Agents, Actions, and Behaviors.}
In \scenic{}, we call objects which take actions over time \emph{dynamic agents}, or simply
\emph{agents}. These are ordinary \scenic{} objects, so we can still use all of the syntax described in the previous section to define their initial positions, orientations, etc.
In addition, we specify their dynamic behavior using a built-in property called \lstinline{behavior}.
Using one of the behaviors defined in Scenic's driving library, we can write for example:

\begin{lstlisting}
model ~scenic.domains.driving.model~
Car &with behavior& FollowLaneBehavior
\end{lstlisting}

A behavior defines a sequence of \emph{actions} for the agent to take, which need not be fixed
but can be probabilistic and depend on the state of the agent or other objects. In
\scenic{}, an \emph{action} is an instantaneous operation executed by an agent, like
setting the steering angle of a car or turning on its headlights. Most actions are
specific to particular application domains, and so different sets of actions are provided
by different simulator interfaces. For example, the \scenic{} driving domain defines a
\texttt{SetThrottleAction} for cars.

To define a behavior, we write a function which runs over the course of the scenario,
periodically issuing actions. \scenic{} uses a discrete notion of time, so at each time
step the function specifies zero or more actions for the agent to take. For example, here
is a very simplified version of the \lstinline{FollowLaneBehavior} above:

\begin{lstlisting}
behavior FollowLaneBehavior():
    while True:
        throttle, steering = ...    # compute controls
        take SetThrottleAction(throttle), SetSteerAction(steering)
\end{lstlisting}

We intend this behavior to run for the entire scenario, so we use an infinite loop. In
each step of the loop, we compute appropriate throttle and steering controls, then use
the \lstinline{take} statement to take the corresponding actions. When that statement is
executed, \scenic{} pauses the behavior until the next time step of the simulation, whereupon the
function resumes and the loop repeats.
\end{paragraph}

\begin{paragraph}{Execution of Behaviors.}
When there are multiple agents, all of their behaviors run in parallel, as illustrated in Fig.~\ref{fig:scenic-sim}; each time step,
Scenic sends their selected actions to the simulator to be executed and advances the
simulation by one step. It then reads back the state of the simulation, updating the \lstinline{position}, \lstinline{speed}, etc. of each object.

\begin{figure}[tb]
\centering
\includegraphics[width=0.5\columnwidth]{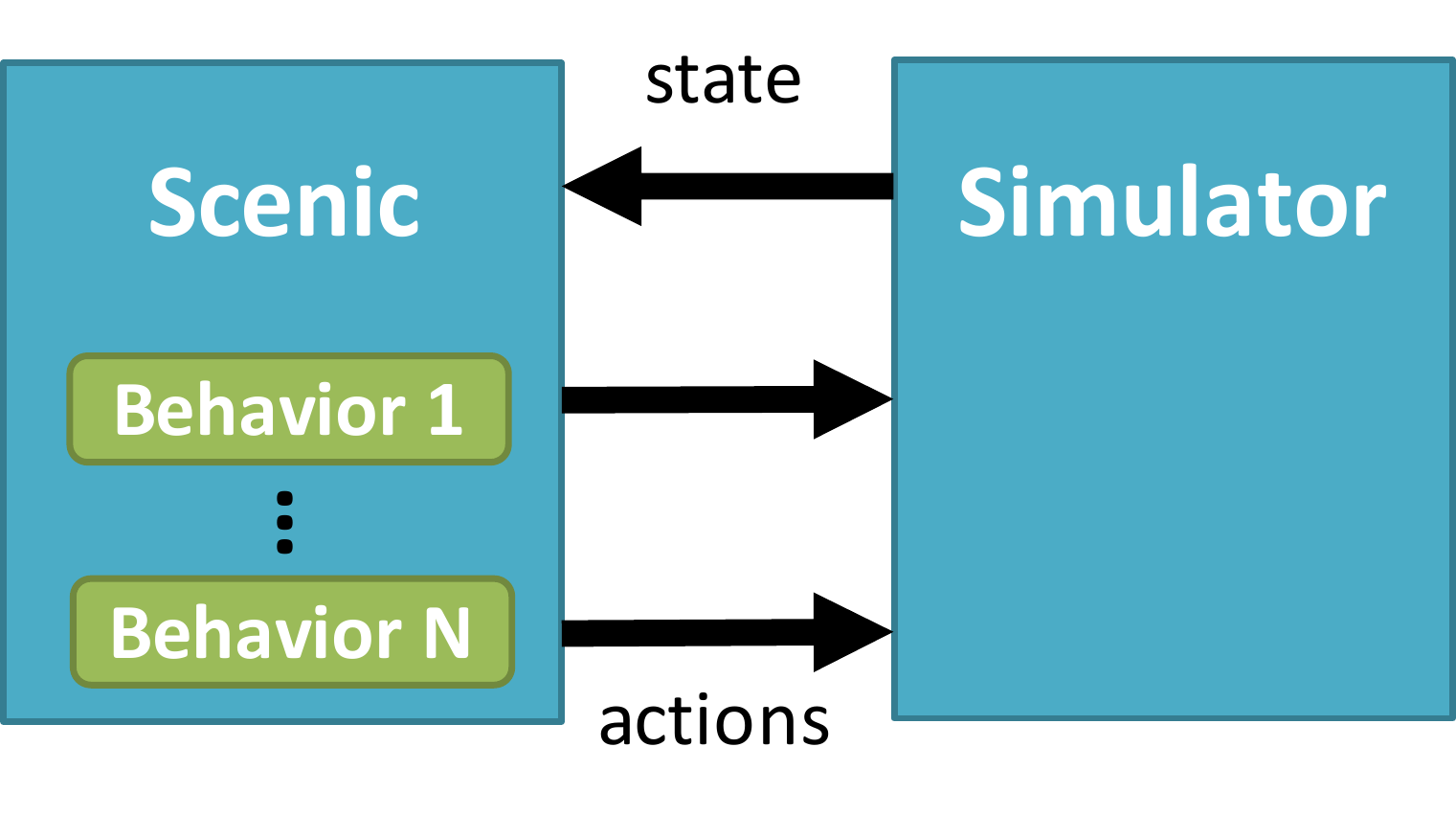}
\caption{Diagram showing interaction between \scenic{} and a simulator during the execution of a dynamic scenario.}
\label{fig:scenic-sim}
\end{figure}

Since behaviors run dynamically during simulations, they can access the current state of the world to decide what actions to take.
Consider the following behavior:

\begin{lstlisting}
behavior WaitUntilClose(threshold=15):
    while (`distance from` self to ego) > threshold:
        wait
    do FollowLaneBehavior()
\end{lstlisting}

Here, we repeatedly query the distance from the agent running the behavior (\lstinline{self})
to the ego car; as long as it is above a threshold, we use the \lstinline{wait} statement, to take no action.
Once the threshold is met, we start driving by using the \lstinline{do} statement to invoke the \lstinline{FollowLaneBehavior}
we saw above.
Since \lstinline{FollowLaneBehavior} runs forever, we will
never return to the \lstinline{WaitUntilClose} behavior.
\end{paragraph}

\begin{paragraph}{Behavior Arguments and Random Parameters.}
The example above also shows how behaviors may take arguments, like any \scenic{} function.
Here, \lstinline{threshold} is an argument to the behavior which has default value 15 but can be
customized, so we could write for example:

\begin{lstlisting}
ego = Car
carB = Car visible, &with behavior& WaitUntilClose
carC = Car visible, &with behavior& WaitUntilClose(20)
\end{lstlisting}

Both \lstinline{carB} and \lstinline{carC} will use the \lstinline{WaitUntilClose} behavior, but independent
copies of it with thresholds of 15 and 20 respectively.

Unlike ordinary \scenic{} code, control flow constructs such as \lstinline{if} and \lstinline{while} are
allowed to depend on random variables inside a behavior. Any distributions defined inside
a behavior are sampled at simulation time, not during scene sampling. Consider the
following behavior:

\begin{lstlisting}
behavior AvoidPedestrian():
    threshold = Range(4, 7)
    while True:
        if self.distanceToClosest(Pedestrian) < threshold:
            strength = TruncatedNormal(0.8, 0.02, 0.5, 1)
            take SetBrakeAction(strength), SetThrottleAction(0)
        else:
            take SetThrottleAction(0.5), SetBrakeAction(0)
\end{lstlisting}

Here, the value of \lstinline{threshold} is sampled only once, at the beginning of the scenario
when the behavior starts running. The value \lstinline{strength}, on the other hand, is sampled
every time control reaches line 5, so that every time step when the car is braking we use
a slightly different braking strength (0.8 on average, but with Gaussian noise added with
standard deviation 0.02, truncating the possible values to between 0.5 and 1).
\end{paragraph}

\begin{paragraph}{Interrupts.}
It is frequently useful to take an existing behavior and add a complication to it; for
example, suppose we want a car that follows a lane, stopping whenever it encounters an
obstacle. \scenic{} provides a concept of \emph{interrupts} which allows us to reuse the basic
\lstinline{FollowLaneBehavior} without having to modify it.

\begin{lstlisting}
behavior FollowAvoidingObstacles():
    try:
        do FollowLaneBehavior()
    interrupt when self.distanceToClosest(Object) < 5:
        take SetBrakeAction(1)
\end{lstlisting}

This \texttt{try-interrupt} statement has similar syntax to the Python \emph{try} statement (and in fact allows \lstinline{except} clauses to catch exceptions just as in
Python, as we'll see later), and begins in the same way: at first, the code block after the \lstinline{try:} (the \emph{body}) is executed. At the start of every time step during its execution, the condition
from each \lstinline{interrupt} clause is checked; if any are true, execution of the body is
suspended and we instead begin to execute the corresponding \emph{interrupt handler}. In the
example above, there is only one interrupt, which fires when we come within 5 meters of
any object. When that happens, \lstinline{FollowLaneBehavior} is paused and we instead apply full
braking for one time step. In the next step, we will resume \lstinline{FollowLaneBehavior} wherever
it left off, unless we are still within 5 meters of an object, in which case the
interrupt will fire again.

If there are multiple \lstinline{interrupt} clauses, successive clauses take precedence over
those which precede them. Furthermore, such higher-priority interrupts can fire even
during the execution of an earlier interrupt handler. This makes it easy to model a
hierarchy of behaviors with different priorities; for example, we could implement a car
which drives along a lane, passing slow cars and avoiding collisions, along the
following lines:

\begin{lstlisting}
behavior Drive():
    try:
        do FollowLaneBehavior()
    interrupt when self.distanceToNextObstacle() < 20:
        do PassingBehavior()
    interrupt when self.timeToCollision() < 5:
        do CollisionAvoidance()
\end{lstlisting}

Here, the car begins by lane following, switching to passing if there is a car or other
obstacle too close ahead. During \emph{either} of those two sub-behaviors, if the time to
collision gets too low, we switch to collision avoidance. Once the \lstinline{CollisionAvoidance}
behavior completes, we will resume whichever behavior was interrupted earlier. If we were
in the middle of \lstinline{PassingBehavior}, it will run to completion (possibly being
interrupted again) before we finally resume \lstinline{FollowLaneBehavior}.

As this example illustrates, when an interrupt handler completes, by default we resume
execution of the interrupted code. If this is undesired, the \lstinline{abort} statement can be
used to cause the entire try-interrupt statement to exit. For example, to run a behavior
until a condition is met without resuming it afterward, we can write:

\begin{lstlisting}
behavior ApproachAndTurnLeft():
    try:
        do FollowLaneBehavior()
    interrupt when (`distance from` self to intersection) < 10:
        abort    # cancel lane following
    do WaitForTrafficLightBehavior()
    do TurnLeftBehavior()
\end{lstlisting}

This is a common enough use case of interrupts that \scenic{} provides a shorthand notation:

\begin{lstlisting}
behavior ApproachAndTurnLeft():
    do FollowLaneBehavior() until (`distance from` self to intersection) < 10
    do WaitForTrafficLightBehavior()
    do TurnLeftBehavior()
\end{lstlisting}

\scenic{} also provides a shorthand for interrupting a behavior after a certain period of
time:

\begin{lstlisting}
behavior DriveForAWhile():
    do FollowLaneBehavior() for 30 seconds
\end{lstlisting}

The alternative form \lstinline{do $\nt{behavior}$ for $\nt{n}$ steps} uses time steps instead of real
simulation time.

Finally, note that when try-interrupt statements are nested, interrupts of the outer
statement take precedence. This makes it easy to build up complex behaviors in a modular
way. For example, the behavior \lstinline{Drive} we wrote above is relatively complicated, using
interrupts to switch between several different sub-behaviors. We would like to be able to
put it in a library and reuse it in many different scenarios without modification.
Interrupts make this straightforward; for example, if for a particular scenario we want a
car that drives normally but suddenly brakes for 5 seconds when it reaches a certain
area, we can write:

\begin{lstlisting}
behavior DriveWithSuddenBrake():
    haveBraked = False
    try:
        do Drive()
    interrupt when self in targetRegion and not haveBraked:
        do StopBehavior() for 5 seconds
        haveBraked = True
\end{lstlisting}

With this behavior, \lstinline{Drive} operates as it did before, interrupts firing as appropriate
to switch between lane following, passing, and collision avoidance. But during any of
these sub-behaviors, if the car enters the \lstinline{targetRegion} it will immediately brake for
5 seconds, then pick up where it left off.
\end{paragraph}

\begin{paragraph}{Stateful Behaviors.}
As the last example shows, behaviors can use local variables to maintain state, which is
useful when implementing behaviors which depend on actions taken in the past. To
elaborate on that example, suppose we want a car which usually follows the \lstinline{Drive}
behavior, but every 15-30 seconds stops for 5 seconds. We can implement this behavior as
follows:

\begin{lstlisting}
behavior DriveWithRandomStops():
    delay = Range(15, 30) seconds
    last_stop = 0
    try:
        do Drive()
    interrupt when simulation.currentTime - last_stop > delay:
        do StopBehavior() for 5 seconds
        delay = Range(15, 30) seconds
        last_stop = simulation.currentTime
\end{lstlisting}

Here \lstinline{delay} is the randomly-chosen amount of time to run \lstinline{Drive} for,
and \lstinline{last_stop} keeps track of the time when we last started to run it. When the time
elapsed since \lstinline{last_stop} exceeds \lstinline{delay}, we interrupt \lstinline{Drive} and
stop for 5 seconds. Afterwards, we pick a new \lstinline{delay} before the next stop, and save
the current time in \lstinline{last_stop}, effectively resetting our timer to zero.
\end{paragraph}

\begin{paragraph}{Requirements and Monitors.}
Just as you can declare spatial constraints on scenes using the \lstinline{require} statement,
you can also impose constraints on dynamic scenarios. For example, if we don't want to
generate any simulations where \lstinline{carA} and \lstinline{carB} are simultaneously visible from the
ego car, we could write:

\begin{lstlisting}
require always not ((ego can see carA) and (ego can see carB))
\end{lstlisting}

The \lstinline{require always $\nt{condition}$} statement enforces that the given condition must
hold at every time step of the scenario; if it is ever violated during a simulation, we
reject that simulation and sample a new one. Similarly, we can require that a condition
hold at \emph{some} time during the scenario using the \lstinline{require eventually} statement:

\begin{lstlisting}
require eventually ego in intersection
\end{lstlisting}

You can also use the ordinary \lstinline{require} statement inside a behavior to require that a
given condition hold at a certain point during the execution of the behavior. For
example, here is a simple elaboration of the \lstinline{WaitUntilClose} behavior we saw above:

\begin{lstlisting}
behavior WaitUntilClose(threshold=15):
    while (`distance from` self to ego) > threshold:
        require self.distanceToClosest(Pedestrian) > threshold
        wait
    do FollowLaneBehavior()
\end{lstlisting}

The requirement ensures that no pedestrian comes close to \lstinline{self} until the ego does;
after that, we place no further restrictions.

To enforce more complex temporal properties like this one without modifying behaviors,
you can define a \emph{monitor}. Like behaviors, monitors are functions which run in parallel
with the scenario, but they are not associated with any agent and any actions they take
are ignored. Here is a monitor
for the property ``\lstinline{carA} and \lstinline{carB} enter the intersection before \lstinline{carC}'':

\begin{lstlisting}
monitor CarCEntersLast:
    seenA, seenB = False, False
    while not (seenA and seenB):
        require carC not in intersection
        if carA in intersection:
            seenA = True
        if carB in intersection:
            seenB = True
        wait
\end{lstlisting}

We use the variables \lstinline{seenA} and \lstinline{seenB} to remember whether we have seen \lstinline{carA}
and \lstinline{carB} respectively enter the intersection. The loop will iterate as long as at
least one of the cars has not yet entered the intersection, so if \lstinline{carC} enters before
either \lstinline{carA} or \lstinline{carB}, the requirement on line 4 will fail and we will reject the
simulation. Note the necessity of the \lstinline{wait} statement on line 9: if we omitted it, the
loop could run forever without any time actually passing in the simulation.
\end{paragraph}

\begin{paragraph}{Preconditions and Invariants.}

Even general behaviors designed to be used in multiple scenarios may not operate
correctly from all possible starting states: for example, \lstinline{FollowLaneBehavior} assumes
that the agent is actually in a lane rather than, say, on a sidewalk. To model such
assumptions, \scenic{} provides a notion of \emph{guards} for behaviors. Most simply, we can
specify one or more \emph{preconditions}:

\begin{lstlisting}
behavior MergeInto(newLane):
    precondition: self.lane is not newLane and self.road is newLane.road
    ...
\end{lstlisting}

Here, the precondition requires that whenever the \lstinline{MergeInto} behavior is executed by
an agent, the agent must not already be in the destination lane but should be on the same
road. We can add any number of such preconditions; like ordinary requirements, violating
any precondition causes the simulation to be rejected.

Since behaviors can be interrupted, it is possible for a behavior to resume execution in
a state it doesn't expect: imagine a car which is lane following, but then swerves onto
the shoulder to avoid an accident; na\"ively resuming lane following, we find we are no
longer in a lane. To catch such situations, \scenic{} allows us to define \emph{invariants} which
are checked at every time step during the execution of a behavior, not just when it
begins running. These are written similarly to preconditions:

\begin{lstlisting}
    behavior FollowLaneBehavior():
        invariant: self in road
        ...
\end{lstlisting}

While the default behavior for guard violations is to reject the simulation, in some
cases it may be possible to recover from a violation by taking some additional actions.
To enable this kind of design, \scenic{} signals guard violations by raising a
\lstinline{GuardViolation} exception which can be caught like any other exception; the simulation
is only rejected if the exception propagates out to the top level. So to model the
lane-following-with-collision-avoidance behavior suggested above, we could write code
like this:

\begin{lstlisting}
behavior Drive():
    while True:
        try:
            do FollowLaneBehavior()
        interrupt when self.distanceToClosest(Object) < 5:
            do CollisionAvoidance()
        except InvariantViolation:   # FollowLaneBehavior has failed
            do GetBackOntoRoad()
\end{lstlisting}

When any object comes within 5 meters, we suspend lane following and switch to collision
avoidance. When the latter completes, \lstinline{FollowLaneBehavior}
will be resumed; if its invariant fails because we are no longer on the road, we catch
the resulting \lstinline{InvariantViolation} exception and run a \lstinline{GetBackOntoRoad} behavior to
restore the invariant. The whole \lstinline{try} statement then completes, so the outermost loop
iterates and we begin lane following once again.
\end{paragraph}

\begin{paragraph}{Terminating the Scenario.}

By default, scenarios run forever, unless a time limit is specified when running the \scenic{} tool.
However, scenarios can also define termination criteria using the
\lstinline{terminate when} statement; for example, we could decide to end a scenario as soon as
the ego car travels at least a certain distance:

\begin{lstlisting}
start = Point on road
ego = Car at start
terminate when (distance to start) >= 50
\end{lstlisting}

Additionally, the \lstinline{terminate} statement can be used inside behaviors and monitors: if
it is ever executed, the scenario ends. For example, we can use a monitor to terminate
the scenario once the ego spends 30 time steps in an intersection:

\begin{lstlisting}
monitor StopAfterTimeInIntersection:
    totalTime = 0
    while totalTime < 30:
        if ego in intersection:
            totalTime += 1
        wait
    terminate
\end{lstlisting}
\end{paragraph}

\subsection{Compositional Scenarios}

The previous two sections showed how \scenic{} allows us to model both the spatial and temporal aspects of a scenario.
\scenic{} also provides facilities for defining scenarios as reusable modules and \emph{composing} them in various ways.
These features make it possible to write a library of simple scenarios which can then be used as building blocks to construct many more complex scenarios.

\begin{paragraph}{Modular Scenarios.}
To define a named, reusable scenario, optionally with tunable parameters, \scenic{} provides the \lstinline{scenario} statement.
For example, here is a scenario which creates a parked car on the shoulder of the \lstinline{ego}'s current lane (assuming there is one), using some APIs from the driving library:

\begin{lstlisting}
scenario ParkedCar(gap=0.25):
    precondition: ego.laneGroup._shoulder != None
    setup:
        spot = OrientedPoint on `visible` ego.laneGroup.curb
        parkedCar = Car left of spot by gap
\end{lstlisting}

The \lstinline{setup} block contains \scenic{} code which executes when the scenario is instantiated, and which can define classes, create objects, declare requirements, etc. as in any of the example scenarios we saw above.
Additionally, we can define preconditions and invariants, which operate in the same way as for dynamic behaviors.
Having now defined the \lstinline{ParkedCar} scenario, we can use it in a more complex scenario, potentially multiple times:

\begin{lstlisting}
scenario Main():
    setup:
        ego = Car
    compose:
        do ParkedCar(), ParkedCar(0.5)
\end{lstlisting}

Here our \lstinline{Main} scenario itself only creates the ego car; then its \lstinline{compose} block orchestrates how to run other modular scenarios.
In this case, we invoke two copies of the \lstinline{ParkedCar} scenario in parallel, specifying in one case that the gap between the parked car and the curb should be 0.5 m instead of the default 0.25.
So the scenario will involve three cars in total, and as usual \scenic{} will automatically ensure that they are all on the road and do not intersect.
\end{paragraph}

\begin{paragraph}{Parallel and Sequential Composition.}
The scenario above is an example of \emph{parallel} composition, where we use the \lstinline{do} statement to run two scenarios at the same time.
We can also use \emph{sequential} composition, where one scenario begins after another ends.
This is done the same way as in dynamic behaviors: in fact, the \lstinline{compose} block of a scenario is executed in essentially the same way as a monitor, and allows all the same control-flow constructs.
For example, we could write a \lstinline{compose} block as follows:

\begin{lstlisting}
while True:
    do ParkedCar(gap=0.25) for 30 seconds
    do ParkedCar(gap=0.5) for 30 seconds
\end{lstlisting}

Here, a new parked car is created every 30 seconds\footnote{In a real implementation, we would probably want to require that the parked car is not initially visible from the \lstinline{ego}, to avoid the sudden appearance of cars out of nowhere.}, with the distance to the curb alternating between 0.25 and 0.5 m.
Note that without the \lstinline{for 30 seconds} qualifier, we would never get past line 2, since the \lstinline{ParkedCar} scenario does not define any termination conditions using \lstinline{terminate when} (or \lstinline{terminate}) and so runs forever by default.
If instead we want to create a new car only when the \lstinline{ego} has passed the current one, we can use a \lstinline{do}-\lstinline{until} statement:

\begin{lstlisting}
while True:
    subScenario = ParkedCar(gap=0.25)
    do subScenario until `distance past` subScenario.parkedCar > 10
\end{lstlisting}

Note how we can refer to the \lstinline{parkedCar} variable created in the \lstinline{ParkedCar} scenario as a property of the scenario.
Combined with the ability to pass objects as parameters of scenarios, this is convenient for reusing objects across scenarios.
\end{paragraph}

\begin{paragraph}{Interrupts, Overriding, and Initial Scenarios.}
The \lstinline{try}-\lstinline{interrupt} statement used in behaviors can also be used in \lstinline{compose} blocks to switch between scenarios.
For example, suppose we already have a scenario where the \lstinline{ego} is following a \lstinline{leadCar}, and want to elaborate it by adding a parked car which suddenly pulls in front of the lead car.
We could write a \lstinline{compose} block as follows:

\begin{lstlisting}
try:
    ~following~ = FollowingScenario()
    do ~following~
interrupt when `distance to` ~following~.leadCar < 10:
    do ParkedCarPullingAheadOf(~following~.leadCar)
\end{lstlisting}

If the \lstinline{ParkedCarPullingAheadOf} scenario is defined to end shortly after the parked car finishes entering the lane, the interrupt handler will complete and \scenic{} will resume executing \lstinline{FollowingScenario} on line 3 (unless the \lstinline{ego} is still within 10 m of the lead car).

Suppose that we want the lead car to behave differently while  \lstinline{ParkedCarPullingAheadOf} is running; for example, perhaps the behavior for the lead car defined in \lstinline{FollowingScenario} does not handle a parked car suddenly pulling in.
To enable changing the \lstinline{behavior} or other properties of an object in a sub-scenario, \scenic{} provides the \lstinline{override} statement, which we can use as follows:

\begin{lstlisting}
scenario ParkedCarPullingAheadOf(target):
    setup:
        override target &with behavior& FollowLaneAvoidingCollisions
        parkedCar = Car left of ...
\end{lstlisting}

Here we override the \lstinline{behavior} property of \lstinline{target} for the duration of the scenario, reverting it back to its original value (and thereby continuing to execute the old behavior) when the scenario terminates.
The \lstinline{override $\nt{object}$ $\csl{\nt{specifier}}$} statement has the same syntax as an object definition, and can specify any properties of the object except for dynamic properties like \lstinline{position} or \lstinline{speed} which are updated every time step by the simulator (and can only be indirectly controlled by taking actions).

In order to allow writing scenarios which can both stand on their own and be invoked during another scenario, \scenic{} provides a special conditional statement testing whether we are inside the \emph{initial scenario}, i.e., the very first scenario to run.

\begin{lstlisting}
scenario TwoLanePedestrianScenario():
    setup:
        if in initial scenario:  # create ego car on random 2-lane road
            ~roads~ = filter(lambda r: len(r.lanes) == 2, network.roads)
            road = Uniform(*roads)  # pick uniformly from list
            ego = Car on road
        else:  # use existing ego car; require it is on a 2-lane road
            require len(ego.road.lanes) == 2
            road = ego.road
        Pedestrian on visible road.sidewalkRegion, with behavior ...
\end{lstlisting}
\end{paragraph}

\begin{paragraph}{Random Selection of Scenarios.}
For very general scenarios, like ``driving through a city, encountering typical human traffic'', we may want a variety of different events and interactions to be possible.
We saw above how we can write behaviors for individual agents which choose randomly between possible actions; \scenic{} allows us to do the same with entire scenarios.
Most simply, since scenarios are first-class objects, we can write functions which operate on them, perhaps choosing a scenario from a list of options based on some complex criterion:

\begin{lstlisting}
chosenScenario = pickNextScenario(ego.position, ...)
do chosenScenario
\end{lstlisting}

However, some scenarios may only make sense in certain contexts; for example, a scenario involving a car running a red light can take place only at an intersection.
To facilitate modeling such situations, \scenic{} provides variants of the \lstinline{do} statement which choose scenarios to run randomly amongst only those whose preconditions are satisfied:

\begin{lstlisting}
do choose RedLightRunner, Jaywalker, ParkedCar
do choose {RedLightRunner: 2, Jaywalker: 1, ParkedCar: 1}
do shuffle RedLightRunner, Jaywalker, ParkedCar
\end{lstlisting}

Here, line 1 checks the preconditions of the three given scenarios, then executes one (and only one) of the enabled scenarios. If for example the current road has no shoulder, then \lstinline{ParkedCar} will be disabled and we will have a 50/50 chance of executing either \lstinline{RedLightRunner} or \lstinline{Jaywalker} (assuming their preconditions are satisfied).
If \emph{none} of the three scenarios are enabled, \scenic{} will reject the simulation.
Line 2 shows a non-uniform variant, where \lstinline{RedLightRunner} is twice as likely to be chosen as each of the other scenarios (so if only \lstinline{ParkedCar} is disabled, we will pick \lstinline{RedLightRunner} with probability $2/3$; if none are disabled, $2/4$).
Finally, line 3 is a shuffled variant, where \emph{all three} scenarios will be executed, but in random order\footnote{Respecting preconditions, so in particular the simulation will be rejected if at some point none of the remaining scenarios to execute are enabled.}.
\end{paragraph}

\myparagraph{}
All of the examples we have seen above illustrate the versatility of \scenic{} in modeling a wide range of interesting scenarios.
Complete \scenic{} code for the bumper-to-bumper scenario of Fig.~\ref{figure:bumper-to-bumper}, the Mars rover scenario of Fig.~\ref{fig:mars}, as well as other scenarios used as examples in this section or in our experiments, along with images of generated scenes, can be found in Appendix~\ref{sec:gallery}.

\section{Syntax of \scenic{}\label{sec:scenario_def_language}}


\scenic\ is an object-oriented PPL, with programs consisting of sequences of statements built with standard imperative constructs including conditionals, loops, functions, and methods (which we do not describe further, focusing on the new elements).
Compared to other imperative PPLs, the major restriction of \scenic{}, made in order to allow more efficient sampling, is that conditional branching may not depend on random variables (except in behaviors).
The novel syntax, outlined above, is largely devoted to expressing spatiotemporal relationships in a concise and flexible manner.
Figure~\ref{figure:grammar} gives a formal grammar for \scenic, which we now describe in detail.

\subsection{Data Types}

\scenic\ provides several primitive data types:
\begin{description}
\item[Booleans] expressing truth values.
\item[Scalars] floating-point numbers, which can be sampled from various distributions (see Table~\ref{table:distributions}).
\item[Vectors] representing positions and offsets in space, constructed from coordinates in meters with the syntax \mbox{\lstinline{(X, Y)}} \footnote{The Smalltalk-like~\cite{smalltalk} syntax \lstinline{X @ Y} used in earlier versions of \scenic{} is also legal.}.
\item[Headings] representing orientations in space.
Conveniently, in 2D these are a single angle (in radians, anticlockwise from North).
By convention the heading of a local coordinate system is the heading of its $y$-axis, so, for example, \lstinline{(-2, 3)} means 2 meters left and 3 ahead.
\item[Vector Fields] associating an orientation to each point in space.
For example, the shortest paths to a destination or (in our case study) the nominal traffic direction.
\item[Regions] representing sets of points in space.
These can have an associated vector field giving points in the region preferred orientations
(e.g. the surface of an object could have normal vectors, so that objects placed randomly on the surface face outward by default).
\end{description}

\begin{figure}[tb]
\centering
\texttt{
\setlength\tabcolsep{2pt}
\begin{tabular}{rl}
\nt{program} \Is& \star{\nt{statement}} \\
\nt{boolean} \Is& True \Or False \Or \nt{booleanOp} \\
\nt{scalar} \Is& \nt{number} \Or \nt{distrib} \Or \nt{scalarOp} \\
\nt{distrib} \Is& \nt{baseDist} \Or resample(\nt{distrib}) \\
\nt{vector} \Is& (\nt{scalar}, \nt{scalar}) \Or \nt{Point} \\
&\quad \Or \nt{vectorOp} \\
\nt{heading} \Is& \nt{scalar} \Or \nt{OrientedPoint} \\
&\quad \Or \nt{headingOp} \\
\nt{direction} \Is& \nt{heading} \Or \nt{vectorField} \\
\nt{value} \Is& \nt{boolean} \Or \nt{scalar} \Or \nt{vector} \\
&\quad \Or \nt{direction} \Or \nt{region} \\
&\quad \Or \nt{object} \Or \nt{object}.\nt{property} \\
\nt{classDef} \Is& class \nt{class}\opt{(\nt{superclass})}: \\
&\quad \star{\nt{property}: \nt{value}} \\
\nt{object} \Is& \nt{class} \csl{\nt{specifier}} \\
\nt{specifier} \Is& with \nt{property} \nt{value} \\
&\quad \Or \nt{posSpec} \Or \nt{headSpec} \\
\end{tabular}
\begin{tabular}{rl}
\nt{behavior} \Is& behavior \nt{name}(\nt{params}):\\
&\quad \star{precondition: \nt{boolean}} \\
&\quad \star{invariant: \nt{boolean}} \\
&\quad \star{\nt{statement}} \\
\nt{try} \Is& try:\\
&\quad \star{\nt{statement}} \\
&\star{interrupt when \nt{boolean}:\\&\quad \star{\nt{statement}}} \\
&\star{except \nt{exception}:\\&\quad \star{\nt{statement}}} \\
\nt{scenario} \Is& scenario \nt{name}(\nt{params}):\\
&\quad \star{precondition: \nt{boolean}} \\
&\quad \star{invariant: \nt{boolean}} \\
&\quad \opt{setup:\\&\qquad \star{\nt{statement}}} \\
&\quad \opt{compose:\\&\qquad \star{\nt{statement}}}
\end{tabular}
}
\caption{Simplified \scenic\ grammar.
\nt{Point} and \nt{OrientedPoint} are instances of the corresponding classes.
See Tab.~\ref{table:statements} for statements, Fig.~\ref{figure:operators} for operators, Tab.~\ref{table:distributions} for \nt{baseDist}, and Tables~\ref{table:position-specs} and \ref{table:heading-specs} for \nt{posSpec} and \nt{headSpec}.}
\label{figure:grammar}
\end{figure}

\begin{table}[tb]
\centering
\caption{Built-in distributions. All parameters are \nt{scalar}s except \nt{value}.}
\label{table:distributions}
\texttt{
\begin{tabular}{ll}
\toprule
\textrm{Syntax} & \textrm{Distribution} \\
\midrule
Range(\nt{low}, \nt{high}) & \textrm{uniform on continuous interval} \\
Uniform(\csl{\nt{value}}) & \textrm{uniform over discrete values} \\
Discrete(\{\csl{\nt{value}:~\nt{weight}}\}) & \textrm{discrete with weights} \\
Normal(\nt{mean}, \nt{stdDev}) & \textrm{normal (Gaussian)} \\
TruncatedNormal(\nt{mean}, \nt{stdDev}, \nt{low}, \nt{high}) & \textrm{normal, truncated to the given window} \\
\bottomrule
\end{tabular}
}
\end{table}

In addition, \scenic\ provides \emph{objects}, organized into single-inheritance \emph{classes} specifying a set of properties their instances must have, together with corresponding default values (see Fig.~\ref{figure:grammar}).
Default value expressions are evaluated each time an object is created.
Thus if we write \lstinline{weight: Range(1, 5)} when defining a class then each instance will have a \lstinline{weight} drawn \emph{independently} from \lstinline{Range(1, 5)}.
Default values may use the special syntax \lstinline{self.$\nt{property}$} to refer to one of the other properties of the object, which is then a \emph{dependency} of this default value.
In our case study, for example, the \lstinline{width} and \lstinline{length} of a \lstinline{Car} are by default derived from its \lstinline{~model~}.

Physical objects in a scene are instances of \lstinline{Object}, which is the default superclass when none is specified.
\lstinline{Object} descends from the two other built-in classes: its superclass is \lstinline{OrientedPoint}, which in turn subclasses \lstinline{Point}.
These represent locations in space, with and without an orientation respectively, and so provide the fundamental properties \lstinline{heading} and \lstinline{position}.
\lstinline{Object} extends them by defining a bounding box with the properties \lstinline{width} and \lstinline{length}, as well as temporal information like \lstinline{speed} and \lstinline{~behavior~}.
Table~\ref{table:properties} lists the properties of these classes and their default values.

\begin{table}[tb]
\caption{Properties of the built-in classes \texttt{Point}, \texttt{OrientedPoint}, and \texttt{Object}.}
\label{table:properties}
\centering
\texttt{
\begin{tabular}{lcl}
\toprule
\textrm{Property} & \textrm{Default} & \textrm{Meaning} \\
\midrule
position & $(0,0)$ & \textrm{position in global coordinates} \\
viewDistance & $50$ & \textrm{distance for `\texttt{can see}' predicate} \\
mutationScale & $0$ & \textrm{overall scale of mutations} \\
positionStdDev & $1$ & \textrm{mutation $\sigma$ for \texttt{position}} \\
\midrule
heading & $0$ & \textrm{heading in global coordinates} \\
viewAngle & $360^\circ$ & \textrm{angle for `\texttt{can see}' predicate} \\
headingStdDev & $5^\circ$ & \textrm{mutation $\sigma$ for \texttt{heading}} \\
\midrule
width & $1$ & \textrm{width of bounding box} \\
length & $1$ & \textrm{length of bounding box} \\
speed & $0$ & \textrm{speed of object} \\
velocity & $(0, 0)$ & \textrm{velocity (default from \texttt{speed}, \texttt{heading})} \\
angularSpeed & $0$ & \textrm{angular speed (in rad/s)} \\
behavior & None & \textrm{dynamic behavior, if any} \\
allowCollisions & \textrm{false} & \textrm{collisions allowed} \\
requireVisible & \textrm{true} & \textrm{must be visible from \tm{ego}} \\
regionContainedIn & None & \textrm{region object must be contained in} \\
\bottomrule
\end{tabular}
}
\end{table}

To allow cleaner notation, \lstinline{Point} and \lstinline{OrientedPoint} are automatically interpreted as vectors or headings in contexts expecting these (as shown in Fig.~\ref{figure:grammar}).
For example, we can write \lstinline{taxi offset by (1, 2)} and \lstinline{30 deg relative to taxi} instead of \lstinline{taxi.position offset by (1, 2)} and \lstinline{30 deg relative to taxi.heading}.
Ambiguous cases, e.g. \lstinline{taxi relative to limo}, are illegal (caught by a simple type system); the more verbose syntax must be used instead.

\subsection{Expressions}

\scenic's expressions are mostly straightforward, largely consisting of the arithmetic, boolean, and geometric operators shown in Fig.~\ref{figure:operators}.
The meanings of these operators are largely clear from their syntax, so we defer complete definitions of their semantics to the Appendix~\cite{scenic-full}.
Figure~\ref{figure:operator-examples} illustrates several of the geometric operators (as well as some specifiers, which we will discuss in the next section).
Various points to note:
\begin{myitemize}
\item \lstinline{$\nt{X}$ can see $\nt{Y}$} uses a simple model where a \lstinline{Point} can see a certain distance, and an \lstinline{OrientedPoint} restricts this to the sector along its \lstinline{heading} with a certain angle (see Table~\ref{table:properties}).
An \lstinline{Object} is visible iff its bounding box is.
\item \lstinline{$\nt{X}$ relative to $\nt{Y}$} interprets \nt{X} as an offset in a local coordinate system defined by \nt{Y}.
Thus \lstinline{(-3, 0) relative to $\nt{Y}$} yields 3 m West of \nt{Y} if \nt{Y} is a vector, and 3 m \emph{left} of \nt{Y} if \nt{Y} is an \lstinline{OrientedPoint}.
If defining a heading inside a specifier, either \nt{X} or \nt{Y} can be a vector field, interpreted as a heading by evaluating it at the \lstinline{position} of the object being specified.
So we can write for example \lstinline{Car at (120, 70), facing 30 deg relative to roadDirection}.
\item \lstinline{`visible` $\nt{region}$} yields the part of the region visible from the \lstinline{ego}, so we can write for example \lstinline{Car on `visible` road}.
The form \lstinline{$\nt{region}$ `visible from` $\nt{X}$} uses \nt{X} instead of \lstinline{ego}.
\item \lstinline{`front of` $\nt{Object}$}, \lstinline{`front left of` $\nt{Object}$}, etc. yield the corresponding points on the bounding box of the object, oriented along the object's \lstinline{heading}.
\end{myitemize}

\begin{figure}[tb]
\centering
\begin{tikzpicture}[every node/.append style={font=\small\ttfamily},scale=0.94]
 \draw[fill] (0,0) circle (0.075) node[below=4pt,xshift=-3pt] {\large \lstinline{ego}};
 \draw (0,1) -- (-1,0) -- (0,-1) -- (1,0) -- cycle;
 \draw[line width=2pt, ->] (0,0) -- (0.5, 0.5);
 \draw[dashed, ->, >=latex] (0,0) -- (2,2);
 \draw[dashed, ->, >=latex] (-1.25, 1.25) -- (1.25, -1.25);
 \draw[fill] (-0.5, 0.5) circle (0.05);
 \draw[line width=1.5pt, ->] (-0.5,0.5) -- (0, 1);
 \draw[->] (-1.25, 0.5) node[left] {\lstinline{`left of` ego}} to (-0.7, 0.5);
 \draw[fill] (0, -1) circle (0.05);
 \draw[line width=1.5pt, ->] (0,-1) -- (0.5, -0.5);
 \draw[->] (-0.5, -1) node[left] {\lstinline{`back right of` ego}} to (-0.2, -1);
 
 \coordinate (OB) at (2.121, 0.707);
 \draw[fill] (OB) circle (0.05);
 \dimline[extension start length=0, extension end length=0]{(1.414,1.414)}{(OB)}{1};
 \dimline[extension start length=0, extension end length=0]{(0.707,-0.707)}{(OB)}{2};
 \draw[->] (3.5, 0) node[below, align=center] {\lstinline{Point offset by (1, 2)}\\\textrm{\emph{or}}\\\lstinline{(1, 2) relative to ego}} to[in=0, out=90] (2.3, 0.707);
 
 \coordinate (P) at (1,3);
 \draw[fill] (P) circle (0.075) node[right=3pt, yshift=6pt] {\large P};
 \draw[line width=2pt, ->] (P) -- (0.293, 3);
 \draw[dashed, ->, >=latex] (0,0) -- (1.666,5);
 \draw[dashed, ->, >=latex] (-1.5, 3.833) -- (1.5, 2.833);
 
 \draw[dashed] (P) -- (4.5,3);
 \draw[fill] (3,3) circle (0.05);
 \draw[line width=1.5pt, ->] (3,3) -- (2.293, 3);
 \draw[->] (2.5, 4) node[above, xshift=1cm] {\lstinline{P offset by (0, -2)}} to[out=-90, in=-225] (2.85,3.15);
 \dimline[extension start path={(1,2.5) (1,3)}, extension end path={(3,2.5) (3,3)}]{(1, 2.5)}{(3, 2.5)}{2};
 
 \coordinate (BY) at (-0.581, 4.581);
 \draw[fill] (BY) circle (0.05);
 \dimline[extension start length=0, extension end length=0]{(1.316,3.949)}{(BY)}{2};
 \dimline[extension start length=0, extension end length=0]{(-0.897,3.632)}{(BY)}{1};
 \draw[->] (-0.25, 5) node[above] {\lstinline{Point beyond P by (-2, 1)}} to (-0.5, 4.7);
 
 \draw (3, 3.5) -- (3, 2.5) -- (4, 2.5) -- (4, 3.5) -- cycle;
 \draw[->] (3.5, 1.8) node[below, align=center] {\lstinline{Object behind P by 2}} to (3.5, 2.4);
 
 \draw (1.125, 3.375) to[out=162, in=90] (0.605, 3);
 \draw[->] (-1.1, 2.65) node[below] {\lstinline{`apparent heading of` P}} to[out=90, in=135] (0.65, 3.35);
\end{tikzpicture}
\caption{Various \scenic{} operators and specifiers applied to the \lstinline{ego} object and an \lstinline{OrientedPoint} \texttt{P}.
Instances of \lstinline{OrientedPoint} are shown as bold arrows.}
\label{figure:operator-examples}
\end{figure}
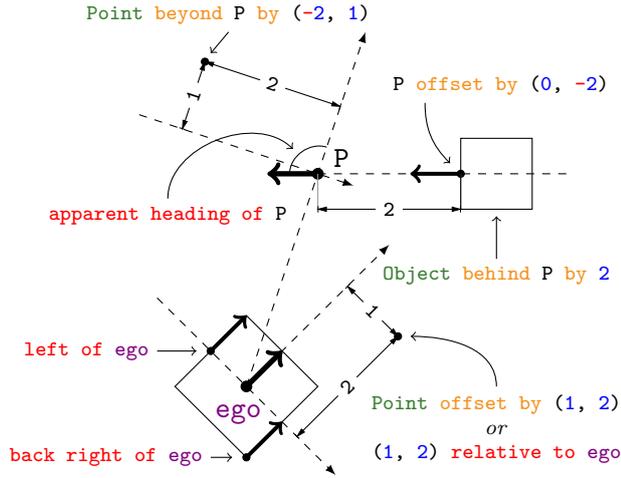

\begin{figure}[tb]
\centering
\texttt{
\setlength{\tabcolsep}{0pt}
\begin{tabular}{l}
\nt{scalarOperator} \Is \; max(\csl{\nt{scalar}}) \Or min(\csl{\nt{scalar}}) \\
\quad\Or -\nt{scalar} \Or abs(\nt{scalar}) \Or \nt{scalar} \either{+}{*} \nt{scalar} \\
\quad\Or relative heading of \nt{heading} \opt{from \nt{heading}} \\
\quad\Or apparent heading of \nt{OrientedPoint} \opt{from \nt{vector}} \\
\quad\Or distance to \nt{vector} \opt{from \nt{vector}} \\
\quad\Or distance \opt{of \nt{OrientedPoint}} past \nt{vector} \\
\quad\Or angle \opt{from \nt{vector}} to \nt{vector} \\
\nt{booleanOperator} \Is \; not \nt{boolean} \\
\quad\Or \nt{boolean} \either{and}{or} \nt{boolean} \\
\quad\Or \nt{scalar} \textrm{(}== \Or != \Or < \Or > \Or <= \Or >=\textrm{)} \nt{scalar} \\
\quad\Or \either{\nt{Point}}{\nt{OrientedPoint}} can see \either{\nt{vector}}{\nt{Object}} \\
\quad\Or \either{\nt{vector}}{\nt{Object}} in \nt{region} \\
\nt{headingOperator} \Is \; \nt{scalar} deg \\
\quad\Or \nt{vectorField} at \nt{vector} \\
\quad\Or \nt{direction} relative to \nt{direction} \\
\nt{vectorOperator} \Is \; \nt{vector} relative to \nt{vector} \\
\quad\Or \nt{vector} offset by \nt{vector} \\
\quad\Or \nt{vector} offset along \nt{direction} by \nt{vector} \\
\nt{regionOperator} \Is \; visible \nt{region} \\
\quad\Or \nt{region} visible from \either{\nt{Point}}{\nt{OrientedPoint}} \\
\nt{orientedPointOperator} \Is \\
\quad\; \nt{vector} relative to \nt{OrientedPoint} \\
\quad\Or \nt{OrientedPoint} offset by \nt{vector} \\
\quad\Or\textrm{(}front \Or back \Or left \Or right\textrm{)} of \nt{Object} \\
\quad\Or \either{front}{back} \either{left}{right} of \nt{Object}
\end{tabular}
}
\caption{Operators by result type.}
\label{figure:operators}
\end{figure}

Two types of \scenic{} expressions are more complex: distributions and object definitions.
As in a typical imperative probabilistic programming language, a distribution evaluates to a \emph{sample} from the distribution.
Thus the program
\begin{lstlisting}
x = Range(0, 1)
y = (x, x)
\end{lstlisting}
does not make \tm{y} uniform over the unit box, but rather over its diagonal.
For convenience in sampling multiple times from a primitive distribution, \scenic\ provides a \lstinline{resample($D$)} function returning an independent\footnote{Conditioned on the values of the distribution's parameters (e.g. \nt{low} and \nt{high} for a uniform interval), which are not resampled.} sample from $D$, one of the distributions in Tab.~\ref{table:distributions}.
\scenic{} also allows defining custom distributions beyond those in the Table.

The second type of complex \scenic{} expressions are object definitions.
These are the only expressions with a side effect, namely creating an object in the generated scene.
More interestingly, properties of objects are specified using the system of \emph{specifiers} discussed above, which we now detail.

\subsection{Specifiers}

As shown in the grammar in Fig.~\ref{figure:grammar}, an object is created by writing the class name followed by a (possibly empty) comma-separated list of specifiers.
The specifiers are combined, possibly adding default specifiers from the class definition, to form a complete specification of all properties of the object.
Arbitrary properties (including user-defined properties with no meaning in \scenic) can be specified with the generic specifier \lstinline{with $\nt{property}$ $\nt{value}$}, while \scenic{} provides many more specifiers for the built-in properties \lstinline{position} and \lstinline{heading}, shown in Tables~\ref{table:position-specs} and \ref{table:heading-specs} respectively.

In general, a specifier is a function taking in values for zero or more properties, its \emph{dependencies}, and returning values for one or more other properties, some of which can be specified \emph{optionally}, meaning that other specifiers will override them.
For example, \tm{on \nt{region}} specifies \lstinline{position} and optionally specifies \lstinline{heading} if the given region has a preferred orientation.
If \lstinline{road} is such a region, as in our case study, then \lstinline{Object on road} will create an object at a position uniformly random in \lstinline{road} and with the preferred orientation there.
But since \lstinline{heading} is only specified optionally, we can override it by writing \lstinline{Object on road, facing 20 deg}.

\begin{table*}[tb]
\caption{Specifiers for \texttt{position}. Those in the second group also optionally specify \texttt{heading}.}
\label{table:position-specs}
\centering
\texttt{
\begin{tabular}{lc}
\toprule
\textrm{Specifier} & \textrm{Dependencies} \\
\midrule
at \nt{vector} & \noreqs \\
offset by \nt{vector} & \noreqs \\
offset along \nt{direction} by \nt{vector} & \noreqs \\
\either{left}{right} of \nt{vector} \opt{by \nt{scalar}} & heading\textrm{, }width \\
\either{ahead of}{behind} \nt{vector} \opt{by \nt{scalar}} & heading\textrm{, }length \\
beyond \nt{vector} by \nt{vector} \opt{from \nt{vector}} & \noreqs \\
visible \opt{from \either{\nt{Point}}{\nt{OrientedPoint}}} & \noreqs \\
\midrule
\either{in}{on} \nt{region} & \noreqs \\
\either{left}{right} of \either{\nt{OrientedPoint}}{\nt{Object}} \opt{by \nt{scalar}} & width \\
\either{ahead of}{behind} \either{\nt{OrientedPoint}}{\nt{Object}} \opt{by \nt{scalar}} & length \\
following \nt{vectorField} \opt{from \nt{vector}} for \nt{scalar} & \noreqs \\
\bottomrule
\end{tabular}
}
\end{table*}

{
\setlength{\tabcolsep}{4.5pt}
\begin{table}[tb]
\caption{Specifiers for \texttt{heading}.}
\label{table:heading-specs}
\centering
\texttt{
\begin{tabular}{lcc}
\toprule
\textrm{Specifier} & \textrm{Deps.} \\
\midrule
facing \nt{heading} & \noreqs \\
facing \nt{vectorField} & position \\
facing \either{toward}{away from} \nt{vector} & position \\
apparently facing \nt{heading} \opt{from \nt{vector}} & position \\
\bottomrule
\end{tabular}
}
\end{table}
}

Specifiers are combined to determine the properties of an object by evaluating them in an order ensuring that their dependencies are always already assigned.
If there is no such order or a single property is specified twice, the scenario is ill-formed.
The procedure by which the order is found, taking into account properties that are optionally specified and default values, will be described in the next section.

As the semantics of the specifiers in Tables~\ref{table:position-specs} and \ref{table:heading-specs} are largely evident from their syntax, we defer exact definitions to the Appendix~\cite{scenic-full}.
We briefly discuss some of the more complex specifiers, referring to the examples in Fig.~\ref{figure:operator-examples}:
\begin{myitemize}
\item \lstinline{behind $\nt{vector}$} means the object is placed with the midpoint of its front edge at the given vector, and similarly for \lstinline{ahead/left/right of $\nt{vector}$}.
\item \lstinline{beyond $\nt{A}$ by $\nt{O}$ &from& $\nt{B}$} means the position obtained by treating \nt{O} as an offset in the local coordinate system at \nt{A} oriented along the line of sight from \nt{B}.
In this and other specifiers, if the \lstinline{&from& $\nt{B}$} is omitted, the ego object is used by default.
So for example \lstinline{beyond taxi by (0, 3)} means 3 m directly behind the taxi as viewed by the camera (see Fig.~\ref{figure:operator-examples} for another example).
\item The \texttt{heading} optionally specified by \lstinline{left of $\nt{OrientedPoint}$}, etc. is that of the \lstinline{OrientedPoint} (thus in Fig.~\ref{figure:operator-examples}, \lstinline{P offset by (0, -2)} yields an \lstinline{OrientedPoint} facing the same way as \lstinline{P}).
Similarly, the \lstinline{heading} optionally specified by the \lstinline{following $\nt{vectorField}$} specifier is that of the vector field at the specified \lstinline{position}.
\item \lstinline{apparently facing $\nt{H}$} means the object has heading \nt{H} with respect to the line of sight from \lstinline{ego}.
For example, \lstinline{apparently facing 90 deg} would orient the object so that the camera views its left side head-on.
\end{myitemize}

\subsection{Statements}
\label{sec:statements}

Finally, we discuss \scenic's statements, listed in Table~\ref{table:statements}.
Class and object definitions have been discussed above, and variable assignment behaves in the standard way.

\begin{table}[tb]
\caption{Statements (excluding \texttt{if}, \texttt{while}, \texttt{def}, \texttt{import}, etc. from Python). Those in the second group are only legal inside behaviors, monitors, and \texttt{compose} blocks.}
\label{table:statements}
\centering
\texttt{
\begin{tabular}{ll}
\toprule
\textrm{Syntax} & \textrm{Meaning} \\
\midrule
model \nt{name} & \textrm{select world model} \\
\nt{name} = \nt{value} & \textrm{variable assignment} \\
param \csl{\nt{name} = \nt{value}} & \textrm{global parameter assignment} \\
\nt{classDefn} \textrm{(see Fig.~\ref{figure:grammar})} & \textrm{class definition} \\
\nt{object} & \textrm{object definition} \\
\nt{behavior} & \textrm{behavior definition} \\
\nt{monitor} & \textrm{monitor definition} \\
\nt{scenario} & \textrm{(modular) scenario definition} \\
require \nt{boolean} & \textrm{hard requirement} \\
require[\nt{number}] \nt{boolean} & \textrm{soft requirement} \\
require always \nt{boolean} & \textrm{always dynamic requirement} \\
require eventually \nt{boolean} & \textrm{eventually dynamic requirement} \\
terminate when \nt{boolean} & \textrm{termination condition} \\
mutate \csl{\nt{name}} \opt{by \nt{number}} & \textrm{enable mutation} \\
\midrule
take \csl{\nt{action}} & \textrm{invoke action(s)} \\
wait & \textrm{invoke no actions this step} \\
terminate & \textrm{end scenario immediately} \\
do \csl{\nt{name}} & \textrm{invoke sub-behavior(s)/sub-scenario(s)} \\
do \csl{\nt{name}} for \nt{scalar} \either{seconds}{steps} & \textrm{invoke with time limit} \\
do \csl{\nt{name}} until \nt{boolean} & \textrm{invoke until condition} \\
\nt{try} \textrm{(see Fig.~\ref{figure:grammar})} & \textrm{try-interrupt statement} \\
abort & \textrm{abort try-interrupt statement} \\
override \nt{name} \csl{\nt{specifier}} & \textrm{override object properties dynamically} \\
\bottomrule
\end{tabular}
}
\end{table}

\myparagraph{Selecting a World Model.}
The \lstinline{model $\nt{name}$} statement specifies that the \scenic{} program is written for the given \scenic{} world model.
It is equivalent to the statement \lstinline{from $\nt{name}$ import *} (as in Python), importing everything from the given \scenic{} module, but can be overridden from the command-line when running the \scenic{} tool.
This enables writing cross-platform scenarios using abstract domains like \texttt{scenic.domains.driving}, then executing them in particular simulators by overriding the model with a more specific module (e.g.~\texttt{scenic.simulators.carla.model}).

\myparagraph{Global Parameters.}
The statement \lstinline{param $\csl{\nt{name} \texttt{ = } \nt{value}}$} assigns values to global parameters of the scenario.
These have no semantics in \scenic{} but provide a general-purpose way to encode arbitrary global information.
For example, in our case study we used parameters \tm{time} and \tm{weather} to put distributions on the time of day and the weather conditions during the scene.

\myparagraph{Behaviors and Monitors.}
The \lstinline{behavior} statement (see Fig.~\ref{figure:grammar}) defines a dynamic behavior.
A behavior definition has the same structure as a function definition, except: 1) it may begin with any number of \lstinline{precondition: $\nt{boolean}$} and \lstinline{invariant: $\nt{boolean}$} lines defining preconditions and invariants; 2) it may use the statements in the second section of Tab.~\ref{table:statements}, which are not allowed in ordinary functions.
The \lstinline{monitor} statement has the same structure as a \lstinline{behavior} statement but defines a monitor.

\myparagraph{Modular Scenarios.}
The \lstinline{scenario} statement (see Fig.~\ref{figure:grammar}) defines a modular scenario which can be invoked from another scenario.
Scenario definitions begin like behavior definitions, with a name, parameters, preconditions, and invariants.
However, the body of a scenario consists of two parts, either of which can be omitted: a \lstinline{setup} block and a \lstinline{compose} block.
The \lstinline{setup} block contains code that runs once when the scenario begins to execute, and is a list of statements like a top-level \scenic{} program\footnote{In fact, a top-level \scenic{} program is equivalent to an unnamed scenario definition with no parameters, preconditions, invariants, or \lstinline{compose} block, and whose \lstinline{start} block consists of the whole program.}.
The \lstinline{compose} block orchestrates the execution of sub-scenarios during a dynamic scenario, and may use \lstinline{do} and any of the other statements allowed inside behaviors (except \lstinline{take}, which only makes sense for an individual agent).

\myparagraph{Requirements.}
The \lstinline{require $\nt{boolean}$} statement requires that the given condition hold in all generated scenarios (equivalently to \emph{observe} statements in other probabilistic programming languages; see e.g.~\cite{blog,claret2013bayesian}).
The variant \lstinline{require[$p$] $\nt{boolean}$} adds a \emph{soft} requirement that need only hold with some probability $p$ (which must be a constant).
We will discuss the semantics of these in the next section.
The \lstinline{require always} and \lstinline{require eventually} variants define requirements that must hold in \emph{every} and \emph{some} time step of the scenario respectively.

\myparagraph{Mutation.}
The \lstinline{mutate $\csl{\nt{instance}}$ ^by^ $\nt{number}$} statement adds Gaussian noise with the given standard deviation (default 1) to the \lstinline{position} and \lstinline{heading} properties of the listed objects (or every \lstinline{Object}, if no list is given).
For example, \lstinline{mutate taxi ^by^ 2} would add twice as much noise as \lstinline{mutate taxi}.
The noise can be controlled separately for \lstinline{position} and \lstinline{heading}, as we discuss in the next section.

\myparagraph{Termination Conditions.}
The \lstinline{terminate when $\nt{boolean}$} statement defines a condition which is monitored as in \lstinline{require eventually}, but which when true causes the scenario to end.
The \lstinline{terminate} statement can be called inside a behavior, monitor, or \lstinline{compose} block to end the scenario immediately.

\myparagraph{Actions.}
The \lstinline{take $\csl{\nt{action}}$} statement can be used inside behaviors to select one or more actions\footnote{The statement will accept lists and tuples of actions, in order to support taking a number of actions that is not fixed, i.e., if \lstinline{myActions} is a list of actions, we can write \lstinline{take myActions}.} for the agent to take in the current time step.
The \lstinline{wait} statement means no actions are taken in this time step (which makes sense inside monitors and \lstinline{compose} blocks).
When either of these statements is executed, the behavior is suspended until one time step has elapsed; then its invariants are checked (raising an \lstinline{InvariantViolation} exception if any are violated) and it is resumed.

\myparagraph{Invoking Other Behaviors and Scenarios.}
The \lstinline{do $\csl{\nt{name}}$} statement has the same structure as the \lstinline{take} statement, but invokes one or more behaviors (if in a behavior) or scenarios (if in a \lstinline{compose} block).
It does not return until the sub-behavior/sub-scenario terminates, so multiple time steps may pass (unlike \lstinline{take}).
Early termination can be enabled by adding a \lstinline{for $\nt{scalar}$ seconds/steps} clause, which enforces a maximum time limit, or an \lstinline{until $\nt{boolean}$} clause, which adds an arbitrary termination criterion.
When the \lstinline{do} statement returns, the invariants of the calling behavior/scenario are checked as above.

\myparagraph{Interrupts.}
The \lstinline{try} statement (see Fig.~\ref{figure:grammar}) consists of a \lstinline{try:} block and one or more \lstinline{interrupt when $\nt{boolean}$:} and \lstinline{except $\nt{exception}$:} blocks, each containing arbitrary lists of statements.
As described in Sec.~\ref{sec:dynamic-scenarios}, when a \lstinline{try} statement executes, the conditions for each \lstinline{interrupt when} block are checked at each time step.
While none of them are true, the \lstinline{try} block executes.
When an interrupt condition becomes true, the body of the corresponding block is executed (with lower blocks preempting those above), suspending any behaviors/scenarios that were executing in the \lstinline{try} block until the interrupt handler completes (at which point the invariants of the suspended behavior/scenario are checked as usual).
Any exceptions raised in the \lstinline{try} block or any interrupt handler can be caught by \lstinline{except} blocks as in the Python \lstinline{try} statement.
Additionally, any block may execute the \lstinline{abort} statement to immediately terminate the entire \lstinline{try} statement.

\myparagraph{Overrides.}
The \lstinline{override $\nt{name}$ $\csl{\nt{specifier}}$} statement may be used inside a scenario definition to override properties of an object during a dynamic scenario.
It has the same structure as an object definition, with \lstinline{override} and the name of the object replacing the class, so for example given an object \lstinline{taxi} we could write \lstinline{override taxi &with aggression& 3} to set the \lstinline{aggression} property of \lstinline{taxi} to $3$.
Dynamic properties read back from the simulator at every time step, like \lstinline{position}, cannot be overridden since they are controlled using actions and not direct assignments.
Properties overridden by a scenario revert to their original values when the scenario terminates.
When the \lstinline{~behavior~} property is overridden, the original behavior is suspended, then resumed at the end of the scenario.

\section{Semantics and Scenario Generation\label{sec:improvisation}}


\subsection{Semantics of \scenic}

The output of a \scenic{} program has two parts: first, a \emph{scene} consisting of an assignment to all the properties of each \lstinline{Object} defined in the scenario, plus any global parameters defined with \lstinline{param}.
For dynamic scenarios, this scene forms the initial state of the scenario, which then changes after each time step according to the actions taken by the agents.
Since actions and their effects are domain-specific (consider for example the different physics involved for aerial, ground, and underwater vehicles), dynamic \scenic{} scenarios do not directly define trajectories for objects.
Instead, the second part of the output of a \scenic{} program is a \emph{policy}, a function mapping the history of past scenes to the choice of actions for the agents in the current time step\footnote{In fact the policy is a probabilistic function, since behaviors can make random choices, and it can also return special values indicating that the scenario should terminate or that it has violated a requirement and should be discarded, as we discuss below.}.
This pair of a scene and a policy is what we mean formally by the \emph{scenario} generated by a \scenic{} program.

Since \scenic{} is a probabilistic programming language, the semantics of a program is actually a \emph{distribution} over possible outputs, here scenarios.
As for other imperative PPLs, the semantics can be defined operationally as a typical interpreter for an imperative language but with two differences.
First, the interpreter makes random choices when evaluating distributions~\cite{saheb-djahromi}.
For example, the \scenic{} statement \lstinline{x = Range(0, 1)} updates the state of the interpreter by assigning a value to \lstinline{x} drawn from the uniform distribution on the interval $(0, 1)$.
In this way every possible run of the interpreter has a probability associated with it.
Second, every run where a \lstinline{require} statement (the equivalent of an ``observation'' in other PPLs) is violated gets discarded, and the run probabilities appropriately normalized (see, e.g., \cite{prob-prog}).
For example, adding the statement \lstinline{require x > 0.5} above would yield a uniform distribution for \lstinline{x} over the interval $(0.5, 1)$.

\scenic{} uses the standard semantics for assignments, arithmetic, loops, functions, and so forth.
Below, we define the semantics of the main constructs unique to \scenic{}.
See the Appendix~\cite{scenic-full} for a more formal treatment.

\myparagraph{Soft Requirements.}
The statement \lstinline{require[p] B} is interpreted as \lstinline{require B} with probability \lstinline{p} and as a no-op otherwise: that is, it is interpreted as a hard requirement that is only checked with probability \lstinline{p}.
This ensures that the condition \lstinline{B} will hold with probability at least \lstinline{p} in the induced distribution of the \scenic{} program, as desired.

\paragraph{Specifiers and Object Definitions.}
As we saw above, each specifier defines a function mapping values for its dependencies to values for the properties it specifies.
When an object of class $C$ is constructed using a set of specifiers $S$, the object is defined as follows (see the Appendix~\cite{scenic-full} for details):
\begin{mylist}
\item If a property is specified (non-optionally) by multiple specifiers in $S$, an ambiguity error is raised.
\item The set of properties $P$ for the new object is found by combining the properties specified by all specifiers in $S$ with the properties inherited from the class $C$.
\item Default value specifiers from $C$ are added to $S$ as needed so that each property in $P$ is paired with a unique specifier in $S$ specifying it, with precedence order: non-optional specifier, optional specifier, then default value.
\item The dependency graph of the specifiers $S$ is constructed. If it is cyclic, an error is raised.
\item The graph is topologically sorted and the specifiers are evaluated in this order to determine the values of all properties $P$ of the new object.
\end{mylist}

\myparagraph{Mutation.}
The \lstinline{mutate X by N} statement sets the special \lstinline{mutationScale} property to \lstinline{N} (the \lstinline{mutate X} form sets it to 1).
At the end of evaluation of the \scenic{} program, but before requirements are checked, Gaussian noise is added to the \lstinline{position} and \lstinline{heading} properties of objects with nonzero \lstinline{mutationScale}.
The standard deviation of the noise is the value of the \lstinline{positionStdDev} and \lstinline{headingStdDev} property respectively (see Table~\ref{table:properties}), multiplied by \lstinline{mutationScale}.

\myparagraph{Dynamic Constructs.}
As suggested in Sec.~\ref{sec:statements}, behaviors and monitors are coroutines: they usually execute like ordinary functions, but are suspended when they \lstinline{take} an action (or \lstinline{wait}) until one time step has passed.
Scenarios behave similarly: in their \lstinline{compose} blocks, using \lstinline{wait} causes them to wait for one step, and any sub-scenarios they invoke using \lstinline{do} run recursively; scenarios without \lstinline{compose} blocks do nothing in a time step other than check whether any of their \lstinline{terminate when} conditions have been met or their \lstinline{require always} conditions violated.

The output of the policy of a dynamic \scenic{} program is defined according to the following procedure:
\begin{enumerate}
    \item Run the \lstinline{compose} blocks of all currently-running scenarios for one time step.
    If any \lstinline{require} conditions fail, discard the simulation.
    If instead the top-level scenario finishes its \lstinline{compose} block (if any), one of its \lstinline{terminate when} conditions is true, or it executes \lstinline{terminate}, set a flag to remember this (we use a flag rather than terminating immediately since we need to ensure that all requirements are satisfied before terminating).

    \item Check all \lstinline{require always} conditions of currently-running scenarios; if any fail, discard the simulation.
    
    \item \label{step:monitors} Run all monitors of currently-running scenarios for one time step.
    As above, discard the simulation if any \lstinline{require} conditions fail, and set the terminate flag if the \lstinline{terminate} statement is executed.
    
    \item \label{step:termination} If the flag is set, check that all \lstinline{require eventually} conditions were satisfied at some time step: if so, terminate the simulation; otherwise, discard it.
    
    \item Run all the behaviors of dynamic agents for one time step, gathering their actions and discarding the simulation or setting the terminate flag as in (\ref{step:monitors}).
    
    \item Repeat (\ref{step:termination}) to check the terminate flag.
    
    \item Return the choice of actions selected by the dynamic agents.
\end{enumerate}

\myparagraph{}
The problem of sampling scenes from the distribution defined by a \scenic{} program is essentially a special case of the sampling problem for imperative PPLs with observations (since soft requirements can also be encoded as observations).
While we could apply general techniques for such problems\footnote{Note however that the presence of dynamic agents complicates the use of standard PPL techniques, since the fact that the physics relating actions to their effects on the world is not modeled in \scenic{} means that the program effectively contains an unknown, black-box function.}, the domain-specific design of \scenic{} enables specialized sampling methods, which we discuss below.
We also note that the scenario generation problem is closely related to \emph{control improvisation}, an abstract framework capturing various problems requiring synthesis under hard, soft, and randomness constraints~\cite{fremont-fsttcs15}.
\emph{Scenario improvisation} from a \scenic{} program can be viewed as an extension with a more detailed randomness constraint given by the imperative part of the program.

\subsection{Domain-Specific Sampling Techniques} \label{sec:sampling-techniques}

The geometric nature of the constraints in \scenic{} programs, together with \scenic{}'s lack of conditional control flow outside behaviors, enable domain-specific sampling techniques inspired by robotic path planning methods.
Specifically, we can use ideas for constructing configuration spaces to prune parts of the sample space where the objects being positioned do not fit into the workspace.
Furthermore, by combining spatial and temporal constraints, we can prune some initial scenes by proving that they \emph{force} a requirement to be violated at some future point during a dynamic scenario.
We describe several pruning techniques below, deferring formal statements of the algorithms to the Appendix~\cite{scenic-full}.

\myparagraph{Pruning Based on Containment.}
The simplest technique applies to any object $X$ whose position is uniform in a region $R$ and which must be contained in a region $C$ (e.g.~the road in our case study).
If \nt{minRadius} is a lower bound on the distance from the center of $X$ to its bounding box, then we can restrict $R$ to $R \cap \erode(C, \nt{minRadius)}$.
This is sound, since if $X$ is centered anywhere not in the restriction, then some point of its bounding box must lie outside of $C$.

\myparagraph{Pruning Based on Orientation.}
The next technique applies to scenarios placing constraints on the relative heading and the maximum distance $M$ between objects $X$ and $Y$, which are oriented with respect to a vector field that is constant within polygonal regions (such as our roads).
For each polygon $P$, we find all polygons $Q_i$ satisfying the relative heading constraints with respect to $P$ (up to a perturbation if $X$ and $Y$ need not be exactly aligned to the field), and restrict $P$ to $P \cap \dilate(\cup Q_i, M)$.
This is also sound: suppose $X$ can be positioned at $x$ in polygon $P$.
Then $Y$ must lie at some $y$ in a polygon $Q$ satisfying the constraints, and since the distance from $x$ to $y$ is at most $M$, we have $x \in \dilate(Q, M)$.

\myparagraph{Pruning Based on Size.}
In the setting above of objects $X$ and $Y$ aligned to a polygonal vector field (with maximum distance $M$), we can also prune the space using a lower bound on the width of the configuration.
For example, in our bumper-to-bumper scenario we can infer such a bound from the \lstinline{offset by} specifiers in the program.
We first find all polygons that are not wide enough to fit the configuration according to the bound: call these ``narrow''.
Then we restrict each narrow polygon $P$ to $P \cap \dilate(\cup Q_i, M)$ where $Q_i$ runs over all polygons except $P$.
To see that this is sound, suppose object $X$ can lie at $x$ in polygon $P$.
If $P$ is not narrow, we do not restrict it; otherwise, object $Y$ must lie at $y$ in some other polygon $Q$.
Since the distance from $x$ to $y$ is at most $M$, as above we have $x \in \dilate(Q, M)$.

\myparagraph{Pruning Based on Reachability.}
Finally, we can prune initial positions for objects which make it impossible to reach a goal location within the duration of the scenario; for example, a car which travels down a road and then runs a red light must start sufficiently close to an intersection.
Suppose an object is required to enter a region $R$ within $T$ time (either by an explicit \lstinline{require eventually} statement or a precondition of a behavior or scenario guaranteed to eventually execute) and we have an upper bound $S$ on the object's speed.
Then we can prune away all initial positions of the object which do not lie within a distance $D = S T$ of $R$, i.e., we can restrict its initial positions to $\dilate(R, D)$.
If the object is also required to stay within some containing region $C$ (e.g., a road) for the entire duration of the scenario, we can compute a tighter value of $D$ by considering only paths that lie within $C$.

\myparagraph{}
After pruning the space as described above, our implementation uses rejection sampling, generating scenes from the imperative part of the scenario until all requirements are satisfied.
While this samples from exactly the desired distribution, it has the drawback that a huge number of samples may be required to yield a single valid scene (in the worst case, when the requirements have probability zero of being satisfied, the algorithm will not even terminate).
However, we found in our experiments that all reasonable scenarios we tried required only several hundred iterations at most, yielding a sample within a few seconds.
Furthermore, the pruning methods above could reduce the number of samples needed by a factor of 3 or more (see the Appendix~\cite{scenic-full} for details of our experiments).
In future work it would be interesting to see whether Markov chain Monte Carlo methods previously used for probabilistic programming (see, e.g., \cite{blog,nori2014r2,wood2014new}) could be made effective in the case of \scenic{}.

\section{Experiments\label{sec:experiments}}


\newcommand{\trand}{\ensuremath{T_\mathrm{generic}}}
\newcommand{\tgood}{\ensuremath{T_\mathrm{good}}}
\newcommand{\tbad}{\ensuremath{T_\mathrm{bad}}}
\newcommand{\ttwocar}{\ensuremath{T_\mathrm{twocar}}}
\newcommand{\tover}{\ensuremath{T_\mathrm{overlap}}}
\newcommand{\tmatrix}{\ensuremath{T_\mathrm{matrix}}}

\newcommand{\xgeneric}{\ensuremath{X_\mathrm{generic}}}
\newcommand{\xtwocar}{\ensuremath{X_\mathrm{twocar}}}
\newcommand{\xover}{\ensuremath{X_\mathrm{overlap}}}
\newcommand{\xmatrix}{\ensuremath{X_\mathrm{matrix}}}

\newcommand{\mgeneric}{\ensuremath{M_\mathrm{generic}}}
\newcommand{\mtwocar}{\ensuremath{M_\mathrm{twocar}}}

We demonstrate the three applications of \scenic{} discussed in Sec.~\ref{sec:overview}: testing a system under particular conditions, either a perception component in isolation (\ref{sec:specialized-testing}) or a dynamic closed-loop system (\ref{sec:carla-falsification}), training a system to improve accuracy in hard cases (\ref{sec:experiment-two}), and debugging failures (\ref{sec:generalizing-failure}).
We begin by describing the general experimental setup.

\subsection{Experimental Setup}
\label{sec:experimental_setup}

For our main case study, we generated scenes in the virtual world of the video game Grand Theft Auto V (GTAV)~\cite{gtav}.
We wrote a \scenic{} world model defining \lstinline{Region}s representing the roads and curbs in (part of) this world, as well as a type of object \lstinline{Car} providing two additional properties\footnote{For the full definition of \lstinline{Car}, see the Appendix~\cite{scenic-full}; the definitions of \tm{road}, \tm{curb}, etc. are a few lines loading the corresponding sets of points from a file storing the GTAV map (see the Appendix for how this file was generated).}:
\texttt{model}, representing the type of car, with a uniform distribution over 13 diverse models provided by GTAV,
and
\texttt{color}, representing the car color,
with a default distribution based on real-world car color statistics~\cite{dupont-colors}.
In addition, we implemented two global scene parameters:
\texttt{time}, representing the time of day, and
\texttt{weather}, representing the weather as one of 14 discrete types supported by GTAV (e.g. ``clear'' or ``snow'').

GTAV is closed-source and does not expose any kind of scene description language.
Therefore, to import scenes generated by \scenic{} into GTAV, we wrote a plugin based on DeepGTAV\footnote{\url{https://github.com/aitorzip/DeepGTAV}}.
The plugin calls internal functions of GTAV to create cars with the desired positions, colors, etc., as well as to set the camera position, time of day, and weather.

Our experiments used SqueezeDet~\cite{squeezedet}, a convolutional neural network real-time object detector for autonomous driving\footnote{Used industrially, for example by DeepScale (http://deepscale.ai/).}.
We used a batch size of $20$ and trained all models for 10,000 iterations unless otherwise noted.
Images captured from GTAV with resolution $1920 \times 1200$ were resized to $1248 \times 384$, the resolution used by SqueezeDet and the standard KITTI benchmark~\cite{KITTI}.
All models were trained and evaluated on NVIDIA TITAN XP GPUs.

We used standard metrics 
\emph{precision} and \emph{recall} to measure the accuracy of detection on a particular image set.
The accuracy is computed based on how well the network predicts 
the correct bounding box, score, and category of objects in the image set.
Details are in the Appendix~\cite{scenic-full}, but in brief,
precision is defined as $tp / (tp + fp)$ and recall as $tp / (tp + fn)$, where
\emph{true positives} $tp$ is the number of correct detections,
\emph{false positives} $fp$ is the number of predicted boxes that
do not match any ground truth box, and \emph{false negatives} $fn$ is the 
number of ground truth boxes that are not detected.

\subsection{Testing and Falsification}
\label{sec:5_2}

We begin with the most straightforward application of \scenic{}, namely generating specialized data to test a system under particular conditions.
We demonstrate both using a static scenario to test a perception component, and using a dynamic scenario to falsify a closed-loop system.

\subsubsection{Testing a Perception Module}
\label{sec:specialized-testing}

When testing a model, one may be interested in a particular operation regime.
For instance, an autonomous car manufacturer may be more interested in certain road conditions (e.g. desert vs. forest roads) depending on where its cars will be mainly used.
\scenic{} provides a systematic way to describe scenarios of interest and construct corresponding test sets.

To demonstrate this, we first wrote very general scenarios describing static scenes of 1--4 cars (not counting the camera), specifying only that the cars face within $10^\circ$ of the road direction.
We generated 1,000 images from each scenario, yielding a training set $\xgeneric$ of 4,000 images, and used these to train a model \mgeneric{} as described in Sec.~\ref{sec:experimental_setup}.
We also generated an additional 50 images from each scenario to obtain a generic test set \trand{} of 200 images.

Next, we specialized the general scenarios in opposite directions: scenarios for good/bad road conditions fixing the time to noon/midnight and the weather to sunny/rainy respectively, generating specialized test sets \tgood{} and \tbad{}.

Evaluating \mgeneric{} on \trand{}, \tgood{}, and \tbad{}, we obtained precisions of $83.1\%$, $85.7\%$, and $72.8\%$, respectively,
and recalls of $92.6\%$, $94.3\%$, and $92.8\%$.
This shows that, as might be expected, the model performs better on bright days than on rainy nights.
This suggests there might not be enough examples of rainy nights in the training set, and indeed under our default weather distribution rain is less likely than shine.
This illustrates how specialized test sets can highlight the weaknesses and strengths of a particular model.
In Sec.~\ref{sec:experiment-two}, we go one step further and use \scenic{} to redesign the training set and improve model performance.

\subsubsection{Falsifying a Dynamic Closed-Loop System}
\label{sec:carla-falsification}

Next, we demonstrate how we can use a dynamic \scenic{} scenario to test a closed-loop system, using \verifai{}'s falsification facilities to monitor and analyze counterexamples to a system-level specification.
We tested an autonomous agent\footnote{\url{https://github.com/carla-simulator/carla/blob/dev/PythonAPI/examples/automatic_control.py}} in the CARLA~\cite{Dosovitskiy17} driving simulator, for which we wrote a similar \scenic{} world model as we did for GTAV.
This agent consists of a planner and controller (but no perception components) which implement basic driving behaviors including abiding by traffic lights, lane following, and collision avoidance.

We wrote a \scenic{} program describing a scenario where the ego vehicle (i.e.~the autonomous agent) is performing a right turn at an intersection, yielding to the crossing traffic.
As the ego approaches the intersection, the traffic light turns green, but a crossing car runs the red light.
The ego vehicle has to decide either to yield or make a right turn. The crossing car executes a reactive behavior where it slows down to maintain a minimum distance with any car in front. 

We allowed three environment parameters to vary in this scenario:
\begin{itemize}
    \item The traffic light's transition from red to green is triggered when the distance between the ego and the crossing car reaches a threshold, which was uniformly random between 10--20 m.
    
    \item The crossing car's speed was uniformly random between 5--12 m/s.
    
    \item The scenario takes place at a random 4-way intersection in the CARLA map.
    To demonstrate how \scenic{} programs can be written in a generic, map-agnostic style, we used the same \scenic{} code on two different CARLA maps (Town05 and Town03).
\end{itemize}

We formulated a safety specification for the autonomous agent in Metric Temporal Logic, stating that the distance between the agent and the crossing car must be greater than 5 meters at all times.
Giving this specification and the \scenic{} program to \verifai{}, we generated 2,000 scenarios for each map.
\verifai{} monitored each simulation and computed the \emph{robustness value} $\rho$ of the MTL specification, which measures how strongly the specification was satisfied~\cite{mtl} (negative values meaning it was violated).

\begin{figure}[tb]
\centering
\includegraphics[width=0.45\textwidth]{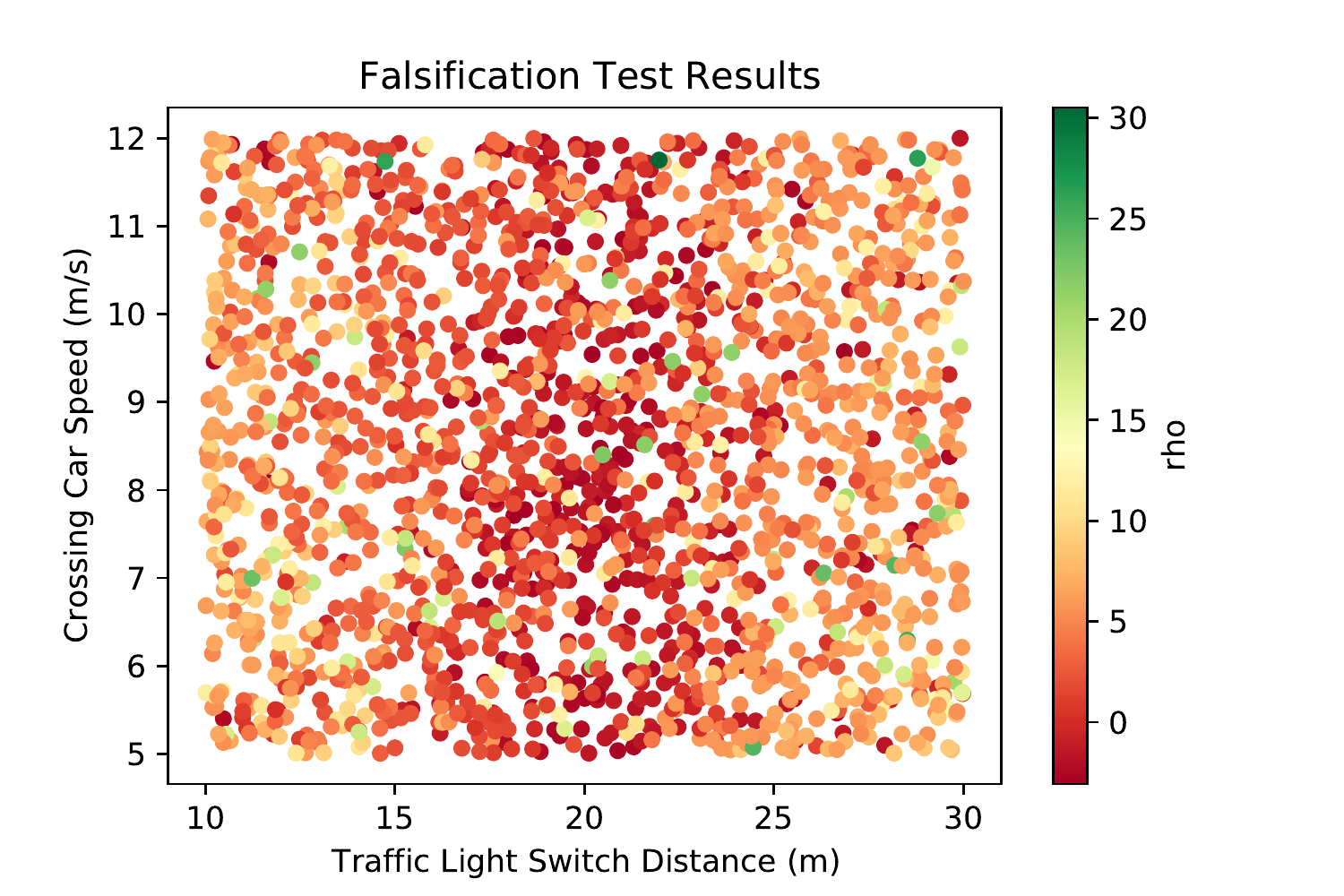}
\includegraphics[width=0.45\textwidth]{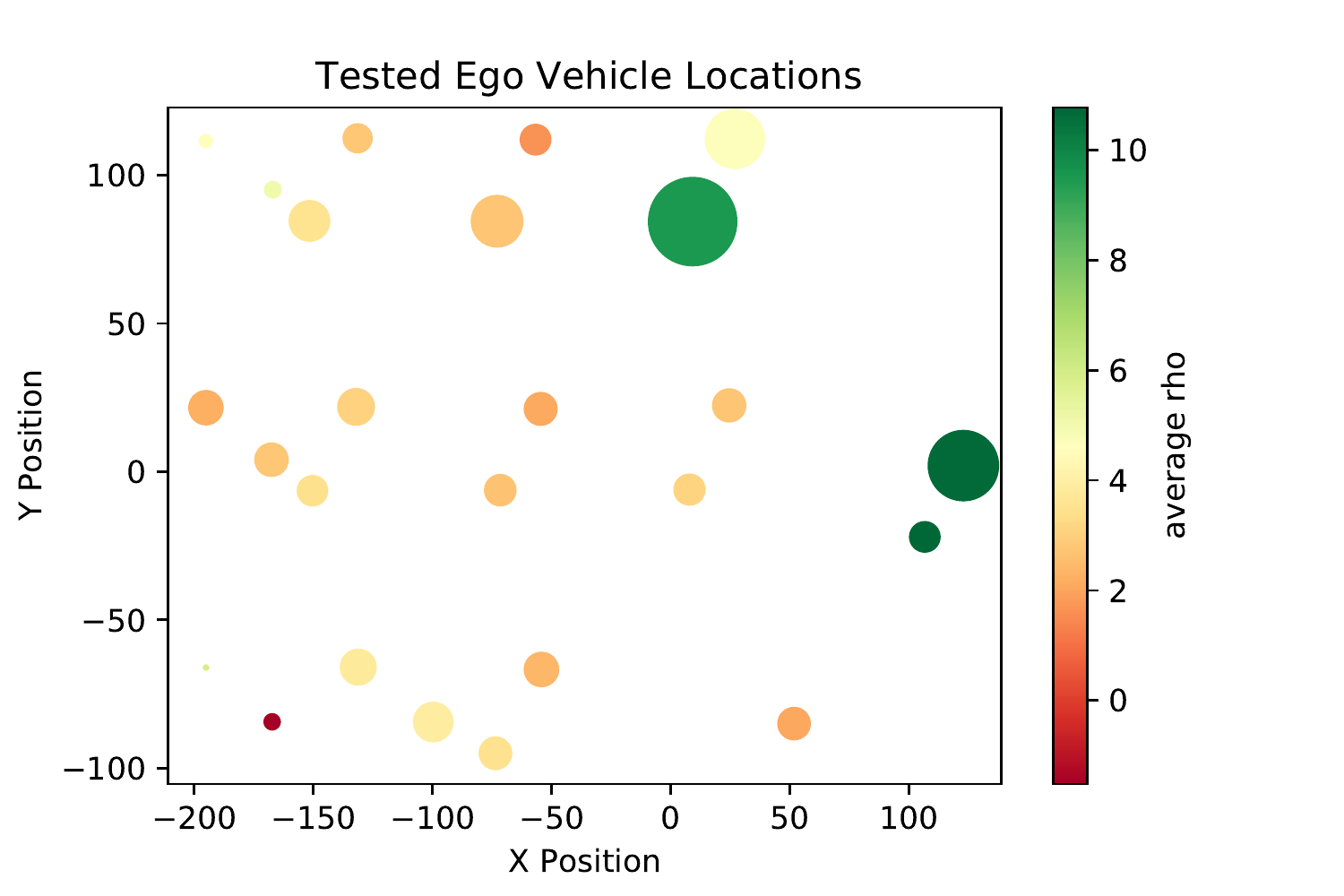}
\includegraphics[width=0.45\textwidth]{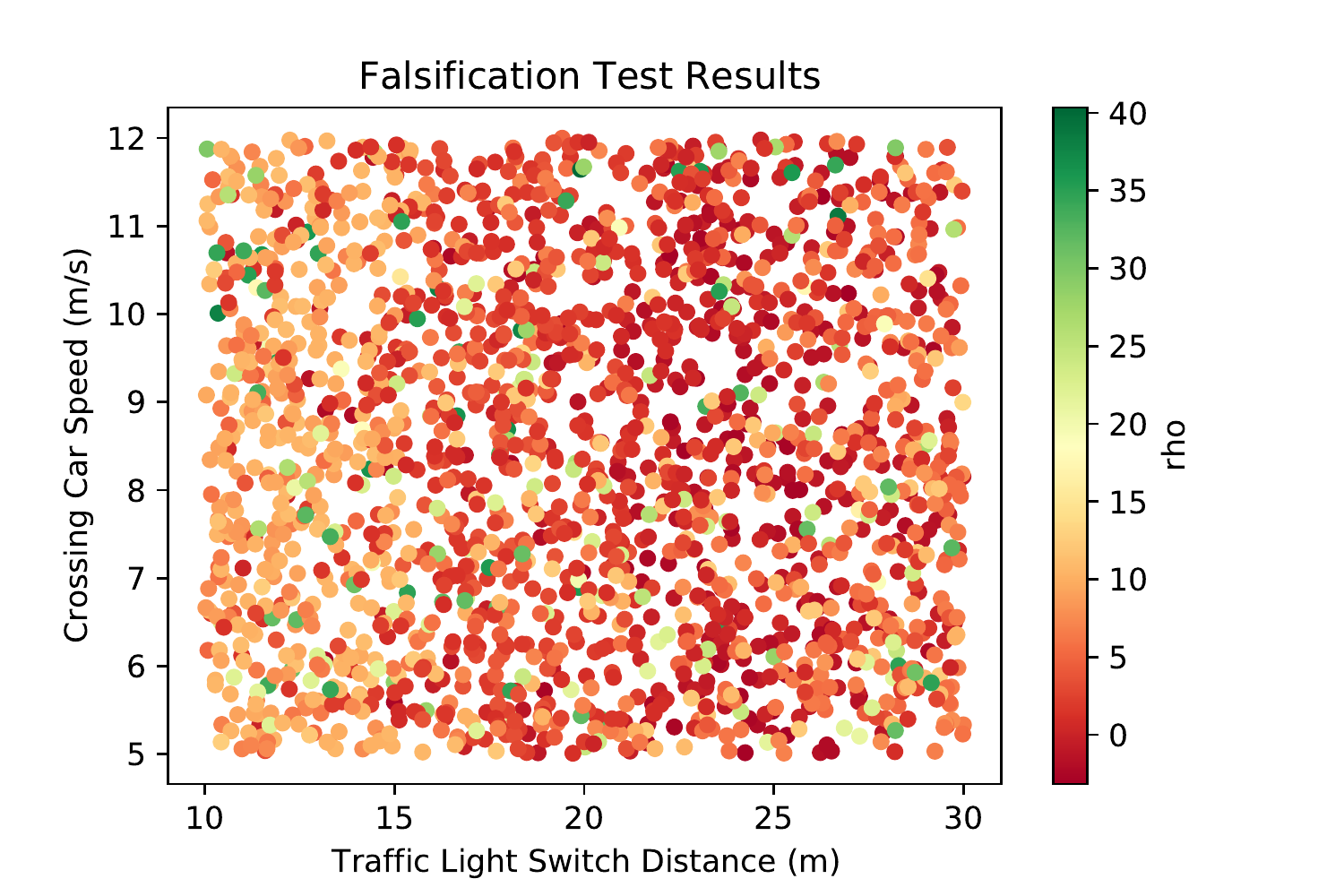}
\includegraphics[width=0.45\textwidth]{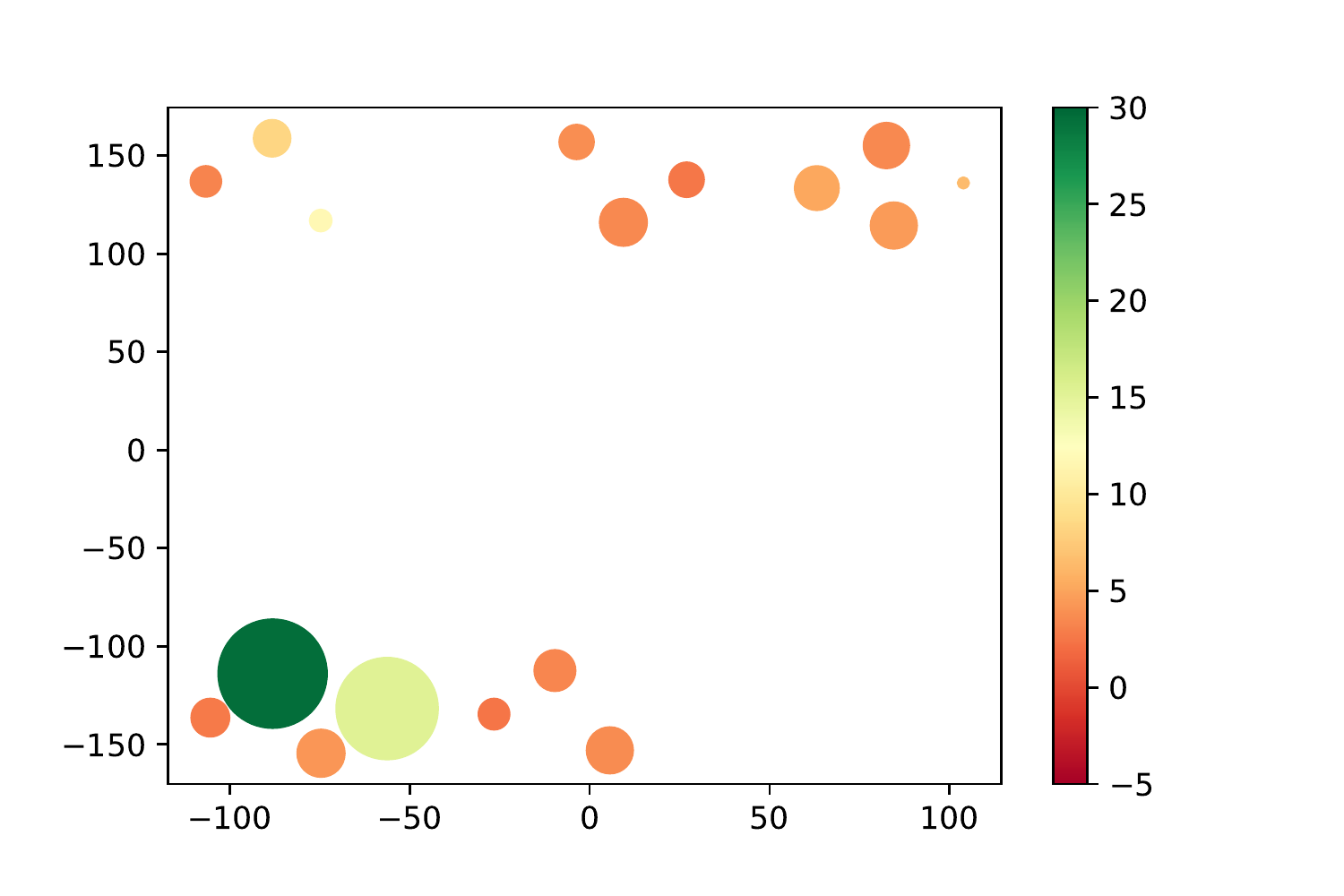}
\caption{Falsification results in CARLA. Top: Town05; bottom: Town03.}
\label{fig:falsification}
\end{figure}

Our results are shown in Fig.~\ref{fig:falsification}.
On the left, we plot $\rho$ as a function of the traffic light trigger threshold and the speed of the crossing car.
Each dot represents one simulation, with redder colors indicating smaller $\rho$, i.e., being closer to violating the safety specification.
We found a significant number of violations, approximately 21\% and 17\% of tests on Town05 and Town03 respectively.
From the plots we observe broadly similar behavior across the two maps, with the distance when the traffic light switch occurs being the dominant factor controlling failures of the autonomous agent (most failures occurring for values of 15--25 m).

On the right side of Fig.~\ref{fig:falsification}, we plot the average value of $\rho$ at each intersection, with color again indicating the average value of $\rho$ and the size of each dot being proportional to its variance.
We can see that some intersections are much easier or harder for the autonomous agent to handle.
Investigating some of the most extreme intersections, we observed that those with 4-lane legs and a turning radius of about 6.5 m caused the agent to fail most frequently.
Re-testing the agent at such intersections, we found that this geometry often created a situation where the agent and the crossing car were merging into the same lane simultaneously, instead of one car completing its maneuver before the other.

These results show how we can use \scenic{} to find scenarios where a closed-loop system violates its specification.
In \ref{sec:generalizing-failure}, we will further show how \scenic{} can help us diagnose the root causes of failures and eliminate them through retraining.

\subsection{Training on Rare Events}
\label{sec:experiment-two}

In the synthetic data setting, we are limited not by data availability but by the cost of training.
The natural question is then how to generate a synthetic data set that as effective as possible given a fixed size.
In this section we show that \emph{over-representing} a type of input that may occur rarely but is difficult for the model can improve performance on the hard case without compromising performance in the typical case.
\scenic{} makes this possible by allowing the user to write a scenario capturing the hard case specifically.

\begin{figure}
\centering
\includegraphics[width=0.49\textwidth]{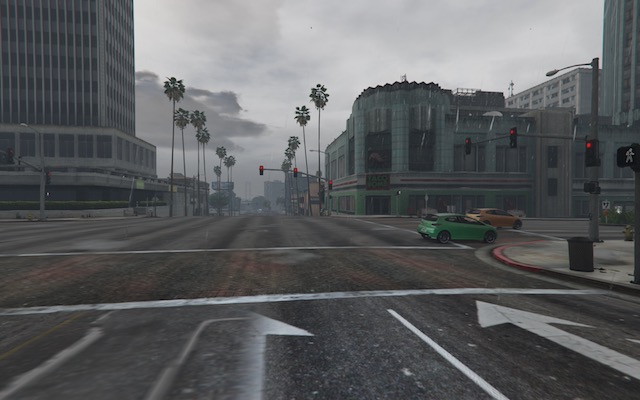}
\includegraphics[width=0.49\textwidth]{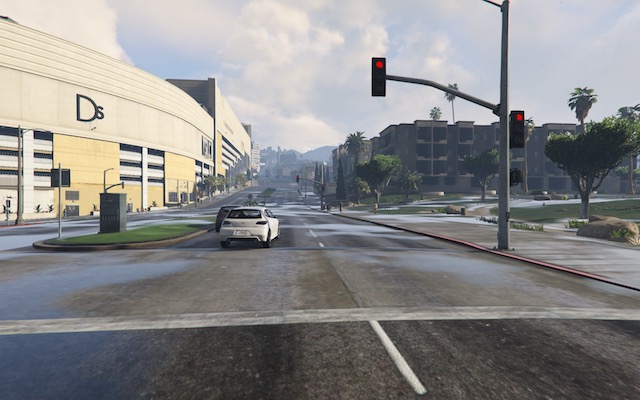}
\caption{Two scenes generated from the partial-occlusion scenario.}
\label{fig:overlapping-images}
\end{figure}

For our car detection task, an obvious hard case is when one car substantially occludes another.
We wrote a simple scenario, shown in Fig.~\ref{fig:overlapping}, which generates such scenes by placing one car behind the other as viewed from the camera, offset left or right so that it is at least partially visible; Fig.~\ref{fig:overlapping-images} shows some of the resulting images.
Generating images from this scenario we obtained a training set \xover{} of 250 images and a test set \tover{} of 200 images.

\begin{figure}
\centering
\begin{lstlisting}
wiggle = Range(-10 deg, 10 deg)
ego = Car &with roadDeviation& wiggle
c = Car visible,
        &with roadDeviation& resample(wiggle)
leftRight = Uniform(1.0, -1.0) * Range(1.25, 2.75)
Car beyond c by (leftRight, Range(4, 10)),
    &with roadDeviation& resample(wiggle)
\end{lstlisting}
\caption{A scenario where one car partially occludes another. The property \tm{roadDeviation} is defined in \lstinline{Car} to mean its \tm{heading} relative to the \tm{roadDirection}.}
\label{fig:overlapping}
\end{figure}

For a baseline training set we used the ``Driving in the Matrix'' synthetic data set~\cite{johnson2017driving}, which has been shown to yield good car detection performance even on real-world images\footnote{We use the ``Matrix'' data set since it is known to be effective for car detection and was not designed by us, making the fact that \scenic{} is able to improve it more striking.
The results of this experiment also hold under the Average Precision (AP) metric used in~\cite{johnson2017driving}, as well as in a similar experiment using the \scenic{} generic two-car scenario from the last section as the baseline.
See Appendix~\cite{scenic-full} for details.}.
Like our images, the ``Matrix'' images were rendered in GTAV; however, rather than using a PPL to guide generation, they were produced by allowing the game's AI to drive around randomly while periodically taking screenshots.
We randomly selected 5,000 of these images to form a training set $\xmatrix$, and 200 for a test set $\tmatrix$.
We trained SqueezeDet for 5,000 iterations on $\xmatrix$, evaluating it on $\tmatrix$ and $\tover$.
To reduce the effect of jitter during training we used a standard technique~\cite{arlot2010}, saving the last 10 models in steps of 10 iterations and picking the one achieving the best total precision and recall.
This yielded the results in the first row of Tab.~\ref{tab:matrix-mixtures}.
Although $\xmatrix$ contains many images of overlapping cars, the precision on $\tover$ is significantly lower than for $\tmatrix$, indicating that the network is predicting lower-quality bounding boxes for such cars\footnote{Recall is much \emph{higher} on $\tover$, meaning the false-negative rate is better; this is presumably because all the $\tover$ images have exactly 2 cars and are in that sense easier than the $\tmatrix$ images, which can have many cars.}.

\begin{table}
	\caption{Performance of models trained on 5,000 images from $\xmatrix$ or a mixture with $\xover$, averaged over 8 training runs with random selections of images from $\xmatrix$.
	\label{tab:matrix-mixtures}}
	\centering
	\begin{tabular}{c  c c  c c}
		\toprule
		Mixture  & \multicolumn{2}{c}{$\tmatrix$} & \multicolumn{2}{c}{$\tover$}\\
		\% 	& Precision & Recall &Precision & Recall\\
		\midrule
		100 / 0 &	$72.9 \pm 3.7$ &	$37.1 \pm 2.1$ &	$62.8 \pm 6.1$ &	$65.7 \pm 4.0$\\
		95 / 5 &	$73.1 \pm 2.3$ &	$37.0 \pm 1.6$ &	$68.9 \pm 3.2$ &	$67.3 \pm 2.4$\\
		\bottomrule
	\end{tabular}
\end{table}

Next we attempted to improve the effectiveness of the training set by mixing in the difficult images produced with \scenic{}.
Specifically, we replaced a random 5\% of $\xmatrix$ (250 images) with images from $\xover$, keeping the overall training set size constant.
We then retrained the network on the new training set and evaluated it as above.
To reduce the dependence on which images were replaced, we averaged over 8 training runs with different random selections of the 250 images to replace.
The results are shown in the second row of Tab.~\ref{tab:matrix-mixtures}.
Even altering only 5\% of the training set, performance on $\tover$ significantly improves.
Critically, the improvement on $\tover$ is not paid for by a corresponding decrease on $\tmatrix$: performance on the original data set remains the same.
Thus, by allowing us to specify and generate instances of a difficult case, \scenic{} enables the generation of more effective training sets than can be obtained through simpler approaches not based on PPLs.

\subsection{Debugging Failures} \label{sec:generalizing-failure}

\begin{figure}
\centering
\includegraphics[width=0.9\columnwidth]{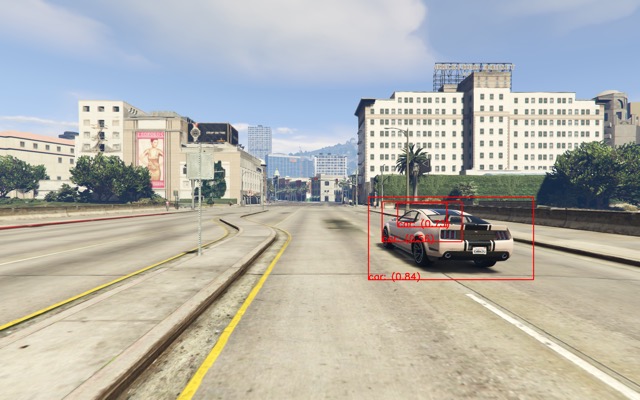}
\caption{The misclassified image, with the predicted bounding boxes.}
\label{figure:misclassified}
\end{figure}

In our final experiment, we show how \scenic{} can be used to generalize a single input on which a model fails, exploring its neighborhood in a variety of different directions and giving insight into which features of the scene are responsible for the failure.
The original failure can then be generalized to a broader scenario describing a class of inputs on which the model misbehaves, which can in turn be used for retraining.
We selected one scene from our first experiment, shown in Fig.~\ref{figure:misclassified}, consisting of a single car viewed from behind at a slight angle, which $\mgeneric$ wrongly classified as three cars (thus having $33.3\%$ precision and $100\%$ recall).
We wrote several scenarios which left most of the features of the scene fixed but allowed others to vary.
Specifically, scenario (1) varied the model and color of the car, (2) left the position and orientation of the car relative to the camera fixed but varied the absolute position, effectively changing the background of the scene, and (3) used the mutation feature of \scenic{} to add a small amount of noise to the car's position, heading, and color.
For each scenario we generated 150 images and evaluated $\mgeneric$ on them.
As seen in Tab.~\ref{exp3}, changing the model and color improved performance the most, suggesting they were most relevant to the misclassification, while local position and orientation were less important and global position (i.e. the background) was least important.

\begin{table}
\centering
\caption{Performance of $\mgeneric$ on different scenarios representing variations of the image in Fig.~\ref{figure:misclassified}.}
\label{exp3}
\begin{tabular}{lcc}
\toprule
Scenario     & \makebox[7ex]{Precision} & Recall \\
\midrule
(1) varying model and color              & \textbf{80.3}      & 100    \\
(2) varying background                 & 50.5      & 99.3    \\
(3) varying local position, orientation                     & 62.8      & 100    \\
\midrule
(4) varying position but staying close                      & 53.1      & 99.3   \\
(5) any position, same apparent angle               & 58.9      & 98.6   \\
(6) any position and angle       & 67.5      & 100    \\
(7) varying background, model, color & 61.3      & 100    \\
\midrule
(8) staying close, same apparent angle        & 52.4      & 100    \\
(9) staying close, varying model             & 58.6      & 100    \\
\bottomrule
\end{tabular}
\end{table}

To investigate these possibilities further, we wrote a second round of variant scenarios, also shown in Tab.~\ref{exp3}.
The results confirmed the importance of model and color (compare (2) to (7)), as well as angle (compare (5) to (6)), but also suggested that being close to the camera could be the relevant aspect of the car's local position.
We confirmed this with a final round of scenarios (compare (5) and (8)), which also showed that the effect of car model is small among scenes where the car is close to the camera (compare (4) and (9)).

Having established that car model, closeness to the camera, and view angle all contribute to poor performance of the network, we wrote broader scenarios capturing these features.
To avoid overfitting, and since our experiments indicated car model was not very relevant when the car is close to the camera, we decided not to fix the car model.
Instead, we specialized the generic one-car scenario from our first experiment to produce only cars close to the camera.
We also created a second scenario specializing this further by requiring that the car be viewed at a shallow angle.

Finally, we used these scenarios to retrain $\mgeneric$, hoping to improve performance on its original test set $\trand$ (to better distinguish small differences in performance, we increased the test set size to 400 images).
To keep the size of the training set fixed as in the previous experiment, we replaced 400 one-car images in $\xgeneric$ (10\% of the whole training set) with images generated from our scenarios.
As a baseline, we used images produced with classical image augmentation techniques implemented in \texttt{imgaug} \cite{imgaug}.
Specifically, we modified the original misclassified image by randomly cropping 10\%--20\% on each side, flipping horizontally with probability 50\%, and applying Gaussian blur with $\sigma \in [0.0, 3.0]$.

The results of retraining $\mgeneric$ on the resulting data sets are shown in Tab.~\ref{augmentation_res}.
Interestingly, classical augmentation actually \emph{hurt} performance, presumably due to overfitting to relatively slight variants of a single image.
On the other hand, replacing part of the data set with specialized images of cars close to the camera significantly reduced the number of false positives like the original misclassification (while the improvement for the ``shallow angle'' scenario was less, perhaps due to overfitting to the restricted angle range).
This demonstrates how \scenic{} can be used to improve performance by generalizing individual failures into scenarios that capture the essence of the problem but are broad enough to prevent overfitting during retraining.

\begin{table}
\centering
\caption{Performance of $\mgeneric$ after retraining, replacing 10\% of $\xgeneric$ with different data.}
\label{augmentation_res}
\begin{tabular}{ccc}
\toprule
Replacement Data & Precision & Recall \\
\midrule
Original (no replacement)                 & 82.9      & 92.7   \\
Classical augmentation & 78.7      & 92.1   \\
\midrule
Close car                 & 87.4      & 91.6   \\
Close car at shallow angle         & 84.0      & 92.1   \\
\bottomrule
\end{tabular}
\end{table}

\section{Related Work} \label{sec:related-work}


\myparagraph{Data Generation and Testing for ML.}
There has been a large amount of work on generating synthetic data for specific applications, including text recognition~\cite{jaderberg2014synthetic}, text localization~\cite{gupta2016synthetic}, robotic object grasping~\cite{tobin2017domain}, and autonomous driving~\cite{johnson2017driving,filipowicz2017learning}.
Closely related is work on \emph{domain adaptation}, which attempts to correct differences between synthetic and real-world input distributions.
Domain adaptation has enabled synthetic data to successfully train models for several other applications including 3D object detection~\cite{liebelt2010multi,stark2010back}, pedestrian detection~\cite{vazquez2014virtual}, and semantic image segmentation~\cite{ros2016synthia}.
Such work provides important context for our paper, showing that models trained exclusively on synthetic data (possibly domain-adapted) can achieve acceptable performance on real-world data.
The major difference in our work is that we provide, through \scenic{}, language-based systematic data generation for \emph{any} cyber-physical system.

Some works have also explored the idea of using adversarial examples
(i.e. misclassified examples) to retrain and improve ML models (e.g.,~\cite{xu2016improved,wong2016understanding,goodfellowSS14}).
In particular, 
Generative Adversarial Networks (GANs)~\cite{goodfellow2014generative}, a particular kind of neural network able to generate synthetic data, have been used to augment training sets~\cite{liang2017recurrent,marchesi2017megapixel}. The difference 
with \scenic{} is that GANs require an initial training set/pretrained model and do not easily incorporate declarative constraints, while \scenic{} produces synthetic data in an explainable, programmatic fashion requiring only a simulator.

\paragraph{Model-Based Test Generation.}
Techniques using a model to guide test generation have long existed~\cite{Broy:2005}.
A popular approach is to provide \emph{example tests}, as in mutational fuzz testing \cite{fuzzing-book} and example-based scene synthesis \cite{scene-synthesis}.
While these methods are easy to use, they do not provide fine-grained control over the generated data.
Another approach is to give \emph{rules} or a \emph{grammar} specifying how the data can be generated, as in generative fuzz testing \cite{fuzzing-book}, procedural generation from shape grammars \cite{procedural-modeling}, and grammar-based scene synthesis \cite{jiang2018configurable}.
While grammars allow much greater control, they do not easily allow enforcing global properties.
This is also true when writing a \emph{program} in a domain-specific language with nondeterminism \cite{concurrit}.
Conversely, \emph{constraints} as in constrained-random verification \cite{crv} allow global properties but can be difficult to write.
\scenic{} improves on these methods by simultaneously providing fine-grained control, enforcement of global properties, specification of probability distributions, and simple imperative syntax.

\paragraph{Probabilistic Programming Languages.}
The semantics (and to some extent, the syntax) of \scenic{} are similar to that of other probabilistic programming languages such as \textsc{Prob}~\cite{prob-prog}, Church~\cite{church}, and BLOG~\cite{blog}.
In probabilistic programming the focus is usually on \emph{inference} rather than \emph{generation} (the main application in our case), and in particular to our knowledge probabilistic programming languages have not previously been used for test generation.
However, the most popular inference techniques are based on sampling and so could be directly applied to generate scenes from \scenic{} programs, as we discussed in Sec.~\ref{sec:improvisation}.

Several probabilistic programming languages have been used to define generative models of objects and scenes: both general-purpose languages such as WebPPL~\cite{webppl} (see, e.g., \cite{ritchie-thesis}) and languages specifically motivated by such applications, namely Quicksand~\cite{quicksand} and Picture~\cite{picture}.
The latter are in some sense the most closely-related to \scenic{}, although neither provides specialized syntax or semantics for dealing with geometry or dynamic behaviors (Picture also was used only for inverse rendering, not data generation).
The main advantage of \scenic{} over these languages is that its domain-specific design permits concise representation of complex scenarios and enables specialized sampling techniques.

\paragraph{Scenario Description Languages for Autonomous Driving.}
Recently, formal dynamic scenario description languages have been proposed for the domain of autonomous driving. The Paracosm language~\cite{paracosm} is used to model dynamic scenarios with a reactive and synchronous model of computation.
However, it is not a PPL, so it lacks probability distributions and declarative constraints; it also does not provide constructs like \scenic{}'s interrupts which allow easy customization of generic behavior models.
The Measurable Scenario Description Language (M-SDL)~\cite{msdl}, introduced after the first version of \scenic{}, does provide declarative constraints, as well as compositional features similar to those we introduced in this paper.
However, compared to both of these languages, \scenic{} has several distinguishing features: (i) it provides a much higher-level, declarative way of specifying geometric constraints; (ii) it is fundamentally a probabilistic programming language (as opposed to M-SDL where distributions are optional), and (iii) it is not specific to the autonomous driving domain (as demonstrated in~\cite{scenic-pldi,fremont-cav20}).

\section{Conclusion\label{sec:conclusion}}


In this paper, we introduced \scenic{}, a probabilistic
programming language for specifying distributions
over configurations of physical objects and the behaviors of dynamic agents.
We showed how \scenic{} can be used to generate
synthetic data sets useful for a variety of tasks in the design of robust ML-based cyber-physical systems. Specifically,
we used \scenic{} to generate specialized test sets and falsify a system,
improve the robustness of a system by emphasizing difficult cases in its training set,
and generalize from individual failure cases to broader scenarios suitable for retraining.
In particular, by training on hard cases generated by \scenic{}, we were able to boost the performance of a car detector neural network (given a fixed training set size) significantly beyond what could be achieved by prior synthetic data generation methods~\cite{johnson2017driving} not based on PPLs.

In future work we plan to conduct experiments applying \scenic{} to a variety of additional domains, applications, and simulators.
As we mentioned in the Introduction, we have already successfully applied \scenic{} to aircraft~\cite{fremont-cav20}, and we are currently investigating applications in further domains including underwater vehicles and indoor robots.
We also plan to extend the \scenic{} language itself in several directions, including allowing user-defined specifiers and describing 3D scenes.
Finally, we are exploring ways to combine \scenic{} with automated analyses: in particular, reducing the human burden of writing \scenic{} programs through algorithms for synthesizing or adapting such programs (e.g.~\cite{kim-cvpr20}), and improving the efficiency of falsification by performing white-box analyses of the system.


\bibliographystyle{spmpsci}
\bibliography{main,biblio_tom,refs-sas}

\clearpage

\appendix
\section{Gallery of Scenarios} \label{sec:gallery}


This section presents \scenic{} code for a variety of scenarios from our autonomous car case study (and the robot motion planning example used in Sec.~\ref{sec:motivating_examples}), along with images rendered from them.
The scenarios range from simple examples used above to illustrate different aspects of the language, to those representing interesting road configurations like platoons and lanes of traffic.

\etocsetnexttocdepth{subsection}
\localtableofcontents

\newpage

\subsection{The \texttt{scenic.simulators.gta.model} Module}

All the scenarios below begin with a line (not shown here) importing the \scenic{} world model for the GTA simulator, \texttt{scenic.simulators.gta.model}, which as explained above contains all definitions specific to our autonomous car case study.
These include the definitions of the regions \lstinline{road} and \lstinline{curb}, as well as the vector field \lstinline{roadDirection} giving the prevailing traffic direction at each point on the road.
Most importantly, it also defines \lstinline{Car} as a type of object:

\begin{lstlisting}
class Car:
    position: Point on road
    heading: (roadDirection at self.position) \
             + self.roadDeviation
    roadDeviation: 0
    width: self.~model~.width
    height: self.~model~.height
    viewAngle: 80 deg
    visibleDistance: 30
    ~model~: CarModel.defaultModel()
    color: CarColor.defaultColor()
\end{lstlisting}

Most of the properties are inherited from \lstinline{Object} or are self-explanatory.
The property \texttt{roadDeviation}, representing the heading of the car with respect to the local direction of the road, is purely a syntactic convenience; the following two lines are equivalent:

\begin{lstlisting}
Car facing 10 deg relative to roadDirection
Car with roadDeviation 10 deg
\end{lstlisting}

The world model also defines a few convenience subclasses of \lstinline{Car} with different default properties. For example, \lstinline{EgoCar} overrides \lstinline{model} with the fixed car model we used for the ego car in our interface to GTA V.

\clearpage

\subsection{The Simplest Possible Scenario}

This scenario, creating a single car with no specified properties, was used as an example in Sec.~\ref{sec:motivating_examples}. \\

\begin{lstlisting}
ego = Car
Car
\end{lstlisting}

\begin{figure*}[b]
\centering
\includegraphics[width=0.49\textwidth]{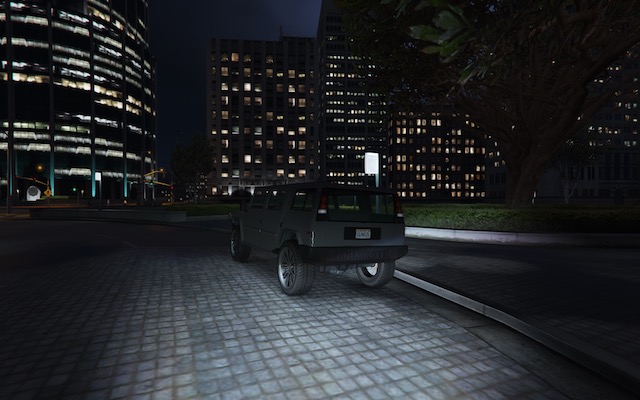}
\includegraphics[width=0.49\textwidth]{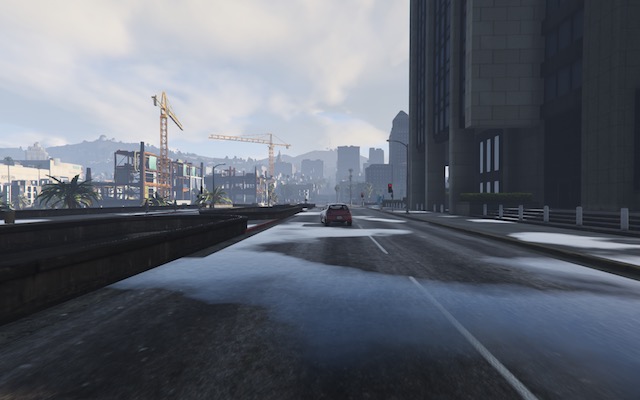}
\includegraphics[width=0.49\textwidth]{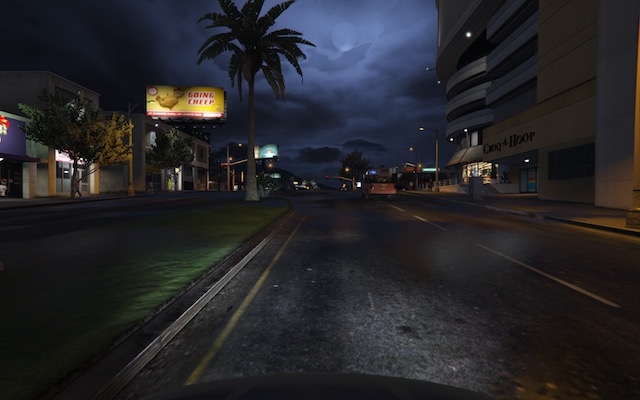}
\includegraphics[width=0.49\textwidth]{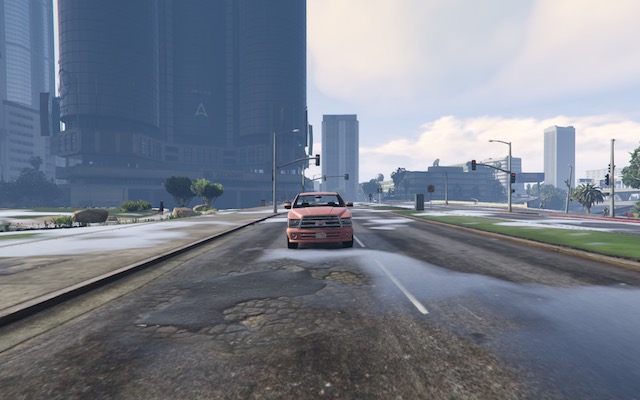}
\caption{Scenes generated from a \scenic{} scenario representing a single car (with reasonable default properties).}
\label{figure:gallery-simplest}
\end{figure*}

\clearpage

\subsection{A Single Car}

This scenario is slightly more general than the previous, allowing the car (and the ego car) to deviate from the road direction by up to 10$^\circ$.
It also specifies that the car must be visible, which is in fact redundant since this constraint is built into \scenic{}, but helps guide the sampling procedure.
This scenario was also used as an example in Sec.~\ref{sec:motivating_examples}. \\

\begin{lstlisting}
wiggle = (-10 deg, 10 deg)
ego = EgoCar with roadDeviation wiggle
Car visible, with roadDeviation resample(wiggle)
\end{lstlisting}

\begin{figure*}[b]
\centering
\includegraphics[width=0.49\textwidth]{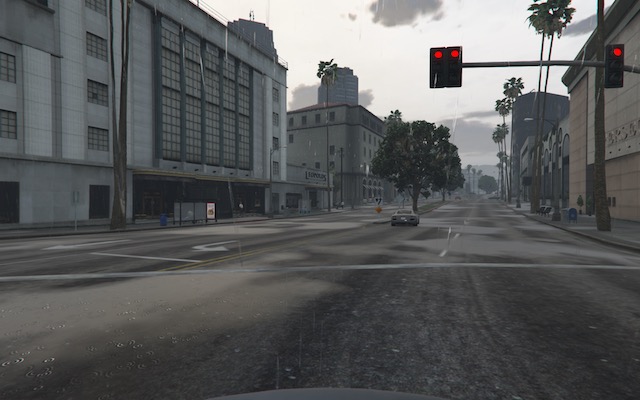}
\includegraphics[width=0.49\textwidth]{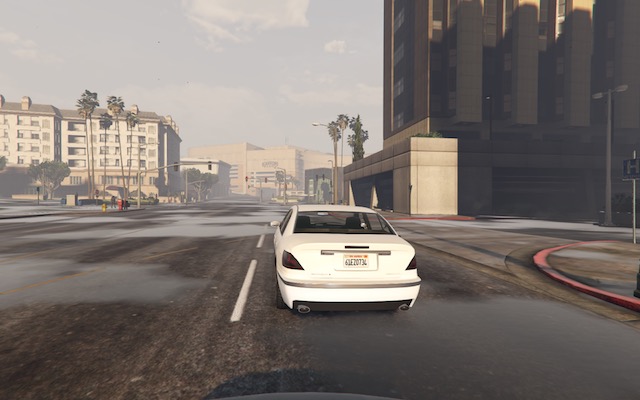}
\includegraphics[width=0.49\textwidth]{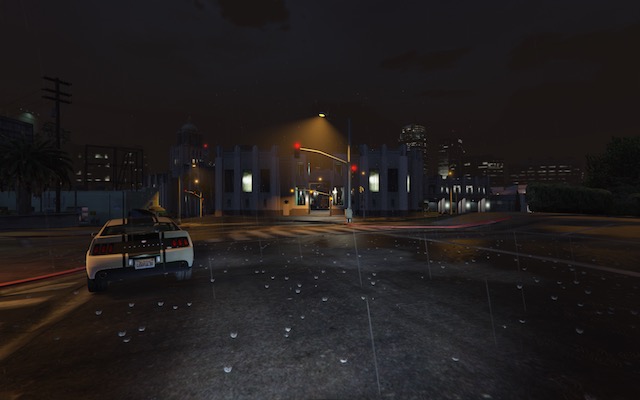}
\includegraphics[width=0.49\textwidth]{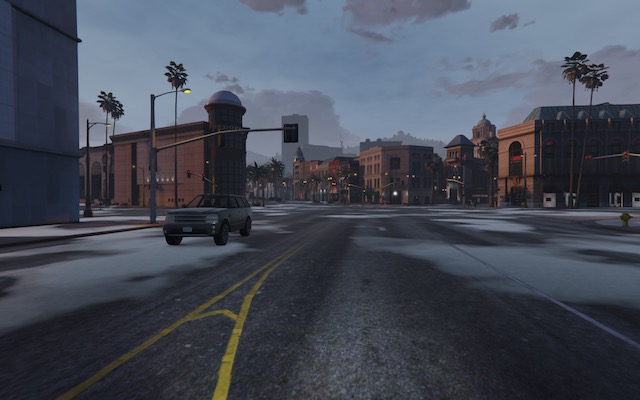}
\caption{Scenes generated from a \scenic{} scenario representing a single car facing roughly the road direction.}
\label{figure:gallery-oneCar}
\end{figure*}

\clearpage

\subsection{A Badly-Parked Car}

This scenario, creating a single car parked near the curb but not quite parallel to it, was used as an example in Sec.~\ref{sec:motivating_examples}. \\

\begin{lstlisting}
ego = Car
spot = OrientedPoint on visible curb
badAngle = Uniform(1.0, -1.0) * (10, 20) deg
Car left of spot by 0.5, facing badAngle relative to roadDirection
\end{lstlisting}

\begin{figure*}[b]
\centering
\includegraphics[width=0.49\textwidth]{badlyParked1.jpg}
\includegraphics[width=0.49\textwidth]{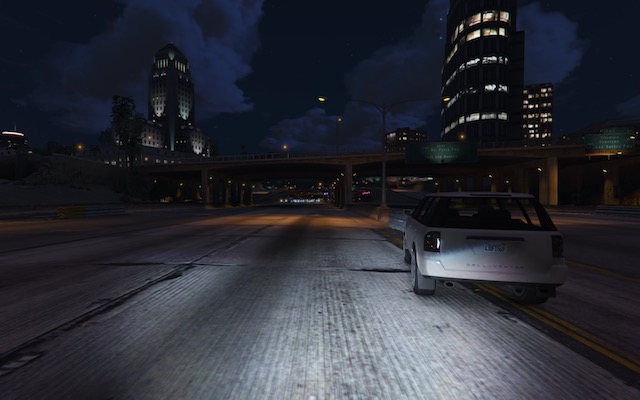}
\includegraphics[width=0.49\textwidth]{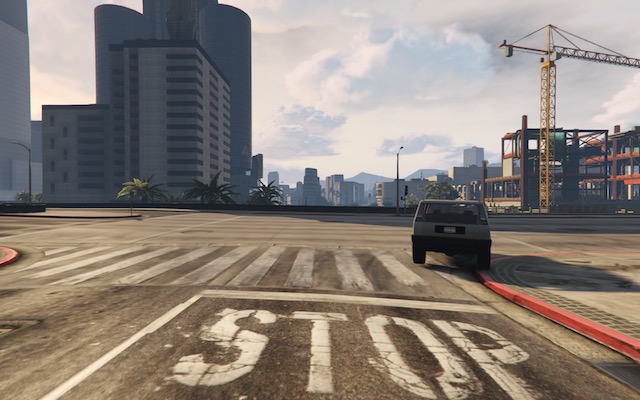}
\includegraphics[width=0.49\textwidth]{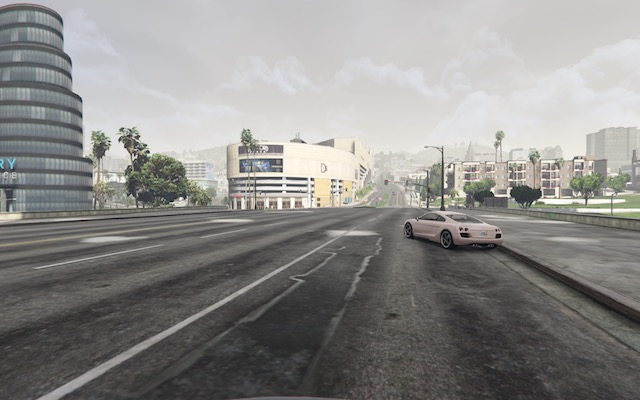}
\caption{Scenes generated from a \scenic{} scenario representing a badly-parked car.}
\label{figure:gallery-badlyParkedCar}
\end{figure*}

\clearpage

\subsection{An Oncoming Car}

This scenario, creating a car 20--40 m ahead and roughly facing towards the camera, was used as an example in Sec.~\ref{sec:motivating_examples}.
Note that since we do not specify the orientation of the car when creating it, the default \texttt{heading} is used and so it will face the road direction.
The \texttt{require} statement then requires that this orientation is also within $15^\circ$ of facing the camera (as the view cone is $30^\circ$ wide). \\

\begin{lstlisting}
ego = Car
car2 = Car offset by (-10, 10) @ (20, 40), with viewAngle 30 deg
require car2 can see ego
\end{lstlisting}

\begin{figure*}[b]
\centering
\includegraphics[width=0.49\textwidth]{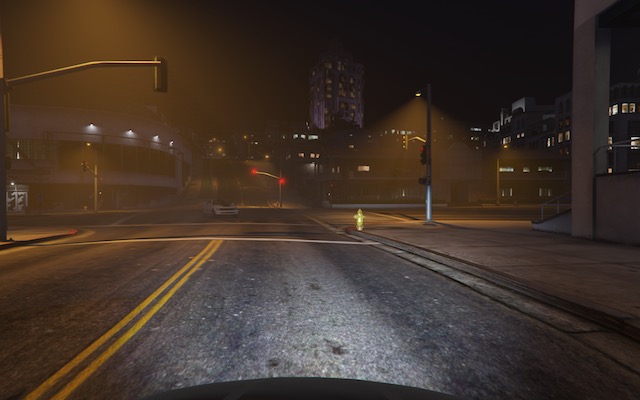}
\includegraphics[width=0.49\textwidth]{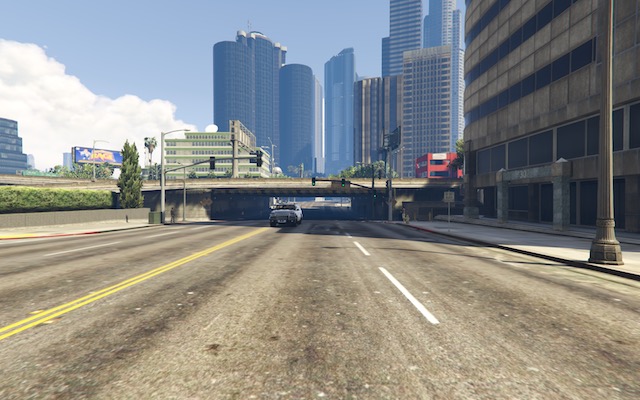}
\includegraphics[width=0.49\textwidth]{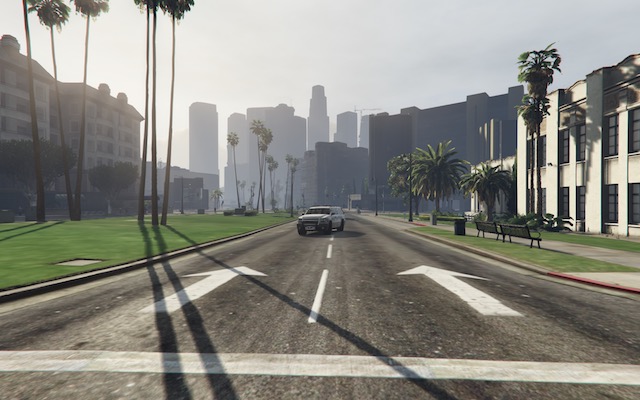}
\includegraphics[width=0.49\textwidth]{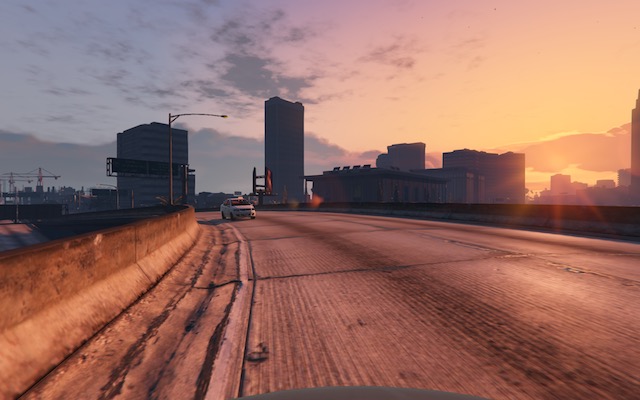}
\caption{Scenes generated from a \scenic{} scenario representing a car facing roughly towards the camera.}
\label{figure:gallery-oncoming}
\end{figure*}

\clearpage

\subsection{Adding Noise to a Scene} \label{sec:noise-code}

\begin{figure}
\centering
\includegraphics[width=0.9\columnwidth]{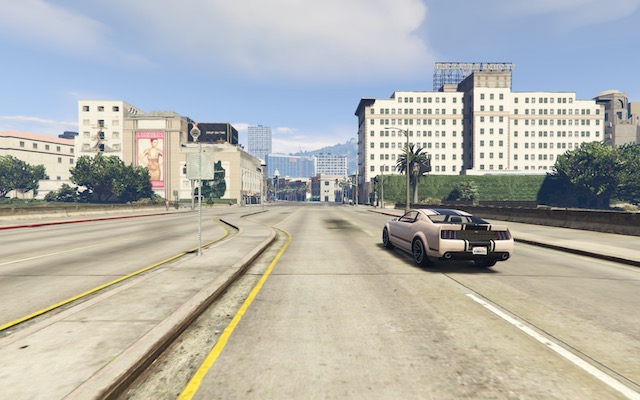}
\caption{The original misclassified image in Sec.~\ref{sec:generalizing-failure}.}
\label{figure:gallery-original}
\end{figure}

This scenario, using \scenic{}'s mutation feature to automatically add noise to an otherwise completely-specified scenario, was used in the experiment in Sec.~\ref{sec:generalizing-failure} (it is Scenario (3) in Table~\ref{exp3}).
The original scene, which is exactly reproduced by this scenario if the \texttt{mutate} statement is removed, is shown in Fig.~\ref{figure:gallery-original}. \\

\begin{lstlisting}
param time = 12 * 60    # noon
param weather = 'EXTRASUNNY'

ego = EgoCar at -628.7878 @ -540.6067,
             facing -359.1691 deg

Car at -625.4444 @ -530.7654, facing 8.2872 deg,
    with ~model~ CarModel.models['DOMINATOR'],
    with color CarColor.byteToReal([187, 162, 157])

mutate
\end{lstlisting}

\begin{figure*}[b]
\centering
\includegraphics[width=0.49\textwidth]{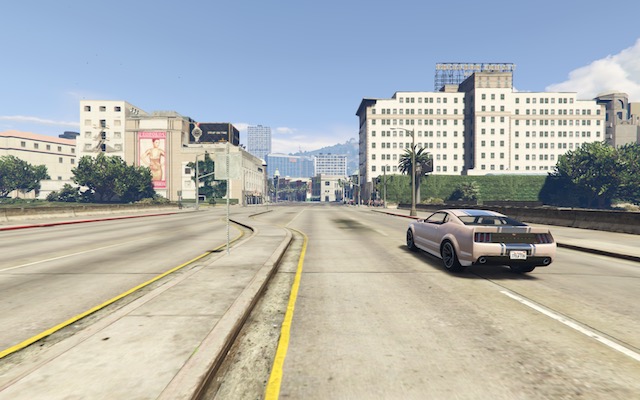}
\includegraphics[width=0.49\textwidth]{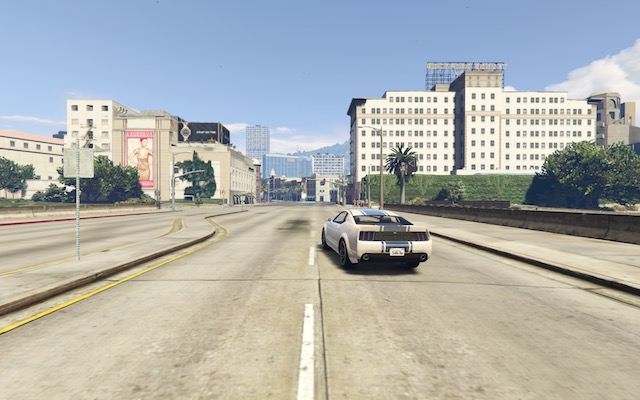}
\includegraphics[width=0.49\textwidth]{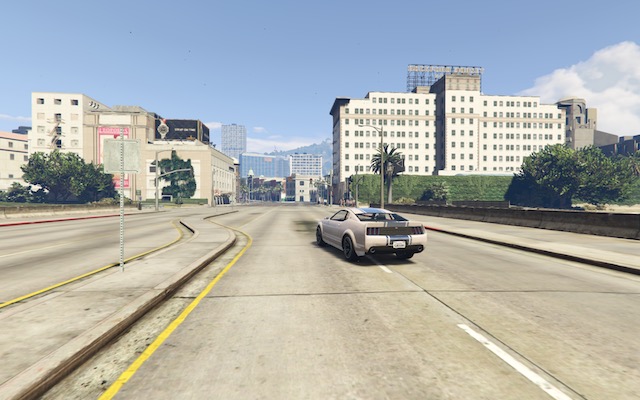}
\includegraphics[width=0.49\textwidth]{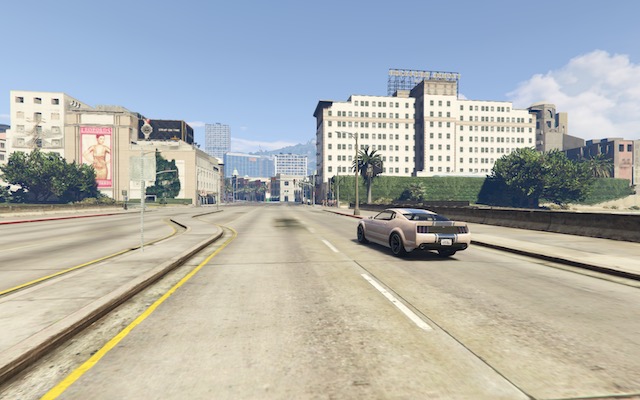}
\caption{Scenes generated from a \scenic{} scenario adding noise to the scene in Fig.~\ref{figure:gallery-original}.}
\label{figure:gallery-noise}
\end{figure*}

\clearpage

\clearpage

\subsection{Two Cars}

This is the generic two-car scenario used in the experiments in Secs.~\ref{sec:5_2} and \ref{sec:experiment-two}. \\

\begin{lstlisting}
wiggle = (-10 deg, 10 deg)
ego = EgoCar with roadDeviation wiggle
Car visible, with roadDeviation resample(wiggle)
Car visible, with roadDeviation resample(wiggle)
\end{lstlisting}

\begin{figure*}[b]
\centering
\includegraphics[width=0.49\textwidth]{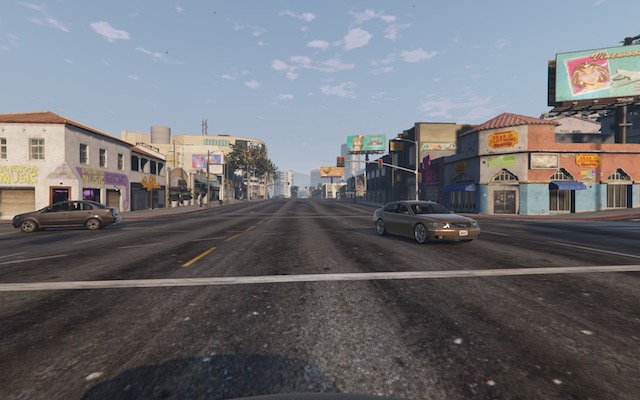}
\includegraphics[width=0.49\textwidth]{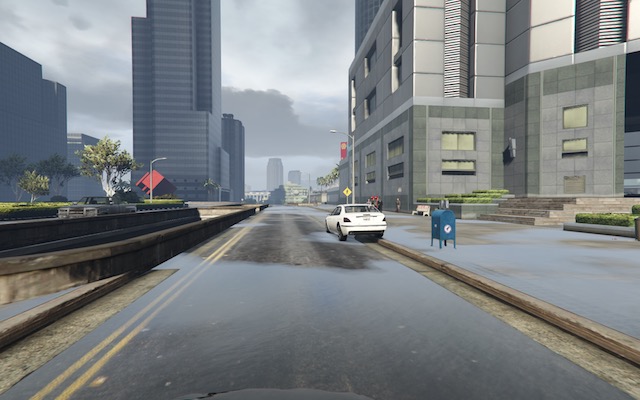}
\includegraphics[width=0.49\textwidth]{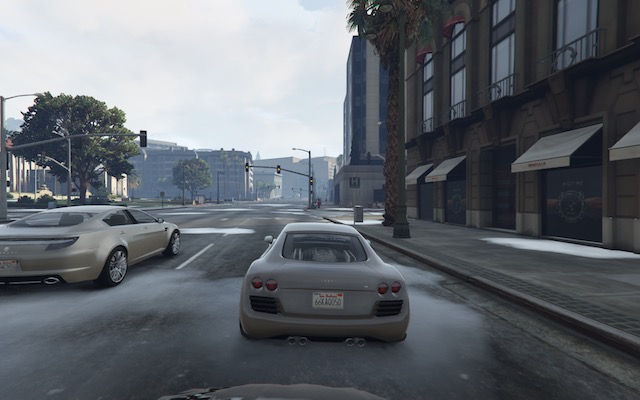}
\includegraphics[width=0.49\textwidth]{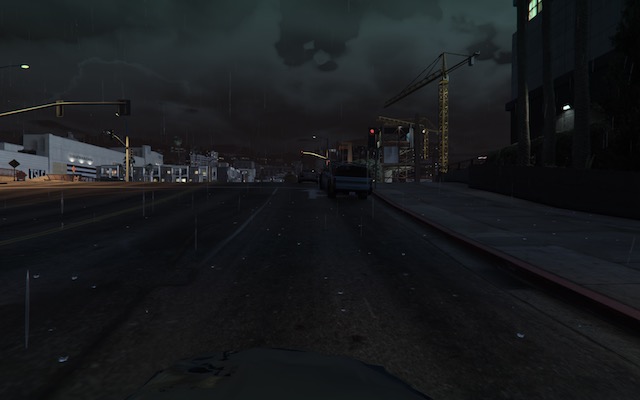}
\caption{Scenes generated from a \scenic{} scenario representing two cars, facing close to the direction of the road.}
\label{figure:gallery-twoCars}
\end{figure*}

\clearpage

\subsection{Two Overlapping Cars} \label{sec:overlapping-code}

This is the scenario used to produce images of two partially-overlapping cars for the experiment in Sec.~\ref{sec:experiment-two}. \\

\begin{lstlisting}
wiggle = (-10 deg, 10 deg)
ego = EgoCar with roadDeviation wiggle

c = Car visible, with roadDeviation resample(wiggle)

leftRight = Uniform(1.0, -1.0) * (1.25, 2.75)
Car beyond c by leftRight @ (4, 10), with roadDeviation resample(wiggle)
\end{lstlisting}

\begin{figure*}[b]
\centering
\includegraphics[width=0.49\textwidth]{overlapping1.jpg}
\includegraphics[width=0.49\textwidth]{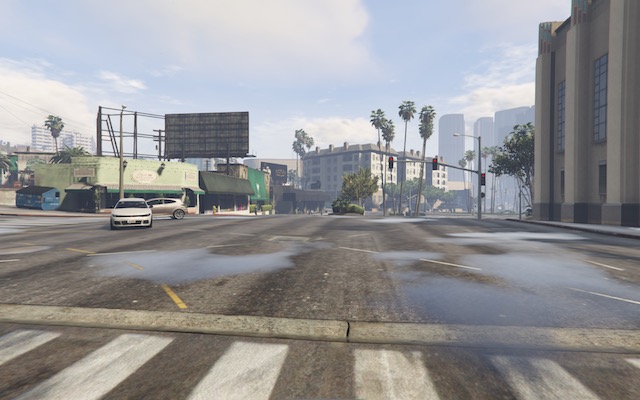}
\includegraphics[width=0.49\textwidth]{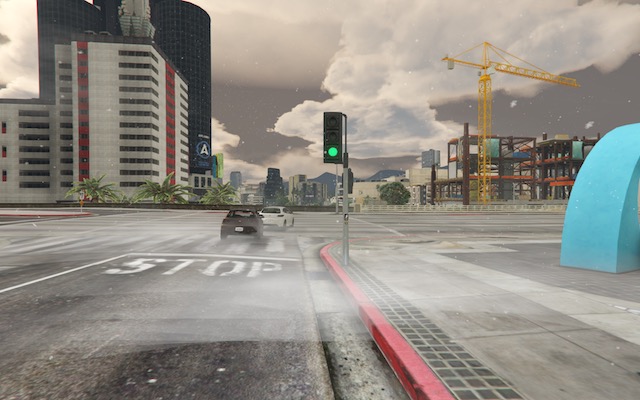}
\includegraphics[width=0.49\textwidth]{overlapping4.jpg}
\caption{Scenes generated from a \scenic{} scenario representing two cars, one partially occluding the other.}
\label{figure:gallery-overlapping}
\end{figure*}

\clearpage

\subsection{Four Cars, in Poor Driving Conditions}

This is the scenario used to produce images of four cars in poor driving conditions for the experiment in Sec.~\ref{sec:5_2}.
Without the first two lines, it is the generic four-car scenario used in that experiment. \\

\begin{lstlisting}
param weather = 'RAIN'
param time = 0 * 60   # midnight

wiggle = (-10 deg, 10 deg)
ego = EgoCar with roadDeviation wiggle
Car visible, with roadDeviation resample(wiggle)
Car visible, with roadDeviation resample(wiggle)
Car visible, with roadDeviation resample(wiggle)
Car visible, with roadDeviation resample(wiggle)
\end{lstlisting}

\begin{figure*}[b]
\centering
\includegraphics[width=0.49\textwidth]{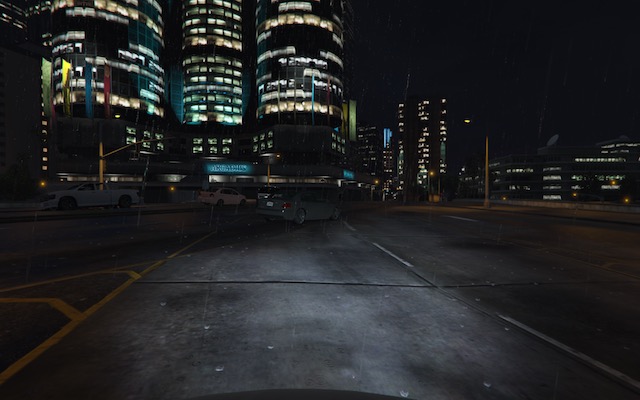}
\includegraphics[width=0.49\textwidth]{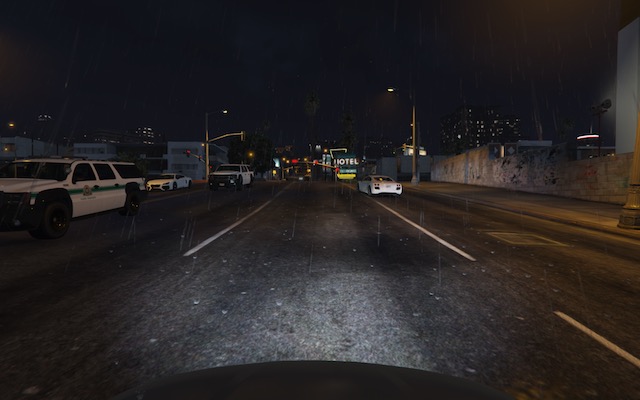}
\includegraphics[width=0.49\textwidth]{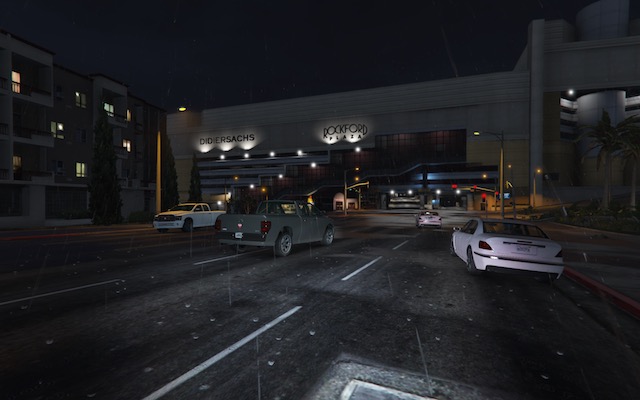}
\includegraphics[width=0.49\textwidth]{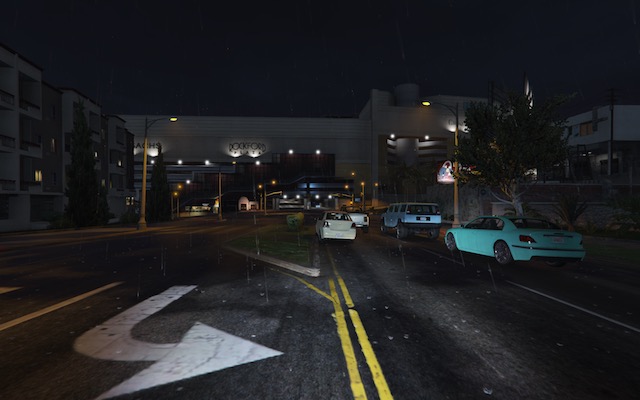}
\caption{Scenes generated from a \scenic{} scenario representing four cars in poor driving conditions.}
\label{figure:gallery-fourBad}
\end{figure*}

\clearpage

\subsection{A Platoon, in Daytime}

This scenario illustrates how \scenic{} can construct structured object configurations, in this case a platoon of cars.
It uses a helper function provided by \texttt{gtaLib} for creating platoons starting from a given car, shown in Fig.~\ref{figure:create-platoon}.
If no argument \texttt{model} is provided, as in this case, all cars in the platoon have the same \texttt{model} as the starting car; otherwise, the given model distribution is sampled independently for each car.
The syntax for functions and loops supported by our \scenic{} implementation is inherited from Python. \\

\begin{lstlisting}
param time = (8, 20) * 60	# 8 am to 8 pm
ego = Car with visibleDistance 60
c2 = Car visible
platoon = createPlatoonAt(c2, 5, dist=(2, 8))
\end{lstlisting}

\begin{figure*}[b]
\begin{lstlisting}
def createPlatoonAt(car, numCars, ~model~=None, dist=(2, 8), shift=(-0.5, 0.5), wiggle=0):
  lastCar = car
  for i in range(numCars-1):
    center = follow roadDirection from (`front of` lastCar) for resample(dist)
    pos = OrientedPoint right of center by shift,
    		    facing resample(wiggle) relative to roadDirection
    lastCar = Car ahead of pos, with model (car.~model~ if ~model~ is None else resample(~model~))
\end{lstlisting}
\caption{Helper function for creating a platoon starting from a given car.}
\label{figure:create-platoon}
\end{figure*}

\begin{figure*}[b]
\centering
\includegraphics[width=0.49\textwidth]{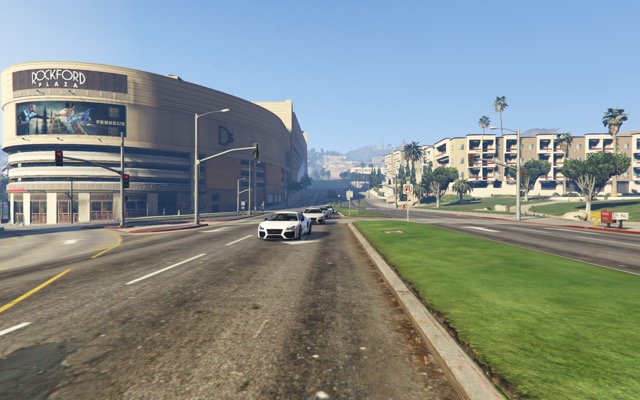}
\includegraphics[width=0.49\textwidth]{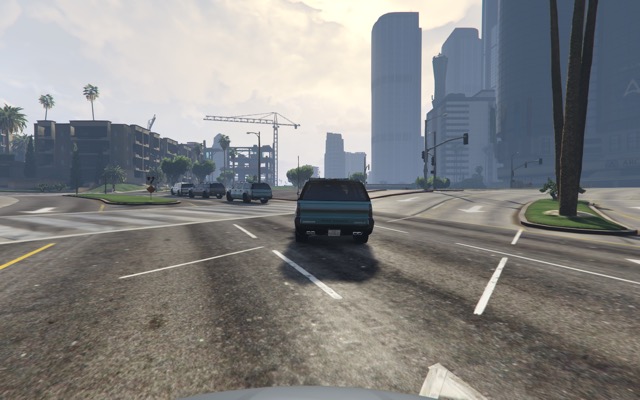}
\includegraphics[width=0.49\textwidth]{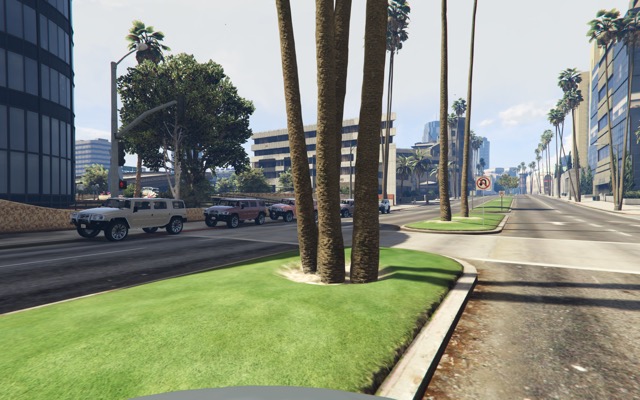}
\includegraphics[width=0.49\textwidth]{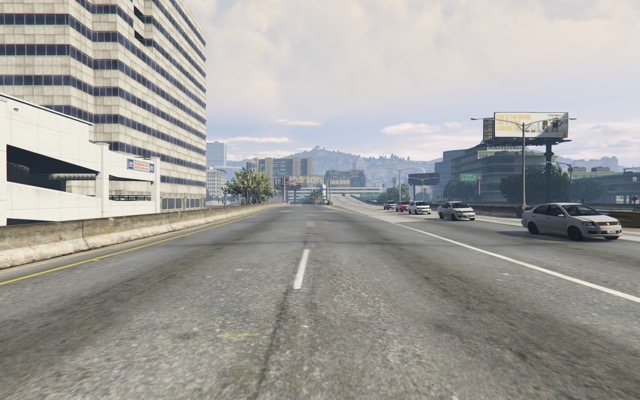}
\caption{Scenes generated from a \scenic{} scenario representing a platoon of cars during daytime.}
\label{figure:gallery-platoon}
\end{figure*}

\clearpage

\subsection{Bumper-to-Bumper Traffic}

This scenario creates an even more complex type of object structure, namely three lanes of traffic.
It uses the helper function \texttt{createPlatoonAt} discussed above, plus another for placing a car ahead of a given car with a specified gap in between, shown in Fig.~\ref{figure:gallery-car-ahead}. \\

\begin{lstlisting}
depth = 4
laneGap = 3.5
carGap = (1, 3)
laneShift = (-2, 2)
wiggle = (-5 deg, 5 deg)
modelDist = CarModel.defaultModel()

def createLaneAt(car):
  createPlatoonAt(car, depth, dist=carGap, wiggle=wiggle, ~model~=modelDist)

ego = Car with visibleDistance 60
leftCar = carAheadOfCar(ego, laneShift + carGap, offsetX=-laneGap, wiggle=wiggle)
createLaneAt(leftCar)

midCar = carAheadOfCar(ego, resample(carGap), wiggle=wiggle)
createLaneAt(midCar)

rightCar = carAheadOfCar(ego, resample(laneShift) + resample(carGap), offsetX=laneGap, wiggle=wiggle)
createLaneAt(rightCar)
\end{lstlisting}

\begin{figure}[h]
\begin{lstlisting}
def carAheadOfCar(car, gap, offsetX=0, wiggle=0):
  pos = OrientedPoint at (`front of` car) offset by (offsetX @ gap),
    		  facing resample(wiggle) relative to roadDirection
  return Car ahead of pos
\end{lstlisting}
\caption{Helper function for placing a car ahead of a car, with a specified gap in between.}
\label{figure:gallery-car-ahead}
\end{figure}

\begin{figure*}[b]
\centering
\includegraphics[width=0.49\textwidth]{btb1.jpg}
\includegraphics[width=0.49\textwidth]{btb2.jpg}
\includegraphics[width=0.49\textwidth]{btb3.jpg}
\includegraphics[width=0.49\textwidth]{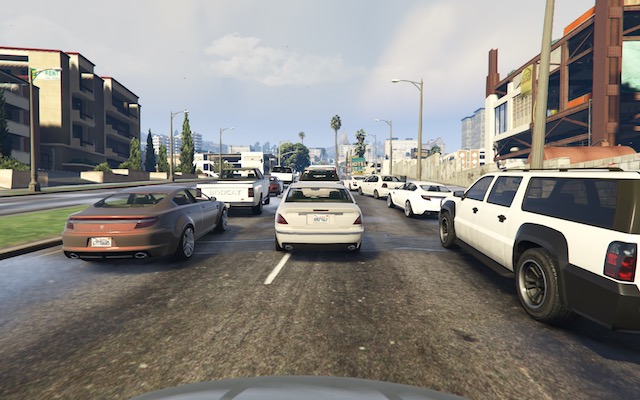}
\caption{Scenes generated from a \scenic{} scenario representing bumper-to-bumper traffic.}
\label{figure:gallery-btb}
\end{figure*}

\clearpage

\subsection{Robot Motion Planning with a Bottleneck}

This scenario illustrates the use of \scenic{} in another domain (motion planning) and with another simulator (Webots~\cite{webots}).
Figure~\ref{fig:mars-scenario} encodes a scenario representing a rubble field of rocks and pipes with a bottleneck between a robot and its goal that forces the path planner to consider climbing over a rock.
The code is broken into four parts: first, we import a small library defining the workspace and the types of objects, then create the robot at a fixed position and the goal (represented by a flag) at a random position on the other side of the workspace.
Second, we pick a position for the bottleneck, requiring it to lie roughly on the way from the robot to its goal, and place a rock there.
Third, we position two pipes of varying lengths which the robot cannot climb over on either side of the bottleneck, with their ends far enough apart for the robot to be able to pass between.
Finally, to make the scenario slightly more interesting we add several additional obstacles, positioned either on the far side of the bottleneck or anywhere at random.
Several resulting workspaces are shown in Fig.~\ref{fig:mars-big}.

\begin{figure*}[b]
\begin{lstlisting}
import mars
ego = Rover at 0 @ -2
goal = Goal at (-2, 2) @ (2, 2.5)

halfGapWidth = (1.2 * ego.width) / 2
bottleneck = OrientedPoint offset by (-1.5, 1.5) @ (0.5, 1.5), facing (-30, 30) deg
require abs((`angle to` goal) - (`angle to` bottleneck)) <= 10 deg
BigRock at bottleneck

leftEnd = OrientedPoint left of bottleneck by halfGapWidth,
    		    facing (60, 120) deg relative to bottleneck
rightEnd = OrientedPoint right of bottleneck by halfGapWidth,
			 facing (-120, -60) deg relative to bottleneck
Pipe ahead of leftEnd, with height (1, 2)
Pipe ahead of rightEnd, with height (1, 2)

BigRock beyond bottleneck by (-0.5, 0.5) @ (0.5, 1)
BigRock beyond bottleneck by (-0.5, 0.5) @ (0.5, 1)
Pipe
Rock
Rock
Rock
\end{lstlisting}
\caption{A \scenic{} representing rubble fields with a bottleneck so that the direct route to the goal requires climbing over rocks.}
\label{fig:mars-scenario}
\end{figure*}

\begin{figure*}[p]
\begin{minipage}{0.646\textwidth}
\centering
\includegraphics[width=\textwidth]{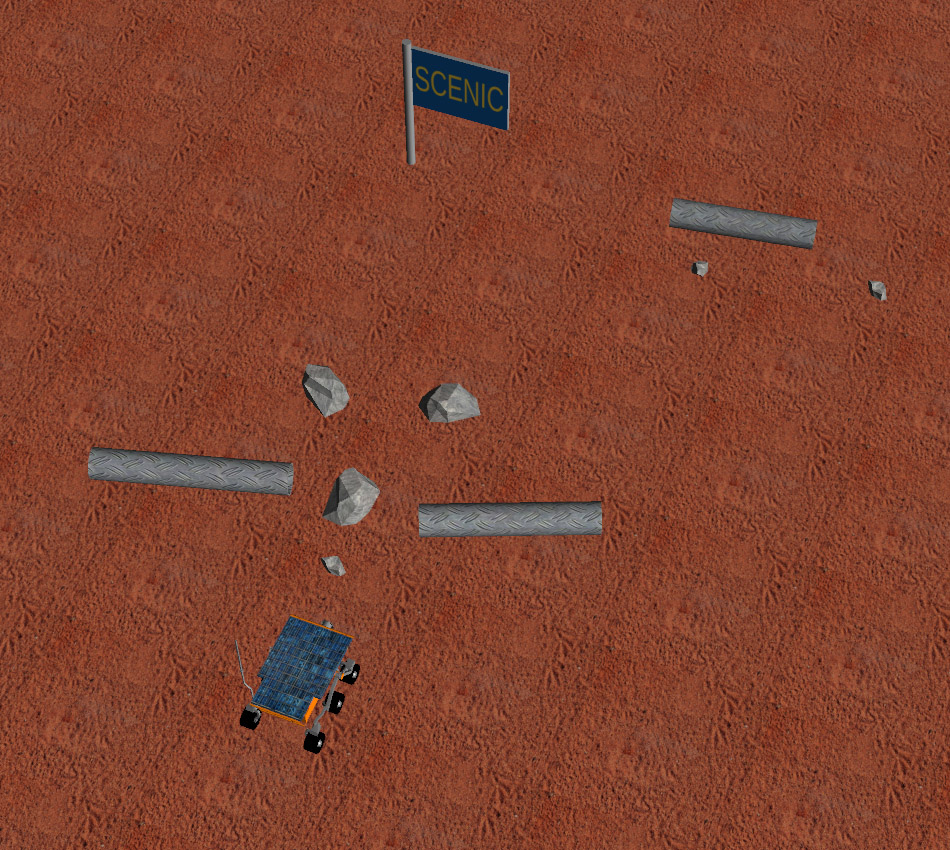}
\end{minipage}
\begin{minipage}{0.321\textwidth}
\centering
\includegraphics[width=\textwidth]{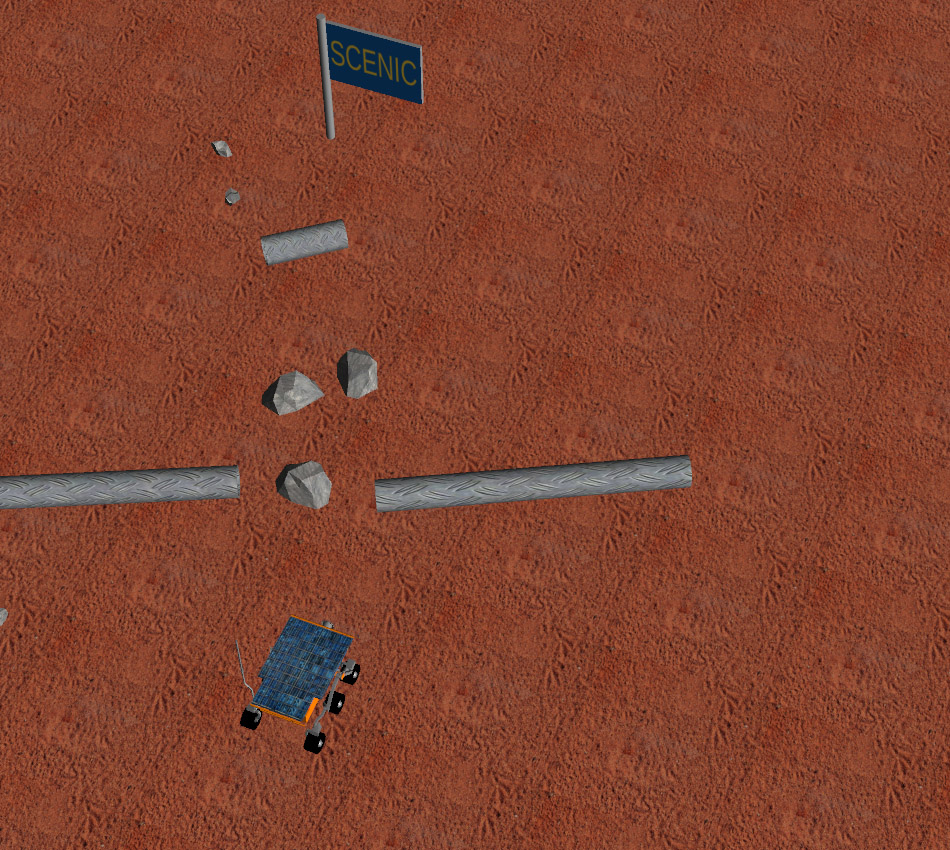}
\includegraphics[width=\textwidth]{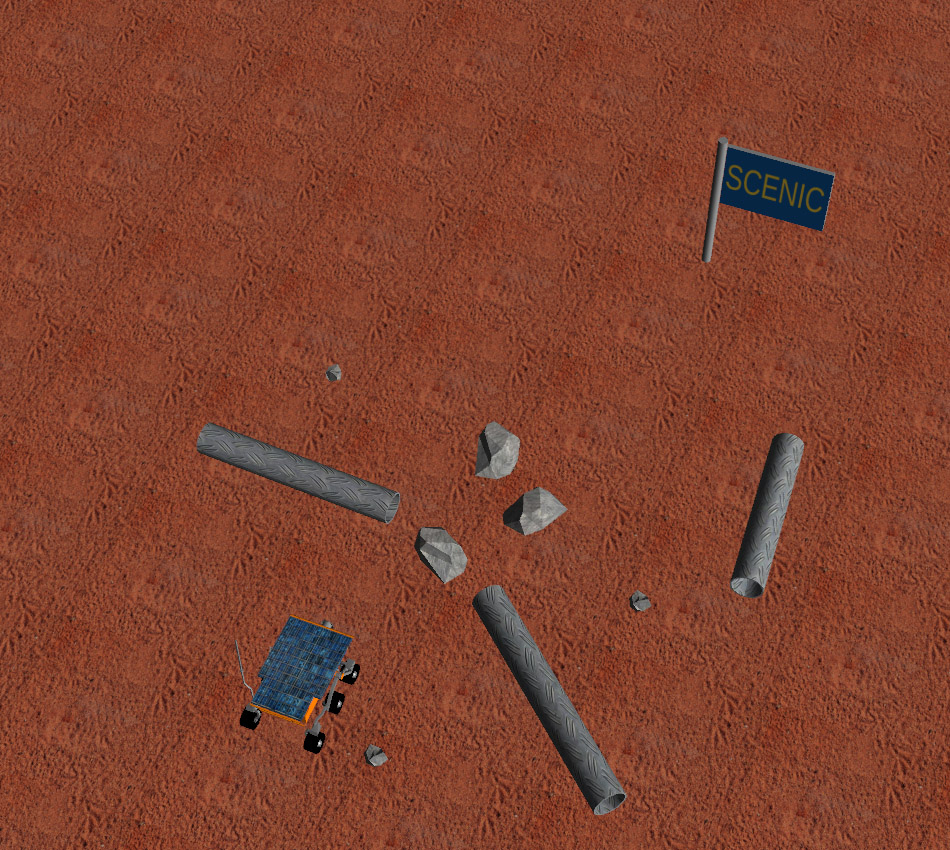}
\end{minipage}
\caption{Workspaces generated from the scenario in Fig.~\ref{fig:mars-scenario}, viewed in Webots from a fixed camera.}
\label{fig:mars-big}
\end{figure*}

\clearpage

\onecolumn

\section{Semantics of \scenic} \label{sec:full-semantics}


In this section we give a precise semantics for \scenic{} expressions and statements, building up to a semantics for a complete program as a distribution over scenes.

\etocsetnexttocdepth{subsection}
\localtableofcontents

\subsection{Notation for State and Semantics}

We will precisely define the meaning of \scenic{} language constructs by giving a small-step operational semantics.
We will focus on the aspects of \scenic{} that set it apart from ordinary imperative languages, skipping standard inference rules for sequential composition, arithmetic operations, etc. that we essentially use without change.
In rules for statements, we will denote a state of a \scenic{} program by $\pstate{s}{\sigma}{\pi}{\objects}$, where $s$ is the statement to be executed, $\sigma$ is the current variable assignment (a map from variables to values), $\pi$ is the current global parameter assignment (for \tm{param} statements), and $\objects$ is the set of all objects defined so far.
In rules for expressions, we use the same notation, although we sometimes suppress the state on the right-hand side of rules for expressions without side effects: $\pexpr{e} \rightarrow v$ means that in the state $(\sigma, \pi, \objects)$, the expression $e$ evaluates to the value $v$ without side effects.

Since \scenic{} is a probabilistic programming language, a single expression can be evaluated different ways with different probabilities.
Following the notation of \cite{saheb-djahromi,claret2013bayesian}, we write $\rightarrow^p$ for a rewrite rule that fires with probability $p$ (probability density $p$, in the case of continuous distributions).
We will discuss the meaning of such rules in more detail below.

\subsection{Semantics of Expressions}

As explained in the previous section, \scenic{}'s expressions are straightforward except for distributions and object definitions. 
As in a typical imperative probabilistic programming language, a distribution evaluates to a \emph{sample} from the distribution, following the first rule in Fig.~\ref{fig:expression-semantics}.
For example, if $\nt{baseDist}$ is a uniform interval distribution and the parameters evaluate to $\nt{low} = 0$ and $\nt{high} = 1$, then the distribution can evaluate to any value in $[0, 1]$ with probability density $1$.

\begin{figure}[b]
\begin{mathpar}
\infer[Distributions]%
{\pexpr{\nt{params}} \rightarrow \theta \\ v \in \dom \nt{baseDist}(\theta) \\ }%
{\pexpr{\nt{baseDist}(\nt{params})} \rightarrow^{P_\theta(v)} v}

\infer[Object Definitions]%
{\nt{resolveSpecifiers}(\nt{class}, \nt{specifiers}) = ((s_1, p_1), \dots, (s_n, p_n)) \\\\
\pstate{s_1}{\sigma\subst{\tm{self}}{\bot}}{\pi}{\objects} \rightarrow^{r_1} \pstate{v_1}{\sigma_1}{\pi}{\objects_1} \\\\
\pstate{s_2}{\sigma_1\subst{\tm{self.}p_1}{v_1(p_1)}}{\pi}{\objects_1} \rightarrow^{r_2} \pstate{v_2}{\sigma_2}{\pi}{\objects_2} \\\\
\vdots \\\\
\pstate{s_n}{\sigma_{n-1}\subst{\tm{self.}p_{n-1}}{v_{n-1}(p_{n-1})}}{\pi}{\objects_{n-1}} \rightarrow^{r_n} \pstate{v_n}{\sigma_n}{\pi}{\objects_n} \\\\
\nt{inst} = \nt{newInstance}(\nt{class}, \sigma_n\subst{\tm{self.}p_n}{v_n(p_n)}(\tm{self}))}%
{\pexpr{\texttt{\nt{class} \nt{specifiers}}} \rightarrow^{r_1 \dots r_n} \pstate{\nt{inst}}{\sigma}{\pi}{\objects_n \cup \{\nt{inst}\}}}

\infer[`\tm{with}' specifier]%
{\pexpr{E} \rightarrow \pstate{v}{\sigma}{\pi}{\objects'}}%
{\pexpr{\tm{with } \nt{property}\tm{ } E} \rightarrow \{\nt{property} \mapsto v\}}

\infer[`\tm{facing }{\normalfont \textit{vectorField}}' specifier]%
{\pexpr{\nt{vectorField}} \rightarrow v \\ \pexpr{\tm{self.position}} \rightarrow p}%
{\pexpr{\tm{facing } \nt{vectorField}} \rightarrow \{\tm{heading} \mapsto v(p)\}}
\end{mathpar}
\caption{Semantics of expressions (excluding operators, defined in Appendix~\ref{sec:operator-semantics}), and two example specifiers. Here \nt{baseDist} is viewed as a function mapping parameters $\theta$ to a distribution with density function $P_\theta$, and $\nt{newInstance}(\nt{class}, \nt{props})$ creates a new instance of a class with the given property values.} \label{fig:expression-semantics}
\end{figure}

The semantics of object definitions are given by the second rule in Fig.~\ref{fig:expression-semantics}.
First note the side effect, namely adding the newly-defined object to the set $\objects$.
The premises of the rule describe the procedure for combining the specifiers to obtain the overall set of properties for the object.
The main step is working out the evaluation order for the specifiers so that all their dependencies are satisfied, as well as deciding for each specifier which properties it should specify (if it specifies a property optionally, another specifier could take precedence).
This is done by the procedure \nt{resolveSpecifiers}, shown formally as Alg.~\ref{alg:resolve-specifiers} and which essentially does the following:

\begin{algorithm}[tb]
\caption{$\nt{resolveSpecifiers}\,(\nt{class}, \nt{specifiers})$} \label{alg:resolve-specifiers}
\begin{algorithmic}[1]
\LineComment gather all specified properties
\State $\nt{specForProperty} \gets \emptyset$
\State $\nt{optionalSpecsForProperty} \gets \emptyset$
\ForAll {specifiers $S$ in \nt{specifiers}}
	\ForAll {properties $P$ specified non-optionally by $S$}
		\If {$P \in \dom \nt{specForProperty}$}
			\State syntax error: property $P$ specified twice
		\EndIf
		\State $\nt{specForProperty}\,(P) \gets S$
	\EndFor
	\ForAll {properties $P$ specified optionally by $S$}
		\State $\nt{optionalSpecsForProperty}\,(P).\nt{append}(S)$
	\EndFor
\EndFor
\LineComment filter optional specifications
\ForAll {properties $P \in \dom \nt{optionalSpecsForProperty}$}
	\If {$P \in \dom \nt{specForProperty}$}
		\State \textbf{continue}
	\EndIf
	\If {$|\nt{optionalSpecsForProperty}\,(P)| > 1$}
		\State syntax error: property $P$ specified twice
	\EndIf
	\State $\nt{specForProperty}\,(P) \gets \nt{optionalSpecsForProperty}\,(P)[0]$
\EndFor
\LineComment add default specifiers as needed
\State $\nt{defaults} \gets \nt{defaultValueExpressions}\,(\nt{class})$
\ForAll {properties $P \in \dom \nt{defaults}$}
	\If {$P \not \in \dom \nt{specForProperty}$}
		\State $\nt{specForProperty}\,(P) \gets \nt{defaults}\,(P)$
	\EndIf
\EndFor
\LineComment build dependency graph
\State $G \gets \text{empty graph on } \dom \nt{specForProperty}$
\ForAll {specifiers $S \in \dom \nt{specForProperty}$}
	\ForAll {dependencies $D$ of $S$}
		\If {$D \not \in \dom \nt{specForProperty}$}
			\State syntax error: missing property $D$ required by $S$
		\EndIf
		\State add an edge in $G$ from $\nt{specForProperty}\,(D)$ to $S$
	\EndFor
\EndFor
\If {$G$ is cyclic}
	\State syntax error: specifiers have cyclic dependencies
\EndIf
\LineComment construct specifier and property evaluation order
\State $\nt{specsAndProps} \gets \text{empty list}$
\ForAll {specifiers $S$ in $G$ in topological order}
	\State $\nt{specsAndProps}.\nt{append}((S, \{ P \;|\; \nt{specForProperty}\,(P) = S \}))$
\EndFor
\State \Return $\nt{specsAndProps}$
\end{algorithmic}
\end{algorithm}

Let $P$ be the set of properties defined in the object's class and superclasses, together with any properties specified by any of the specifiers.
The object will have exactly these properties, and the value of each $p \in P$ is determined as follows.
If $p$ is specified non-optionally by multiple specifiers the scenario is ill-formed.
If $p$ is only specified optionally, and by multiple specifiers, this is ambiguous and we also declare the scenario ill-formed.
Otherwise, the value of $p$ will be determined by its unique non-optional specifier, unique optional specifier, or the most-derived default value, in that order: call this specifier $s_p$.
Construct a directed graph with vertices $P$ and edges to $p$ from each of the dependencies of $s_p$ (if a dependency is not in $P$, then a specifier references a nonexistent property and the scenario is ill-formed).
If this graph has a cycle, there are cyclic dependencies and the scenario is ill-formed (e.g. \lstinline{Car left of 0 @ 0, facing roadDirection}: the \texttt{heading} must be known to evaluate \lstinline{left of $\nt{vector}$}, but \lstinline{facing $\nt{vectorField}$} needs \texttt{position} to determine \texttt{heading}).
Otherwise, topologically sorting the graph yields an evaluation order for the specifiers so that all dependencies are available when needed.

The rest of the rule in Fig.~\ref{fig:expression-semantics} simply evaluates the specifiers in this order, accumulating the results as properties of \tm{self} so they are available to the next specifier, finally creating the new object once all properties have been assigned.
Note that we also accumulate the probabilities of each specifier's evaluation, since specifiers are allowed to introduce randomness themselves (e.g. the \lstinline{on $\nt{region}$} specifier returns a random point in the region).

As noted above the semantics of the individual specifiers are mostly straightforward, and exact definitions are given in Appendix~\ref{sec:operator-semantics}.
To illustrate the pattern we precisely define two specifiers in Fig.~\ref{fig:expression-semantics}: the \lstinline{with $\nt{property}$ $\nt{value}$} specifier, which has no dependencies but can specify any property, and the \lstinline{facing $\nt{vectorField}$} specifier, which depends on \tm{position} and specifies \tm{heading}.
Both specifiers evaluate to maps assigning a value to each property they specify.

\subsection{Semantics of Statements}

The semantics of class and object definitions have been discussed above, while rules for the other statements are given in Fig.~\ref{fig:statement-semantics}.
As can be seen from the first rule, variable assignment behaves in the standard way.
Parameter assignment is nearly identical, simply updating the global parameter assignment $\pi$ instead of the variable assignment $\sigma$.

\begin{figure*}[tb]
\begin{mathpar}
\infer[Variable/Parameter Assignments]%
{\pexpr{E} \rightarrow \pstate{v}{\sigma}{\pi}{\objects'}}%
{\pstate{x \texttt{ = } E}{\sigma}{\pi}{\objects} \rightarrow \pstate{\pass}{\sigma\subst{x}{v}}{\pi}{\objects'} \\\\ \pstate{\texttt{param } x \texttt{ = } E}{\sigma}{\pi}{\objects} \rightarrow \pstate{\pass}{\sigma}{\pi\subst{x}{v}}{\objects'}}
\\
\infer[Hard Requirements]%
{\pexpr{B} \rightarrow \pstate{\tm{True}}{\sigma}{\pi}{\objects'}}%
{\pstate{\tm{require } B}{\sigma}{\pi}{\objects} \rightarrow \pstate{\pass}{\sigma}{\pi}{\objects'}}

\infer[Soft Requirements]%
{}%
{\pstate{\tm{require[}p\tm{] } B}{\sigma}{\pi}{\objects} \rightarrow^p \pstate{\tm{require } B}{\sigma}{\pi}{\objects} \\\\ \pstate{\tm{require[}p\tm{] } B}{\sigma}{\pi}{\objects} \rightarrow^{1-p} \pstate{\pass}{\sigma}{\pi}{\objects}}

\infer[Mutations]%
{}%
{\pstate{\tm{mutate } \nt{obj}_i \tm{ by } s}{\sigma}{\pi}{\objects} \rightarrow \pstate{\pass}{\sigma}{\pi}{\objects\subst{\sigma(\nt{obj}_i)\tm{.mutationScale}}{s}}}

\infer[Termination, Step 1: Apply Mutations]%
{\objects = \{o_1, \dots, o_n\} \\ \forall i \in \{1, \dots, n\}: \\\\ S_i = \objects(o_i\tm{.mutationScale}) \\ ps_i = \objects(o_i\tm{.positionStdDev}) \\ hs_i = \objects(o_i\tm{.headingStdDev}) \\\\ \pexpr{\tm{Normal(0, }S_i \cdot ps_i\tm{)}} \rightarrow^{r_{i,x}} \pexpr{n_{i,x}} \\ \pexpr{\tm{Normal(0, }S_i \cdot ps_i\tm{)}} \rightarrow^{r_{i,y}} \pexpr{n_{i,y}} \\ \pexpr{\tm{Normal(0, }S_i \cdot hs_i\tm{)}} \rightarrow^{r_{i,h}} \pexpr{n_{i,h}} \\\\ pos_i = \objects(o_i\tm{.position}) + (n_{i,x}, n_{i,y}) \\ head_i = \objects(o_i\tm{.heading}) + n_{i,h}}%
{\pstate{\pass}{\sigma}{\pi}{\objects} \rightarrow^{r_{0,x} r_{0,y} r_{0,h} \dots} \pstate{\textsc{Done}}{\sigma}{\pi}{\objects\subst{o_i\tm{.position}}{pos_i}\subst{o_i\tm{.heading}}{head_i}}}

\infer[Termination, Step 2: Check Default Requirements]%
{\objects = \{o_1, \dots, o_n\} \\ \forall i: \nt{boundingBox}\,(o_i) \subseteq \nt{workspace} \\ \forall i \ne j: o_i\tm{.allowCollisions} \lor o_j\tm{.allowCollisions} \lor \nt{boundingBox}\,(o_i) \cap \nt{boundingBox}\,(o_j) = \emptyset \\ \forall i: \lnot o_i\tm{.requireVisible} \lor \pexpr{\tm{ego can see } o_i} \rightarrow \tm{True}}%
{\pstate{\textsc{Done}}{\sigma}{\pi}{\objects} \rightarrow (\pi, \objects)}
\end{mathpar}
\caption{Semantics of statements (excluding class definitions and standard rules for sequential composition).
\textsc{Done} denotes a special state ready for the final termination rule to run.} \label{fig:statement-semantics}
\end{figure*}

As noted above, the \lstinline{require $\nt{boolean}$} statement is equivalent to an \emph{observe} in other languages, and following \cite{claret2013bayesian} we model it by allowing the ``Hard Requirement'' rule in Fig.~\ref{fig:statement-semantics} to only fire when the condition is satisfied (then turning the requirement into a no-op).
If the condition is not satisfied, no rules apply and the program fails to terminate normally.
When defining the semantics of entire \scenic{} scenarios below we will discard such non-terminating executions, yielding a distribution only over executions where all hard requirements are satisfied.

The statement \lstinline{require[$p$] $\nt{boolean}$} requires only that its condition hold with at least probability $p$.
There are a number of ways the semantics of such a soft requirement could be defined: we choose the natural definition that \lstinline{require[$p$] $B$} is equivalent to a hard requirement \lstinline{require $B$} that is only enforced with probability $p$.
This is reflected in the two corresponding rules in Fig.~\ref{fig:statement-semantics}, and clearly ensures that the requirement $B$ will hold with probability at least $p$, as desired.

Since the mutation statement \lstinline{mutate $\csl{\nt{instance}$} ^by^ $\nt{number}$} only causes noise to be added at the end of execution, as discussed above, its rule Fig.~\ref{fig:statement-semantics} simply sets a property on the object(s) indicating that mutation is enabled (and giving the scale of noise to be added).
The noise is actually added by the first of two special rules that apply only once the program has been reduced to $\pass$ and so computation has finished.
This rule first looks up the values of the properties \tm{mutationScale}, \tm{positionStdDev}, and \tm{headingStdDev} for each object.
Respectively, these specify the overall scale of the noise to add (by default zero, i.e.~mutation is disabled) and factors allowing the standard deviation for \tm{position} and \tm{heading} to be adjusted individually.
The rule then independently samples Gaussian noise with the desired standard deviation for each object and adds it to the \tm{position} and \tm{heading} properties.

Finally, after mutations are applied, the last rule in Fig.~\ref{fig:statement-semantics} checks \scenic's three built-in hard requirements.
Similarly to the rule for hard requirements, this last rule can only fire if all the built-in requirements are satisfied, otherwise preventing the program from terminating.
If the rule does fire, the final result is the output of the scenario: the assignment $\pi$ to the global parameters, and the set $\objects$ of all defined objects.

\subsection{Semantics of a \scenic{} Program}

As we have just defined it, every time one runs a \scenic{} program its output is a \emph{scene} consisting of an assignment to all the properties of each \lstinline{Object} defined in the scenario, plus any global parameters defined with \lstinline{param}.
Since \scenic{} allows sampling from distributions, the imperative part of a scenario actually induces a {\em distribution over scenes}, resulting from the probabilistic rules of the semantics described above.
Specifically, for any execution trace the product of the probabilities of all rewrite rules yields a probability (density) for the trace (see e.g.~\cite{claret2013bayesian}).
The declarative part of a scenario, consisting of its \lstinline{require} statements, {\em modifies this distribution}.
As mentioned above, hard requirements are equivalent to ``observations'' in other probabilistic programming languages, \emph{conditioning} the distribution on the requirement being satisfied.
In particular, if we discard all traces which do not terminate (due to violating a requirement), then normalizing the probabilities of the remaining traces yields a distribution over traces, and therefore scenes, that satisfy all our requirements.
This is the distribution defined by the \scenic{} scenario.

\subsection{Sampling Algorithms}

This section gives pseudocode for the domain-specific sampling techniques described in Sec.~\ref{sec:sampling-techniques}.
Algorithm~\ref{alg:prune_head} implements pruning by orientation, pruning a set of polygons \nt{map} given an allowed range of relative headings \nt{A}, a distance bound $M$, and a bound $\delta$ on the heading deviation between an object and the vector field at its position.

\begin{algorithm}[tb]
\caption{$\nt{pruneByHeading}\,(\nt{map, A, M, $\delta$})$}\label{alg:prune_head}
\begin{algorithmic}[1]
\State $\nt{map'} \gets \emptyset$
\ForAll {polygons $P$ in \nt{map}}
	\ForAll {polygons $Q$ in \nt{map}}
		\State $Q' \gets dilate(Q, M)$
		\If {$P \cap Q' \neq \emptyset \wedge relHead(P,Q) \pm 2\delta \in A $}
			\State $\nt{map'} \gets \nt{map'} \cup (Q' \cap P)$
		\EndIf
	\EndFor
\EndFor
\State \Return $\nt{map'}$
\end{algorithmic}
\end{algorithm}

Algorithm~\ref{alg:prune_width} similarly implements pruning by size, given \nt{map} and $M$ as above, plus a bound \nt{minWidth} on the minimum width of the configuration.
Here the subroutine \nt{narrow} finds all polygons which are thinner than this bound.

\begin{algorithm}[tb]
\caption{$\nt{pruneByWidth}\,(\nt{map, M, minWidth})$}\label{alg:prune_width}
\begin{algorithmic}[1]
\State $narrowPolys \gets narrow(map, minWidth)$
\State $\nt{map'} \gets map \setminus narrowPolys$
\ForAll {polygons $P$ in \nt{narrowPolys}}
	\State $U \gets \bigcup_{Q \in \nt{map} \setminus \{P\}} dilate(Q, M)$
	\State $\nt{map'} \gets \nt{map'} \cup (P \cap U)$
\EndFor
\State \Return $\nt{map'}$
\end{algorithmic}
\end{algorithm}

\clearpage

\section{Detailed Semantics of Specifiers and Operators} \label{sec:operator-semantics}


\newcommand{\mkvec}[2]{\left\langle#1, #2\right\rangle}
\newcommand{\angleto}[2]{\arctan\,(#1 - #2)}
\newcommand{\rot}[2]{\textit{rotate}\,(#1, #2)}
\newcommand{\disc}[2]{\textit{Disc}\,(#1, #2)}
\newcommand{\sector}[4]{\textit{Sector}\,(#1, #2, #3, #4)}
\newcommand{\bounding}[1]{\textit{boundingBox}\,(#1)}
\newcommand{\pointin}[1]{\textit{uniformPointIn}\,(#1)}
\newcommand{\visible}[1]{\textit{visibleRegion}\,(#1)}
\newcommand{\orientation}[1]{\textit{orientation}\,(#1)}
\newcommand{\mkop}[2]{\texttt{OrientedPoint}\,(#1, #2)}
\newcommand{\opth}[2]{\mkop{#1}{#2}}
\newcommand{\euler}[3]{\textit{forwardEuler}\,(#1,#2,#3)}
\newcommand{\offsetLocal}[2]{\textit{offsetLocal}\,(#1, #2)}

This section provides precise semantics for \scenic{}'s specifiers and operators, which were informally defined above.

\etocsetnexttocdepth{subsection}
\localtableofcontents

\subsection{Notation}

Since none of the specifiers and operators have side effects, to simplify notation we write $\sem{X}$ for the value of the expression $X$ in the current state (rather than giving inference rules).
Throughout this section, $S$ indicates a \nt{scalar}, $V$ a \nt{vector}, $H$ a \nt{heading}, $F$ a \nt{vectorField}, $R$ a \nt{region}, $P$ a \texttt{Point}, and $OP$ an \texttt{OrientedPoint}.
Figure~\ref{figure:semantics-helper} defines notation used in the rest of the semantics.
In \textit{forwardEuler}, $N$ is an implementation-defined parameter specifying how many steps should be used for the forward Euler approximation when following a vector field (we used $N=4$).

\begin{figure}[h]
\begin{align*}
\mkvec{x}{y} &= \text{point with the given XY coordinates} \\
\rot{\mkvec{x}{y}}{\theta} &= \mkvec{x \cos \theta - y \sin \theta}{x \sin \theta + y \cos \theta} \\
\offsetLocal{OP}{v} &= \sem{\tm{OP.position}} + \rot{v}{\sem{\tm{OP.heading}}} \\
\disc{c}{r} &= \text{set of points in the disc centered at } c \text{ and with radius } r \\
\sector{c}{r}{h}{a} &= \text{set of points in the sector of } \disc{c}{r} \text{ centered along } h \text{ and with angle } a \\
\bounding{O} &= \text{set of points in the bounding box of object } O \\
\visible{X} &= \begin{cases}
\sector{\sem{\tm{X.position}}}{\sem{\tm{X.viewDistance}}}{\\\quad\quad\quad\sem{\tm{X.heading}}}{\sem{\tm{X.viewAngle}}} & X \in \texttt{OrientedPoint} \\
\disc{\sem{\tm{X.position}}}{\sem{\tm{X.viewDistance}}} & X \in \texttt{Point}
\end{cases} \\
\orientation{R} &= \text{preferred orientation of } R \text{ if any; otherwise } \bot \\
\pointin{R} &= \text{a uniformly random point in } R \\
\euler{x}{d}{F} &= \text{result of iterating the map } x \mapsto x + \rot{\mkvec{0}{d / N}}{\sem{F}(x)} \text{ a total of } N \text{ times on } x
\end{align*}
\caption{Notation used to define the semantics.}
\label{figure:semantics-helper}
\end{figure}

\subsection{Specifiers for \tm{position}}

Figure~\ref{figure:semantics-position} gives the semantics of the \texttt{position} specifiers.
The figure writes the semantics as a vector value; the semantics of the specifier itself is to assign the \texttt{position} property of the object being specified to that value.
Several of the specifiers refer to properties of \texttt{self}: as explained in Sec.~\ref{sec:scenario_def_language}, this refers to the object being constructed, and the semantics of object construction are such that specifiers depending on other properties are only evaluated after those properties have been specified (or an error is raised, if there are cyclic dependencies).

\begin{figure}[h]
\begin{align*}
\sem{\tm{at } V} &= \sem{V} \\
\sem{\tm{offset by } V} &= \sem{V \tm{ relative to ego.position}} \\
\sem{\tm{offset along } H \tm{ by } V} &=  \sem{\tm{ego.position offset along } H \tm{ by } V} \\
\sem{\tm{left of } V} &= \sem{\tm{left of } V \tm{ by 0}} \\
\sem{\tm{right of } V} &= \sem{\tm{right of } V \tm{ by 0}} \\
\sem{\tm{ahead of } V} &= \sem{\tm{ahead of } V \tm{ by 0}} \\
\sem{\tm{behind } V} &= \sem{\tm{behind } V \tm{ by 0}} \\
\sem{\tm{left of } V \tm{ by } S} &= \sem{V} + \rot{\mkvec{-\sem{\tm{self.width}}/2 - \sem{S}}{0}}{\sem{\tm{self.heading}}} \\
\sem{\tm{right of } V \tm{ by } S} &= \sem{V} + \rot{\mkvec{\sem{\tm{self.width}}/2 + \sem{S}}{0}}{\sem{\tm{self.heading}}} \\
\sem{\tm{ahead of } V \tm{ by } S} &= \sem{V} + \rot{\mkvec{0}{\sem{\tm{self.height}}/2 + \sem{S}}}{\sem{\tm{self.heading}}} \\
\sem{\tm{behind } V \tm{ by } S} &= \sem{V} + \rot{\mkvec{0}{-\sem{\tm{self.height}}/2 - \sem{S}}}{\sem{\tm{self.heading}}} \\
\sem{\tm{beyond } V_1 \tm{ by } V_2} &= \sem{\tm{beyond } V_1 \tm{ by } V_2 \tm{ from } \tm{ego.position}} \\
\sem{\tm{beyond } V_1 \tm{ by } V_2 \tm{ from } V_3} &= \sem{V_1} + \rot{\sem{V_2}}{\angleto{\sem{V_1}}{\sem{V_3}}} \\
\sem{\tm{visible}} &= \sem{\tm{visible from ego}} \\
\sem{\tm{visible from } P} &= \pointin{\visible{P}}
\end{align*}
\caption{Semantics of \texttt{position} specifiers, given as the value $v$ such that the specifier evaluates to the map $\tm{position} \mapsto v$.}
\label{figure:semantics-position}
\end{figure}

\clearpage

\subsection{Specifiers for \tm{position} and optionally \tm{heading}}

Figure~\ref{figure:semantics-position-heading} gives the semantics of the \texttt{position} specifiers that also optionally specify \texttt{heading}.
The figure writes the semantics as an \texttt{OrientedPoint} value; if this is $OP$, the semantics of the specifier is to assign the \texttt{position} property of the object being constructed to \texttt{OP.position}, and the \texttt{heading} property of the object to \texttt{OP.heading} if \texttt{heading} is not otherwise specified (see Sec.~\ref{sec:scenario_def_language} for a discussion of optional specifiers).

\begin{figure}[h]
\begin{align*}
\sem{\tm{in } R} = \sem{\tm{on } R} &= \begin{cases}
\opth{x}{\sem{\orientation{R}}(x)} & \orientation{R} \ne \bot \\
\opth{x}{\bot} & \text{otherwise}
\end{cases}, \text{ with } x = \pointin{\sem{R}} \\
\sem{\tm{ahead of } O} &= \sem{\tm{ahead of (front of } O \tm{)}} \\
\sem{\tm{behind } O} &= \sem{\tm{behind (back of } O \tm{)}} \\
\sem{\tm{left of } O} &= \sem{\tm{left of (left of } O \tm{)}} \\
\sem{\tm{right of } O} &= \sem{\tm{right of (right of } O \tm{)}} \\
\sem{\tm{ahead of } OP} &= \sem{\tm{ahead of } OP \tm{ by 0}} \\
\sem{\tm{behind } OP} &= \sem{\tm{behind } OP \tm{ by 0}} \\
\sem{\tm{left of } OP} &= \sem{\tm{left of } OP \tm{ by 0}} \\
\sem{\tm{right of } OP} &= \sem{\tm{right of } OP \tm{ by 0}} \\
\sem{\tm{ahead of } OP \tm{ by } S} &= \opth{\offsetLocal{OP}{\mkvec{0}{\sem{\tm{self.height}}/2 + \sem{S}}}}{\sem{\tm{OP.heading}}} \\
\sem{\tm{behind } OP \tm{ by } S} &= \opth{\offsetLocal{OP}{\mkvec{0}{-\sem{\tm{self.height}}/2 - \sem{S}}}}{\sem{\tm{OP.heading}}} \\
\sem{\tm{left of } OP \tm{ by } S} &= \opth{\offsetLocal{OP}{\mkvec{-\sem{\tm{self.width}}/2 - \sem{S}}{0}}}{\sem{\tm{OP.heading}}} \\
\sem{\tm{right of } OP \tm{ by } S} &= \opth{\offsetLocal{OP}{\mkvec{\sem{\tm{self.width}}/2 + \sem{S}}{0}}}{\sem{\tm{OP.heading}}}
\end{align*}
\begin{align*}
\sem{\tm{following } F \tm{ for } S} &= \sem{\tm{following } F \tm{ from } \tm{ego.position} \tm{ for } S} \\
\sem{\tm{following } F \tm{ from } V \tm{ for } S} &= \sem{\tm{follow } F \tm{ from } V \tm{ for } S}
\end{align*}
\caption{Semantics of \texttt{position} specifiers that optionally specify \texttt{heading}. If $o$ is the \tm{OrientedPoint} given as the semantics above, the specifier evaluates to the map $\{\tm{position} \mapsto o\tm{.position}, \tm{heading} \mapsto o\tm{.heading}\}$.}
\label{figure:semantics-position-heading}
\end{figure}

\subsection{Specifiers for \tm{heading}}

Figure~\ref{figure:semantics-heading} gives the semantics of the \texttt{heading} specifiers.
As for the \texttt{position} specifiers above, the figure indicates the heading value assigned by each specifier.

\begin{figure}[h]
\begin{align*}
\sem{\tm{facing } H} &= \sem{H} \\
\sem{\tm{facing } F} &= \sem{F}(\sem{\tm{self.position}}) \\
\sem{\tm{facing toward } V} &= \angleto{\sem{V}}{\sem{\tm{self.position}}} \\
\sem{\tm{facing away from } V} &= \angleto{\sem{\tm{self.position}}}{\sem{V}} \\
\sem{\tm{apparently facing } H} &= \sem{\tm{apparently facing } H \tm{ from ego.position}} \\
\sem{\tm{apparently facing } H \tm{ from } V} &= \sem{H} + \angleto{\sem{\tm{self.position}}}{\sem{V}}
\end{align*}
\caption{Semantics of \texttt{heading} specifiers, given as the value $v$ such that the specifier evaluates to the map $\tm{heading} \mapsto v$.}
\label{figure:semantics-heading}
\end{figure}

\clearpage

\subsection{Operators}

Finally, Figures~\ref{figure:semantics-scalar-ops}--\ref{figure:semantics-op-ops} give the semantics for \scenic{}'s operators, broken down by the type of value they return.
We omit the semantics for ordinary numerical and Boolean operators (\tm{max}, \tm{+}, \tm{or}, \tm{>=}, etc.), which are standard.

\begin{figure}[h]
\begin{align*}
\sem{\tm{relative heading of } H} &= \sem{\tm{relative heading of } H \tm{ from ego.heading}} \\
\sem{\tm{relative heading of } H_1 \tm{ from } H_2} &= \sem{H_1} - \sem{H_2} \\
\sem{\tm{apparent heading of } OP} &= \sem{\tm{apparent heading of } OP \tm{ from ego.position}} \\
\sem{\tm{apparent heading of } OP \tm{ from } V} &= \sem{\tm{OP.heading}} - \angleto{\sem{\tm{OP.position}}}{\sem{V})} \\
\sem{\tm{distance to } V} &= \sem{\tm{distance from ego.position to } V} \\
\sem{\tm{distance from } V_1 \tm{ to } V_2} &= |\sem{V_2} - \sem{V_1}| \\
\sem{\tm{angle to } V} &= \sem{\tm{angle from ego.position to } V} \\
\sem{\tm{angle from } V_1 \tm{ to } V_2} &= \angleto{\sem{V_2}}{\sem{V_1}}
\end{align*}
\caption{Scalar operators.}
\label{figure:semantics-scalar-ops}
\end{figure}

\begin{figure}[h]
\begin{align*}
\sem{P \tm{ can see } O} &= \visible{\sem{P}} \cap \bounding{\sem{O}} \ne \emptyset \\
\sem{V \tm{ is in } R} &= \sem{V} \in \sem{R} \\
\sem{O \tm{ is in } R} &= \bounding{\sem{O}} \subseteq \sem{R}
\end{align*}
\caption{Boolean operators.}
\end{figure}

\begin{figure}[h]
\begin{align*}
\sem{F \tm{ at } V} &= \sem{F}(\sem{V}) \\
\sem{F_1 \tm{ relative to } F_2} &= \sem{F_1}(\sem{\tm{self.position}}) + \sem{F_2}(\sem{\tm{self.position}}) \\
\sem{H \tm{ relative to } F} &= \sem{H} + \sem{F}(\sem{\tm{self.position}}) \\
\sem{F \tm{ relative to } H} &= \sem{H} + \sem{F}(\sem{\tm{self.position}}) \\
\sem{H_1 \tm{ relative to } H_2} &= \sem{H_1} + \sem{H_2}
\end{align*}
\caption{Heading operators.}
\end{figure}

\begin{figure}[h]
\begin{align*}
\sem{V_1 \tm{ offset by } V_2} &= \sem{V_1} + \sem{V_2} \\
\sem{V_1 \tm{ offset along } H \tm{ by } V_2} &= \sem{V_1} + \rot{\sem{V_2}}{\sem{H}} \\
\sem{V_1 \tm{ offset along } F \tm{ by } V_2} &= \sem{V_1} + \rot{\sem{V_2}}{\sem{F}(\sem{V_1})}
\end{align*}
\caption{Vector operators.}
\end{figure}

\begin{figure}[h]
\begin{align*}
\sem{\tm{visible } R} &= \sem{R \tm{ visible from ego}} \\
\sem{R \tm{ visible from } P} &= \sem{R} \cap \visible{\sem{P}}
\end{align*}
\caption{Region operators.}
\end{figure}

\begin{figure}[h]
\begin{align*}
\sem{OP \tm{ offset by } V} &= \sem{V \tm{ relative to } OP} \\
\sem{V \tm{ relative to } OP} &= \mkop{\offsetLocal{OP}{\sem{V}}}{\sem{\tm{OP.heading}}} \\
\sem{\tm{follow } F \tm{ for } S} &= \sem{\tm{follow } F \tm{ from ego.position for } S} \\
\sem{\tm{follow } F \tm{ from } V \tm{ for } S} &= \mkop{y}{\sem{F}(y)} \text{ where } y = \euler{\sem{V}}{\sem{S}}{\sem{F}} \\
\sem{\tm{front of } O} &= \sem{\mkvec{0}{\sem{\tm{O.height}}/2} \tm{ relative to } O} \\
\sem{\tm{back of } O} &= \sem{\mkvec{0}{-\sem{\tm{O.height}}/2} \tm{ relative to } O} \\
\sem{\tm{left of } O} &= \sem{\mkvec{-\sem{\tm{O.width}}/2}{0} \tm{ relative to } O} \\
\sem{\tm{right of } O} &= \sem{\mkvec{\sem{\tm{O.width}}/2}{0} \tm{ relative to } O} \\
\sem{\tm{front left of } O} &= \sem{\mkvec{-\sem{\tm{O.width}}/2}{\sem{\tm{O.height}}/2} \tm{ relative to } O} \\
\sem{\tm{back left of } O} &= \sem{\mkvec{-\sem{\tm{O.width}}/2}{-\sem{\tm{O.height}}/2} \tm{ relative to } O} \\
\sem{\tm{front right of } O} &= \sem{\mkvec{\sem{\tm{O.width}}/2}{\sem{\tm{O.height}}/2} \tm{ relative to } O} \\
\sem{\tm{back right of } O} &= \sem{\mkvec{\sem{\tm{O.width}}/2}{-\sem{\tm{O.height}}/2} \tm{ relative to } O}
\end{align*}
\caption{\texttt{OrientedPoint} operators.}
\label{figure:semantics-op-ops}
\end{figure}

\clearpage

\twocolumn

 \section{Additional Experiments} \label{sec:more-experiments}


This section gives additional details on the experiments
and describes an experiment analogous to that of Sec.~\ref{sec:experiment-two} but using the generic two-car \scenic{} scenario as a baseline.

\subsubsection*{Additional Details on Experimental Setup}

Since GTAV does not provide an explicit representation of its map,
we obtained an approximate map by processing a bird's-eye schematic view of the game world\footnote{\url{https://www.gtafivemap.com/}}.
To identify points on a road, we converted the image to black and white, effectively turning roads white and everything else black.
We then used edge detection to find curbs, and computed the nominal traffic direction by finding for each curb point $X$ the nearest curb point $Y$ on the other side of the road, and assuming traffic flows perpendicular to the segment $XY$ (this was more robust than using the directions of the edges in the image).
Since the resulting road information was imperfect, some generated scenes placed cars in undesired places such as sidewalks or medians, and we had to manually filter the generated images to remove these.
With a real simulator, e.g. Webots, this is not necessary.

We now define in detail the
metrics used to measure the performance of our models.
Let $\hat{\vy} = f(\vx)$ be the prediction of the model $f$ for input $\vx$.
For our task, $\hat{\vy}$ encodes bounding 
boxes, scores, and categories predicted by $f$ for the image $\vx$.
Let $B_{gt}$ be a ground truth box (i.e. a bounding box from the label of a training sample that indicates the position of a particular object) and $B_{\hat{\vy}}$ be a box predicted by the model.
The \emph{Intersection over Union} (IoU) is defined as $\iou{B_{gt}}{B_{\hat{\vy}}} = \area{B_{gt} \cap B_{\hat{\vy}}} / \area{B_{gt} \cup B_{\hat{\vy}}}$, where $\area{X}$ is the area of a set $X$.
IoU is a common evaluation metric used to measure how well
predicted bounding boxes match ground truth boxes.
We adopt the common practice of considering $B_{\hat{\vy}}$ a \emph{detection} for $B_{gt}$ if $\iou{B_{gt}}{B_{\hat{\vy}}} > 0.5$.

\emph{Precision} and \emph{recall} are metrics used to measure the accuracy of a prediction on a particular image. 
Intuitively, precision is the fraction of predicted boxes that are correct, while 
recall is the fraction of objects actually detected.
Formally, precision is defined as $tp / (tp + fp)$ and recall as $tp / (tp + fn)$, where
\emph{true positives} $tp$ is the number of correct detections,
\emph{false positives} $fp$ is the number of predicted boxes that
do not match any ground truth box, and \emph{false negatives} $fn$ is the 
number of ground truth boxes that are not detected.
We use \emph{average precision} and \emph{recall} to evaluate the performance of a model on a collection of images constituting a test set.

\subsubsection*{Overlapping Scenario Experiments}

In Sec.~\ref{sec:experiment-two} we showed how we could improve the performance of squeezeDet trained on the Driving in the Matrix dataset~\cite{johnson2017driving} by replacing part of the training set with images of overlapping cars.
We used the standard precision and recall metrics defined above; however, \cite{johnson2017driving} uses a different metric, AP (which stands for Average Precision, but is \emph{not} simply the average of the precision over the test images).
For completeness, Table~\ref{tab:matrix-mixtures-ap} shows the results of our experiment measured in AP (as computed using \cite{ap-calculator}).
The outcome is the same as before: by using the mixture, performance on overlapping images significantly improves, while performance on the original dataset is unchanged.

\begin{table}[tb]
	\caption{Average precision (AP) results for the experiments in Table~\ref{tab:matrix-mixtures}.}
	\centering
	\label{tab:matrix-mixtures-ap}
	{\setlength{\tabcolsep}{10pt}
	\begin{tabular}{c  c c}
		\toprule
		Training Data & \multicolumn{2}{c}{Testing Data} \\
		$\xmatrix$ / $\xover$ & $\tmatrix$ & $\tover$ \\
		\midrule
		100\% / 0\% & $36.1 \pm 1.1$ & $61.7 \pm 2.2$ \\
		95\% / 5\% & $36.0 \pm 1.0$ & $65.8 \pm 1.2$ \\
		\bottomrule
	\end{tabular}
	}
\end{table}

For a cleaner comparison of overlapping vs. non-overlapping cars, we also ran a version of the experiment in Sec.~\ref{sec:experiment-two} using the generic two-car \scenic{} scenario as a baseline.
Specifically, we generated 1,000 images from that scenario, obtaining a training set \xtwocar{}.
We also generated 1,000 images from the overlapping scenario to get a training set \xover{}.

\begin{figure}[tb]
\begin{tikzpicture}
  \begin{axis}[
        ybar, axis on top,
        height=5cm, width=8.75cm,
        bar width=0.15cm,
        ymajorgrids, tick align=inside,
        enlarge y limits={value=.1,upper},
        ymin=0, ymax=3,
        axis x line*=bottom,
        axis y line*=left,
        y axis line style={opacity=0},
        tickwidth=0pt,
        enlarge x limits=true,
        legend style={
            at={(0.5,-0.15)},
            anchor=north,
            legend columns=-1,
            /tikz/every even column/.append style={column sep=0.25cm}
        },
        ylabel={$\log_{10}(\text{number of images})$},
        xlabel={IOU},
        xtick={0.00, 0.05, 0.10, 0.15, 0.20, 0.25,
           0.30, 0.35, 0.40, 0.45, 0.50},
        xticklabel style={
        		/pgf/number format/fixed,
        		/pgf/number format/precision=2
	}
    ]
    \addplot [draw=none, fill=black!70] coordinates {
      (0.00, 2.93) 
      (0.05, 1.61)
      (0.10, 1.48) 
      (0.15, 1.30) 
      (0.20, 1.08)
      (0.25, 1.08) 
      (0.30, 0.90)
      (0.35, 0.85)
      (0.40, 0.30)
      (0.45, 0.48)
      (0.50, 0.60) };
   \addplot [draw=none, fill=gray!60] coordinates {      
      (0.00, 1.612783857)
      (0.05, 1.908485019)
      (0.10, 2.139879086)
      (0.15, 2.247973266)
      (0.20, 2.23299611)
      (0.25, 2.113943352)
      (0.30, 2.056904851)
      (0.35, 1.875061263)
      (0.40, 1.612783857)
      (0.45, 1.380211242)
      (0.50, 0.602059991)      
      };
    \legend{\xtwocar, \xover}
  \end{axis}
  \end{tikzpicture}
  \caption{Intersection Over Union (IOU) distribution for two-car and overlapping training sets (log scale).\label{fig:iou_dist}}
\end{figure}
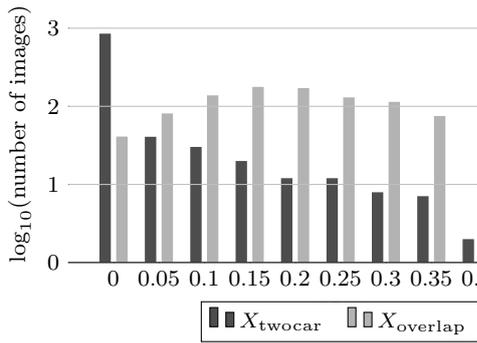

Note that \xtwocar{} did contain images of overlapping cars, since the generic two-car scenario does not constrain the cars' locations.
However, the average overlap was much lower than that of \xover{}, as seen in Fig.~\ref{fig:iou_dist} (note the log scale): thus the overlapping car images are highly ``untypical'' of generic two-car images.
We would like to ensure the network performs well on these difficult images by emphasizing them in the training set.
So, as before, we constructed various mixtures of the two training sets, fixing the total number of images but using different ratios of images from \xtwocar{} and \xover{}.
We trained the network on each of these mixtures and evaluated their performance on 400-image test sets \ttwocar{} and \tover{} from the two-car and overlapping scenarios respectively.

To reduce the effect of randomness in training, we used the maximum precision and recall obtained when training for 4,000 through 5,000 steps in increments of 250 steps.
Additionally, we repeated each training 8 times, using a random mixture each time: for example, for the 90/10 mixture of \xtwocar{} and \xover{}, each training used an independent random choice of which 90\% of \xtwocar{} to use and which 10\% of \xover{}.

As Tab.~\ref{tab:scenic-mixtures} shows, we obtained the same results as in Sec.~\ref{sec:experiment-two}: the model trained purely on generic two-car images has high precision and recall on $\ttwocar$ but has drastically worse recall on $\tover$.
However, devoting more of the training set to overlapping cars gives a large improvement to recall on $\tover$ while leaving performance on $\ttwocar$ essentially the same.
This again demonstrates that we can improve the performance of a network on difficult corner cases by using \scenic{} to increase the representation of such cases in the training set.

\begin{table}[tb]
	\caption{Performance of models trained on mixtures of $\xtwocar$ and $\xover$ and tested on both, averaged over 8 training runs.
	90/10 indicates a 9:1 mixture of $\ttwocar$/$\tover$.
	\label{tab:scenic-mixtures}}
	\begin{tabular}{c  c c  c c}
		\toprule
		& \multicolumn{2}{c}{$\ttwocar$} & \multicolumn{2}{c}{$\tover$}\\
		Mixture	& Precision & Recall &Precision & Recall\\	
		\midrule
		100/0 &	$96.5 \pm 1.0$ &	$95.7 \pm 0.5$ &	$94.6 \pm 1.1$ &	$\mathbf{82.1} \pm 1.4$\\
		90/10 &	$95.3 \pm 2.1$ &	$96.2 \pm 0.5$ &	$93.9 \pm 2.5$ &	$86.9 \pm 1.7$\\
		80/20 &	$96.5 \pm 0.7$ &	$96.0 \pm 0.6$ &	$96.2 \pm 0.5$ &	$\mathbf{89.7} \pm 1.4$\\
		70/30 &	$96.5 \pm 0.9$ &	$96.5 \pm 0.6$ &	$96.0 \pm 1.6$ &	$90.1 \pm 1.8$\\
		\bottomrule
	\end{tabular}
\end{table}

\end{document}